\documentclass[fleqn,usenatbib]{mnras}

\usepackage[T1]{fontenc}
\usepackage{ae,aecompl}

\usepackage{bm}

\usepackage{graphicx}	
\usepackage{amsmath}	
\usepackage{amssymb}	

\usepackage{newtxtext,newtxmath}

\renewcommand{\vec}{\bm}

\newcommand*{\refeq}[1]{Eq.~\ref{#1}}
\newcommand*{\reffig}[1]{Fig.~\ref{#1}}

\newcommand*{\rhs}{right~hand~side}
\newcommand*{\dir}[1]{\ensuremath{\MakeLowercase{#1}}} 
\newcommand*\Nabla{\nabla}
\newcommand*\Laplace{\Delta}
\newcommand*{\e}[1]{\ensuremath{\times\,10^{#1}}}
\newcommand*{\T}[1]{\ensuremath{{t={#1}\,\mathrm{Myr}}}}
\newcommand*{\B}[1]{\ensuremath{{B_0={#1}\,\umu\mathrm{G}}}}
\newcommand*{\CF}[1]{\ensuremath{\text{CF-B}#1}}
\newcommand*{\codename}[1]{\textsc{#1}}

\newcommand*{\rev}[1]{{#1}} 
\newcommand*{\revm}[1]{{#1}} 
\newcommand*{\revb}[1]{{#1}} 
\newcommand*{\revc}[1]{{#1}} 
\newcommand*{\revd}[1]{{#1}} 
\newcommand*{\reve}[1]{{#1}} 

\newcommand*{\simcls}{{\revc{3D-clumps}}}
\newcommand*{\simcos}{{\revc{3D-cores}}}

\title[Properties of clumps/cores in colliding flows]{Properties of Molecular Clumps and Cores in Colliding Magnetized Flows}

\author[Weis et al.]{
M. Weis,$^{1}$\thanks{E-mail: weis@ph1.uni-koeln.de}
S. Walch,$^{1,2}$
D. Seifried$^{1,2}$
S. Ganguly$^{1}$
\\
$^{1}$I. Physikalisches Institut, Universit\"at zu K\"oln, Z\"ulpicher Str. 77, 50937 K\"oln, Germany\\
$^{2}$Center for Data and Simulation science (CDS), University of Cologne, www.cds.uni-koeln.de\\
}

\date{Released 2024}

\pubyear{2024}

\begin{document}
\label{firstpage}
\pagerange{\pageref{firstpage}--\pageref{lastpage}}
\maketitle

\begin{abstract}
We simulate the formation of molecular clouds in colliding flows of warm neutral medium with the \rev{adaptive mesh refinement} code {\sc Flash}
\reve{in eight simulations with varying \rev{initial magnetic field strength,} between {0.01 -- 5~$\mu$G}}.
We include a chemical network to treat heating and cooling and to follow the formation of molecular gas.
The initial magnetic field strength influences the \rev{fragmentation of} the forming cloud because it prohibits motions perpendicular to the field direction and
hence \reve{impacts} the formation of \rev{large-scale} filamentary structures.
Molecular clump and core formation occurs anyhow.
\revc{We identify 3D-clumps and 3D-cores, which are defined as connected, CO-rich regions.
Additionally, 3D-cores are heavily shielded.}
\revd{While we do not claim those 3D-objects to be directly comparable to observations, this enables us to analyse their full virial state.}
With increasing field strength, we find more fragments with a smaller average mass;
yet the \rev{dynamics of the forming clumps and cores only weakly depends on the initial magnetic field strength.}
The molecular clumps are mostly unbound, probably transient objects, which \reve{are} weakly confined by ram pressure or thermal pressure,
indicating that they are swept up by the turbulent flow.
They experience significant fluctuations in the mass flux through their surface,
\reve{such} that the Eulerian reference frame \reve{shows} a dominant time-dependent term due to their \reve{indistinct} nature.
We define the cores to encompass \rev{highly shielded molecular gas}.
Most cores are in gravitational-kinetic equipartition and \reve{are well} described by the common virial parameter $\alpha_\mathrm{vir}$\rev{,}
while some undergo minor dispersion by kinetic surface effects.

\end{abstract}




\section{Introduction}
\label{sec:Intro}
Star formation takes place in dense and cool molecular clouds \citep[MCs; e.g.][]{mckee2007theory}.
Generally, MCs consist of a complex network of filaments \citep{schneider1979catalog,PPVII},
which harbour clumps and cores \citep{andre2014filamentary}.
Understanding the structural and chemical evolution of these clumps and cores is essential for understanding the conditions of star formation.

The empirical description of the cloud substructure using clumps and cores goes back to \citet{williams1999structure}.
Clumps are defined based on their velocity coherence in molecular line observations \citep[for instance by using the bright carbon-monoxide molecule, CO, see e.g.][]{loren1989cobwebs,williams1994determining},
while cores are \rev{rotating,} gravitationally bound, \rev{non-axisymmetric} \citep{Pineda2022PPVIIBubbles} objects,
\revc{usually} with a single maximum in column density \citep{bergin2007cold}.
Cores are hence proposed to gravitationally collapse,
\rev{producing protostellar systems with ringed disks and infalling streamers \citep{Pineda2022PPVIIBubbles},
and subsequently} individual stars or small-N multiples \citep{mckee2007theory,walch2010protostellar,holman2013mapping}.
\revb{An overview of typical properties for clumps and cores can be found in \cite{bergin2007cold}.}
\revb{They define as "gravitationally bound, single-peaked regions out of which individual stars (or simple stellar systems) form".}
\revc{
Recent interferrometric observations of \revd{high-mass cores} draw a geometrically more complex picture,
in which the shape of core scale fragments is related to the choice of the detection method \citep[see e.g.][]{Kong2019CMF, Beuther2023Sf}.
}

MCs are also observed to be very turbulent \citep{zuckerman1974models,larson1981turbulence, ballesteros2007protostars} as well as threaded by magnetic fields.
Most MCs have a comparable turbulent kinetic energy and magnetic energy \citep{crutcher2012magnetic}.
Hence, the impact of both, turbulence and magnetic fields, on the evolution of MCs ought to be considered. 

Traditionally, key parameters to characterise the energetic state of \rev{observed} MC's are their size, velocity dispersion, and surface density.
Although these parameters do not represent impeccable information on a MC's configuration,
they are to some degree observationally accessible.
\citet{larson1981turbulence} describes relations regarding those parameters,
with Larson's linewidth-size relation ($\sigma\propto R^{0.38}$) offering a criterion for kinetic-gravitational equipartition.
Larson's relations were later refined by \citet{heyer2004universality}.
\rev{
\citet[]{heyer2009re} show, supported by data on gravitationally bound clouds, that there is a systematic variation of the scaling coefficient $\sigma R^{-1/2}$ with the surface density.
}

On the other hand, simulated MCs provide complete knowledge on their configuration, including the PPP-space mass distribution as well as velocity, magnetic, and gravitational fields.
This allows for making exhaustive use of the virial theorem, which is a tool to quantify their energetic state \citep{chandrasekhar1953problems}.
It can be written in Lagrangian as well as Eulerian form.
In this paper we will follow the general Eulerian formulation of \citet[][]{mckee1992virial} (see Sec.~\ref{sec:VirTh}), which suits our numerical simulations.

By comparing hierarchical structures identified from PPP- and PPV-space dendrograms, \cite{beaumont2013projection} show that observational projection effects can lead to a misinterpretation of the virial parameter.
Further, \citet{mao2020cloud} investigate the significance of including thermal and magnetic energy when calculating virial parameters of objects defined by density thresholds.
They find that the virial parameter can be unreliable in assessing the boundedness of a structure, as it neglects non-spherical geometry, internal stratification, and tidal forces.

\rev{
Lifetimes of MCs are short, typically  10--30 megayears \citep{Corbelli2017Cycle,chevance2020lifecycle},
while massive star formation therein is limited to an even shorter timespan of a few megayears before dispersal of the cloud \citep{elmegreen2000star,chevance2020lifecycle}.
}
Numerical simulations show that the molecular hydrogen ($\mathrm{H}_{2}$) content of MCs can rapidly form from the interstellar medium (ISM) under supersonically turbulent conditions \citep{glover2007simulating2}.
The so-called colliding-flow scenario \citep[e.g.][]{walder1998formation,heitsch2005formation,audit2005thermal,inoue2008two,heitsch2011flow, kortgen2015impact, valdivia2016h2,joshi2018res}, where two supersonic gas flows collide and form a shocked central layer,
offers a simple model for the rapid creation of thermal and dynamical instabilities driving turbulence \citep{heitsch2008fragmentation}.
It thereby provides a mechanism that generates structures resembling MCs in the collision zone.
In this process, pockets containing very high $\mathrm{H}_{2}$ fractions are formed, 
followed by the formation of dense cores which contain CO molecules \citep{clark2012long}.

Here, we study the formation of MC clumps and cores in colliding flows of warm, atomic gas with different magnetization by means of three-dimensional MHD simulations with the adaptive mesh refinement (AMR) code {\sc Flash}.
We include self-gravity, an external interstellar radiation field (ISRF), which is attenuated using a tree-based radiative transfer method \citep{wunsch2018},
a uniform Cosmic Ray (CR) ionization rate, and a simple chemical network to follow the formation of H$_2$ and CO.
The energy balance of the forming clumps and cores is analysed in detail in order to determine the relative importance of the different physical processes. 

\cite{dib2007virial} have carried out a similar study.
However, their work differs in scope by a more thorough focus on magnetic fragmentation,
while their underlying simulations \citep{vazquez2005lifetimes} are based on isothermal,
supersonically driven turbulent boxes with a resolution of $\sim\,0.0156\,\mathrm{pc}$, and without any gas chemistry.
They observe clumps and cores to be transient, out-of-equilibrium structures, which are dynamically created from their turbulent environment.
\rev{
Some observations investigate the role of external pressure in confining the cores.
\citet{Pattle2015Virial} find gravitational energy and external pressure to be on average of a similar order of magnitude for starless cores for SCUBA-2 souces in Ophiuchus.
}
\citet{Kerr2019Virial} perform a virial analysis on starless cores observed in three nearby star-forming regions,
suggesting that many of those cores require confinement by external pressure to remain bound.
\rev{
\citet{Chen2019Droplets} find small coherent structures ("droplets") in the L1688 region of Ophiuchus and the B18 region in Taurus that are not virially bound,
but pressure confined.
}
\citet{Shimoikura2019Survey} observe molecular cores in M17 SWex.
They find that many massive cores are dynamically stable without external pressure, while some cores are confined and highly influenced by external pressure.
\rev{\citet{Offner2022Phases}} study dense cores in MHD simulations using machine learning techniques,
identifying an successive evolutionary sequence of first forming turbulent density structures,
followed by the dissipation of turbulence and the formation of coherent cores and finally the transition to protostellar cores through gravitational collapse.

Modelling non-equilibrium chemistry enables us to define MC substructures based on the CO content in our simulations.
As CO is also a widely used observational tracer for such structures,
this allows us to define clumps and cores which are likely to resemble their observed counterparts,
while we also retain the information needed to apply the full virial theorem as recapitulated in Section~\ref{sec:VirTh}.
The paper is structured as follows:
In Section~\ref{sec:Methods} we summarise the simulation setup and initial conditions and describe our methods to find the forming clumps and cores.
In Section~\ref{sec:Results} we present the results of analysing those substructures. We investigate the Larson relations and the Virial Theorem.
In Section~\ref{sec:Discussion}, we discuss the implications of the time-dependent component of the Eulerian Virial Theorem for our results and investigate the state of the identified cores.
We conclude in Section~\ref{sec:Conclusions}.

\section{Recapitulation of the Virial Theorem}
\label{sec:VirTh}
We use the full virial theorem to assess the virial state of MC substructures.
For this, we use the Eulerian formulation of the virial theorem given by \cite{mckee1992virial},
as this formulation is applicable for the AMR data of our simulations. This formulation has previously been used by \cite{dib2007virial} to study the energetics of MC cores formed in driven turbulent periodic boxes.

The Eulerian Virial Theorem (EVT) can be written as
\begin{align}
  &\frac{1}{2}\left(\ddot{I}_\mathrm{E}+\dot{\Phi}_I\right) = W\,+E_{\mathrm{MHD}} \, ,
  \label{eq:EVT}
  \\
  &E_{\mathrm{MHD}} = 
  2\left(\varepsilon_\mathrm{kin} -\tau_\mathrm{kin}\right) +2\left(\varepsilon_\mathrm{th} -\tau_\mathrm{th}\right) \,+\left(\varepsilon_\mathrm{mag} +\tau_\mathrm{mag}\right) \, ,
  \label{eq:EMHD}
\end{align}
where $I_\mathrm{E}$ corresponds to the trace of the moment of inertia tensor $I$:
\begin{equation}
  I_\mathrm{E} = \int_V \rho\vec{r}^2 ~\mathrm{d}V = \revc{\frac{1}{2}}~\mathrm{Tr}\left(I\right) \, ,
  \label{eq:I_E}
\end{equation}
with components
\begin{equation}
  I_{ij} = \int_V \rho(\delta_{ij}\vec{r}^2 -r_i r_j) ~\mathrm{d}V \, ,
\end{equation}
where $\rho$ is the mass density and $r_i$ is the $i$-th component of the distance vector relative to the object's centre of mass. Hence, $\ddot{I_\mathrm{E}}$ denotes the second time-derivative of $I_\mathrm{E}$.
Furthermore, $\dot{\Phi}_I$ is the time derivative of $\Phi_I$, which denotes the flux of $I_\mathrm{E}$ through the surface of a given structure with a fixed volume $V$,
\begin{equation}
  \Phi_I = \oiint_{S} \rho\vec{r}^2 v_i \hat{n}_i~\mathrm{d}S \, .
  \label{eq:ES_Phi}
\end{equation}
We emphasise that $\dot{\Phi}_I$ only occurs in \refeq{eq:EVT} due to assuming the Eulerian reference frame.
On the \rhs, $W$ is the gravitational binding energy,
\begin{equation}
    W = -\int_V \revm{\rho}\vec{r}\cdot\Nabla\phi_G~\mathrm{d}V \, ,
    \label{eq:EV_W}
\end{equation}
which is calculated using the global gravitational potential $\phi_G$.
Therefore, in addition to the self-gravity of an object, $W$ also accounts for the tidal forces introduced by its surrounding mass distribution.
Volume energy terms in \refeq{eq:EVT} are identified by $\varepsilon$, where
$\varepsilon_{\mathrm{th}}$, $\varepsilon_{\mathrm{kin}}$, and $\varepsilon_{\mathrm{mag}}$
are the thermal, kinetic, and magnetic volume energies, respectively, given as
\begin{align}
  &\varepsilon_{\mathrm{th}} = \frac{3}{2}\int_V P~\mathrm{d}V\, ,
  \label{eq:EV_th}
\\
  &\varepsilon_{\mathrm{kin}} = \frac{1}{2}\int_V \rho\vec{v}^2 ~\mathrm{d}V \, ,
  \label{eq:EV_kin}
\\
  &\varepsilon_{\mathrm{mag}} = \frac{1}{8\pi}\int_V \vec{B}^2 ~\mathrm{d}V \, .
  \label{eq:EV_mag}
\end{align}
Here, $P$ is the thermal pressure, $\vec{v}$ is the velocity vector, and $\vec{B}$ is the magnetic field vector of each cell inside $V$.
Surface terms in \refeq{eq:EVT} are denoted as $\tau$, where
$\tau_{\mathrm{th}}$, $\tau_{\mathrm{kin}}$, and $\tau_{\mathrm{mag}}$
are the thermal, kinetic, and magnetic surface terms, respectively. They are defined as
\begin{align}
  &\tau_\mathrm{th} = \frac{1}{2}\oiint_{S_{\left(V\right)}} P r_j \hat{n}_j~\mathrm{d}S \, ,
  \label{eq:ES_th}
\\
  &\tau_\mathrm{kin} = \oiint_{S_{\left(V\right)}} r_i K_{ij} \hat{n}_j~\mathrm{d}S \, ,
  \label{eq:ES_kin}
\\
  &\tau_\mathrm{mag} = \oiint_{S_{\left(V\right)}} r_i T_{ij} \hat{n}_j~\mathrm{d}S \, .
  \label{eq:ES_mag}
\end{align}
Here, $\hat{n}_j$ is the outward pointing normal vector to the surface $S$ of $V$.
The kinetic tensor, $K$, in \refeq{eq:ES_kin} is given by
\begin{equation}
K = \frac{1}{2}\rho\vec{v}\otimes\vec{v} \, ,
\end{equation}
where $\otimes$ denotes the outer product. The Maxwell stress tensor in \refeq{eq:ES_mag}, $T$, is given by:
\begin{equation}
T = \frac{1}{4\pi}\left(\vec{B}\otimes\vec{B} -\frac{1}{2}\vec{B}^2\hat{I} \right) \, ,
\end{equation}
with the 3D identity matrix $\hat{I}$.
Since it is easier to compute volume terms on the 3D grid, we convert the surface terms (Eqs.~\ref{eq:ES_Phi}, \ref{eq:ES_th}, \ref{eq:ES_kin}, and~\ref{eq:ES_mag})
to a volume integral formulation by applying Gauss's law to the corresponding surface integrals.
This is done to avoid the potentially intricate evaluation of surface quantities of the cells selected from the adaptive mesh. This conversion yields the following volume integrals:
\begin{align}
  &\Phi_I = \int_V 2\rho\vec{r}\cdot\vec{v} +\vec{r}^2\Nabla\cdot\left(\rho\vec{v}\right) ~\mathrm{d}V \, ,
  \label{eq:gs-SPhi}
\\
  &\tau_\mathrm{th} = \frac{1}{2}\int_V \vec{r}\cdot\Nabla P ~\mathrm{d}V +\varepsilon_\mathrm{th} \, ,
  \label{eq:gs-Sth}
\\
  &\tau_\mathrm{kin} = \int_V r_i \frac{\partial\,K_{ij}}{\partial r_j} ~\mathrm{d}V +\varepsilon_\mathrm{kin} \, ,
  \label{eq:gs-Sk}
\\
  &\tau_\mathrm{mag} = \int_V r_i \frac{\partial\,T_{ij}}{\partial r_j} ~\mathrm{d}V -\varepsilon_\mathrm{mag} \, ,
  \label{eq:gs-Smag}
\end{align}
where the addenda $+\varepsilon_\mathrm{th}$, $+\varepsilon_\mathrm{kin}$, and $-\varepsilon_\mathrm{mag}$ cancel out the respective volume terms in \eqref{eq:EVT}.
Note that in the course of the paper we use the surface term with their respective sign, as given in Eq. \ref{eq:EMHD}.
For example, $-\tau_\mathrm{kin}<0$ implies ram pressure confinement, while $\tau_\mathrm{mag}>0$ implies stabilisation by the magnetic surface term. 

\section{Numerical methods}
\label{sec:Methods}

\subsection{Simulations}
\label{subsec:sim}
Our numerical setup is derived from the colliding flows setup described by \cite{joshi2018res}.

As base architecture, we use the AMR code {\sc Flash} \citep{fryxell2000flash,dubey2008introduction} version 4.3.
We solve the ideal MHD equations using the Bouchut 5-wave MHD solver \citep{bouchut2010multiwave,waagan2011robust}, and divergence cleaning to enforce ${\Nabla\cdot\vec{B}=0}$.
The gravitational potential is calculated by solving the Poisson equation ${\Laplace\Phi_G=4\pi G\rho}$ by means of an octal-spatial tree method \citep{wunsch2018} with an opening angle $\theta_\mathrm{lim}=0.5$.

%
We use a simple chemical network \citep{nelson1997dynamics} 
to track the formation of H$_2$ \citep{glover2007star} and CO \citep{glover2010modelling} from the warm ISM.
In particular, we follow atomic (H), molecular (H$_2$), and ionized (H$^+$) hydrogen, ionized carbon (C$^+$), carbon-monoxide (CO), oxygen \rev{(O)}, and free electrons.
We assume solar fractional abundances for C and O of ${1.4\e{-4}}$ and ${3.2\e{-4}}$, respectively \citep{sembach2000modeling}.
Molecular hydrogen is forming on the surface of dust grains \citep{hollenbach1989molecule}.
We use a simple dust model based on a constant dust-to-gas mass ratio of 1:100.


For the photodissociation of $\mathrm{H}_2$, we use the model described in \citet{walch2015silcc} based on \citet{glover2010modelling}. The strength of the uniform ISRF is ${G_0=1.7}$ times the Habing flux.
The dust and gas temperatures are calculated when evaluating the chemical network. Typically, the dust is much colder than the gas (see below).
The utilized cooling model was introduced by \citet{glover2010modelling} and refined by \citet{glover2012molecular}.
It comprises the fine structure lines of $\mathrm{C}^+$, the rotational and vibrational lines of $\mathrm{H}_2$ and $\mathrm{OH}$,
the Lyman-$\alpha$ line of atomic $\mathrm{H}$, and the energy transfer from gas to dust.
In addition to the dissipation of kinetic energy into heat, the radiative heating model comprises CRs, soft X-rays, H$_2$ formation heating, and photoelectric emission from small grains and polycyclic aromatic hydrocarbons.
The CR ionization rate per hydrogen atom is assumed to be ${\zeta_{\mathrm{H}}=3\e{-17}\,\mathrm{s}^{-1}}$, 
with the resulting heating rate ${\Gamma=20\,\zeta_{\mathrm{H}}\times n\,\mathrm{erg}\,\mathrm{s}^{-1}\,\mathrm{cm}^{-3}}$.

The dust is assumed to be in thermal equilibrium.
The attenuation of the ISRF by dust is accounted for by applying an attenuation factor ${\chi\left(N_{\mathrm{H}}\right)}$
as given by \citet{glover2012molecular}. For this, the column density, $N_{\mathrm{H}}$, is estimated by means of the {\sc TreeRay/OpticalDepth} module \citep{wunsch2018},
\rev{being} a modification of the algorithm proposed by \citep{clark2012treecol}.

\begin{table*}
  \caption{
    Non-exhaustive list of similar simulations from different authors.
  }
  \label{tab:Collflows}
  \centering
  \rev{
  \begin{tabular}{lrrrrcc}
    \hline
    \hline
    Paper & $n_0\,\left[\mathrm{cm}^{-3}\right]$ & $v_\mathrm{inflow}\,\left[c_s\right]$ & $B_0\,\left[\umu\mathrm{G}\right]$ & $l_\mathrm{min}\,\left[\mathrm{pc}\right]$ & Sinks & non-eq. Chemistry \\
    \hline
        \cite{Heitsch2009Magnetic} & 1 & 2.2 & 0, 2.5, 5.0 & 0.18 & n & n \\
        \cite{Banerjee2009Morphology} & 1 & 1.25 & 1 & 0.03 & y & n \\
        \cite{Semadeni2011Evolution} & 1 & 2.44 & 2, 3, 4 & 0.03 & y & n \\
        \cite{clark2012long} & 1 & 1.2, 2.4 & - & SPH & y & NL99, Gl07 \\
        \cite{Inoue2012Formation} & 5.2 & - & 5 & 0.02 & n & $\mathrm{H}_\mathrm{x}$, $\mathrm{C}_\mathrm{x}$, $\mathrm{He}_\mathrm{-}$ \\
        \cite{kortgen2015impact} & 1 & 2 & 3, 4, 5 & 0.0078 & y & n \\
        \cite{valdivia2016h2} & 1 & 2 & 2.5 & 0.05 & n & $\mathrm{H}_\mathrm{2}$ \\
        \cite{Koertgen2016Supernova} & 1 & 2 & 3 & 0.03 & y & n \\
        \cite{joshi2018res} & 0.75 & 2.67 & - & 0.0078 & n & NL99 \\
        \cite{Zamora2018Magnetic} & 2 & 2.42 & 0, 1, 2, 3 & 0.03 & y & n \\
        \cite{Kobayashi2023Metallicity} & 0.57 & 3.2 & 1.0 & 0.02 & n & n \\
        this paper & 1 & 2.13 & 0.01, 1.25, 2.5, 5 & 0.0078 & n & NL97 \\
    \hline
    \hline
  \end{tabular}
  } 
\end{table*}
\rev{
We list the features of a non-exhaustive selection of similar simulations from other authors in Tab.~\ref{tab:Collflows},
\revb{which are frequently used to study the formation of molecular clouds.
There are also examples of prior studies which analyse the substructures found in molecular clouds from similar simulations \citep{Clark2006Clumpy,heitsch2011flow}.
Compared to similar simulations, }we combine a high maximum resolution of ${7.8\e{-3}\,\mathrm{pc}}$ with limited non-equilibrium chemistry.
This enables us to study the formation of molecular clouds as well as follow the formation of substructures (clumps and cores) embedded in these clouds.}
\revc{We note that our setup differs from studies of substructures in magnetized compressed sheets \citep{Chen2015Cores,Abe2021Filaments}
in that our magnetic field is initially not in the plane of the created sheet.
The details of our setup are discussed in Sec. \ref{subsec:IC},
the development of the magnetic field is indicated by Fig. \ref{fig:B-dir} in the appendix.
}
\rev{As we want to focus on the spatial distribution of the gas mass, we do not include a subgrid protostellar model that could remove mass into sink particles or skew the mass distribution inside its accretion range.}
\revc{We stop the simulations shortly after \T{20}.
Maintaining the simulations beyond this point would require us to either 
use unreasonable computational effort to further resolve the gravitational collapse in cores,
or to introduce sink particles and to consequently lose information on the mass distribution in such cores.
}

\subsection{Initial conditions}
\label{subsec:IC}
\begin{figure}
  \centering
  \includegraphics[width=\columnwidth]{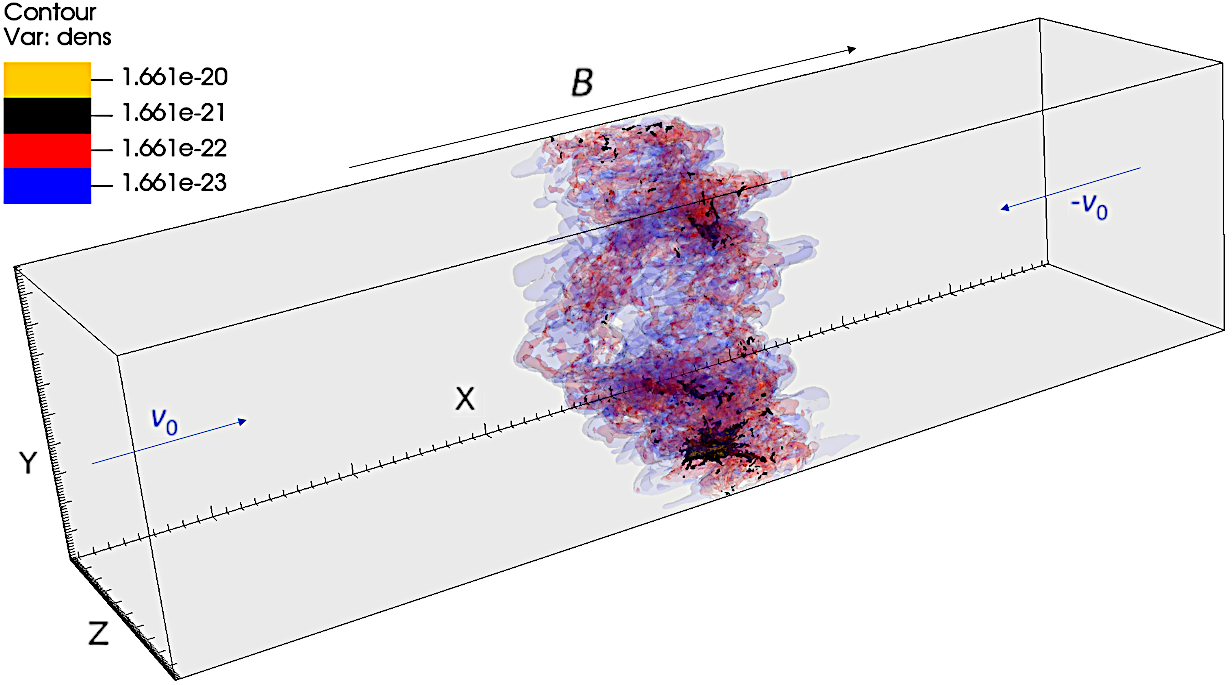}
  \caption{
  Layout sketch of the simulation setup.
  The simulation domain is an elongated cuboid measuring ${128\,\mathrm{pc}\times 32\,\mathrm{pc}\times 32\,\mathrm{pc}}$.
  Inflow and magnetic field are aligned with the \dir{x}-direction.
  The volume-rendering shows the turbulent density structure of the forming molecular cloud near the collision interface.
  The density is displayed in units of $\mathrm{g}\,\mathrm{cm}^{-3}$.
  }
  \label{fig:setup}
\end{figure}
We use the colliding flow setup employed by \cite{joshi2018res}, sketched in \reffig{fig:setup}.
The simulations use a rectangular domain measuring ${128\,\mathrm{pc}\times 32\,\mathrm{pc}\times 32\,\mathrm{pc}}$.
The simulation domain is resolved by an adaptive mesh with a maximum spatial resolution of ${7.8\e{-3}\,\mathrm{pc}}$ (i.e. $\sim$~1615~AU), which, as we have shown in \cite{joshi2018res}, enables a sufficiently converged model of CO formation.
The refinement/derefinement process is controlled by resolving each cells' local Jeans length by at least 8~times the cell size.

The gas filling the simulation domain has a uniform initial density of ${\rho_0=1.67\e{-24}\,\mathrm{g}\,\mathrm{cm}^{-3}}$,
a temperature of ${T_0=5541\,\mathrm{K}}$, and a chemical composition being in equilibrium at these conditions.
The initial conditions are designed to approximate the warm neutral ISM.

We apply periodic boundary conditions along the \dir{y}- and \dir{z}-directions for MHD as well as gravity.
Along the \dir{x}-direction, we apply isolated boundary conditions for gravity and an inflow MHD condition.
Hereby, a steady inflow of mass is supplied through the respective boundary surfaces.
The inflow and the corresponding initial \dir{x}-velocity inside the simulation domain are oriented \rev{inwards} towards a collision interface, separating the flow directions.
We set the initial velocity as well as the constant inflow velocity at the \dir{x}-boundaries to ${\left|v_{x,0}\right|=13.6\,\mathrm{km}\,\mathrm{s}^{-1}}$.
The initial collision interface is constructed from a plane at $\dir{x}=0$, that is then warped by the following $\dir{x}^\prime$-displacement to facilitate flow instabilities in its vicinity:
\begin{equation}
\dir{x}^\prime=A\left[\cos(2-\tilde{y}\tilde{z})\cos(k_y\tilde{y}+\phi)
+\cos(\frac{1}{2}-\tilde{y}\tilde{z})\sin(k_z\tilde{z})\right] \rm{.}
\label{eq:interface}
\end{equation}
\rev{
We execute two sets of simulations, one using low spatial frequencies of ${k_y=2}$, ${k_z=1}$ (set a) and the other one using increased spatial frequencies of ${k_y=10}$, ${k_z=5}$ (set b),
for the initial interface, each with the amplitude $A=1.63\,\mathrm{pc}$, and with set a corresponding to interface I5 in \citet{joshi2018res}.
}

\begin{table}
  \caption{
  \rev{List of initial conditions.}
  }
  \label{tab:sim-IC}
  \centering
  \rev{ 
  \begin{tabular}{lcc}
    \hline
    \hline
    Condition & Symbol & Initial Value\\
    \hline
    Density & $\rho_0$ & $1.67\e{-24}\,\mathrm{g}\,\mathrm{cm}^{-3}$\\
    Temperature & $T_0$ & $5541\,\mathrm{K}$ \\
    Speed of sound & $c_s$ & $6.04\,\mathrm{km}\,\mathrm{s}^{-1}$ \\
    Velocity & $v_{x,0}$ & $\pm 13.6\,\mathrm{km}\,\mathrm{s}^{-1}$ \\
    Magnetic Field & $B_{x,0}$ & $0.01,~1.25,~2.5,~5\,\umu\mathrm{G}$ \\
    Peak Resolution & $l_{\text{min}}$ & $7.8\e{-3}\,\mathrm{pc}$ \\    
    \hline
    \hline
  \end{tabular}
  }
\end{table}

\begin{table}
  \caption{
  \rev{
  List of simulations.
  The strength of the initial magnetic field is $B_0$.
  The time required by the central sheet to acquire sufficient mass to
  reach a critical mass-to-flux ratio is given by}
  $\tau_{\mathrm{B,crit}}$ \rev{(\refeq{eq:MFRrate})}.
  }
  \label{tab:sim-runs}
  \centering
  \rev{
  \begin{tabular}{lrrrr}
    \hline
    \hline
    Run & $B_0\,\left[\umu\mathrm{G}\right]$ & $k_y$ & $k_z$ & $\tau_{\mathrm{B,crit}}\,\left[\mathrm{Myr}\right]$\\
    \hline
    \CF{0.01}a & $0.01$ &  $2$ & $1$ &  $0.04$\\
    \CF{0.01}b & $0.01$ & $10$ & $5$ &  $0.04$\\
    \CF{1.25}a & $1.25$ &  $2$ & $1$ &  $5.37$\\
    \CF{1.25}b & $1.25$ & $10$ & $5$ &  $5.37$\\
    \CF{2.50}a & $2.50$ &  $2$ & $1$ & $10.74$\\
    \CF{2.50}b & $2.50$ & $10$ & $5$ & $10.74$\\
    \CF{5.00}a & $5.00$ &  $2$ & $1$ & $21.49$\\
    \CF{5.00}b & $5.00$ & $10$ & $5$ & $21.49$\\
    \hline
    \hline
  \end{tabular}
  }
\end{table}
\revb{
We apply an initially uniform magnetic field that is aligned with the \dir{x}-direction,
which results in it being initially aligned perpendicular to the collisional sheet created at the centre.
}
\rev{For each set of simulations,}
multiple \rev{runs} covering a range of initial magnetic field strengths from \B{0.01} to \B{5.0} are executed (see Table~\ref{tab:sim-runs}).
\rev{The initial critical field strength is given by ${B_\mathrm{crit}=2\pi\rho_0}L_{z}G^{1/2}=1.07\mu G$.}
\revb{
During the simulation, the magnetic field develops self-consistently, as solved by the MHD-equations (see \citet{bouchut2010multiwave}).
We discuss the development of the magnetic field in App. \ref{sec:magdir}.
\revd{
The resulting magnetic field is highly unordered in the core-scale high-density regime and therein in general 
neither parallel to the initial direction nor aligned to the surrounding field order.
This is similar to the behaviour observed by e.g. \citet{Arzoumanian2021Magnetic}.
}}
The magnetic flux in our simulated volume remains constant, while the mass is increasing due to the inflow boundary conditions.
The mass inflow per unit time is ${\dot{M}_0=2 v_{x,0}\cdot\rho_0\cdot (32\,\mathrm{pc})^2=703\,\mathrm{M}_\odot\,\mathrm{Myr}^{-1}}$.
The mass contained by the central dense sheet around the collisional interface grows approximately with the rate of inflow, as the sheet's mass is supplied from the inflow.
The critical mass-to-flux ratio of a flattened structure is ${\left(M/\Phi_{B}\right)_\mathrm{crit}=\left(4\pi^2 G\right)^{-1/2}}$ \citep{nakano1978gravitational}, where $\Phi_{B} = (32\,\mathrm{pc})^2 B_0$ is the magnetic flux in our simulations.
Hence, the time scale on which a magnetically critical amount of mass is acquired by the sheet via the mass inflow is
\begin{equation}
    \label{eq:MFRrate}
    \tau_{\mathrm{B,crit}} = \frac{\left(M/\Phi_{B}\right)_{\mathrm{crit}}}{ \left(\dot{M}_0/\Phi_{B}\right)} \approx 4.3\,\mathrm{Myr}\,\left(\frac{B_0}{1 \mu\mathrm{G}}\right)\,.
\end{equation}

\subsection{\revc{3D} Clump and core detection}
\label{subsec:Detection}

We consider two different types of \revc{three dimensional} molecular cloud substructures,
which we term \simcls{} and \simcos{}, the latter being a substructure of the first.

To identify clumps, we first select a set of suitable candidate cells from the \rev{simulations} AMR data.
All clump candidate cells must have a local fractional abundance of CO of ${\chi_{_\mathrm{CO}}>1\e{-4}}$, which corresponds to approximately 70\% of the carbon atoms being incorporated in CO molecules.
\revb{
In analogy to the Friends-of-Friends algorithm which is frequently used to identify clusters of particles in data from SPH simulations (see e.g. \citet{Walch2013Clumps,Dobbs2019M33}),
we apply a Neighbour-of-Neighbour algorithm on our selected AMR grid cells} to identify connected subsets amongst all clump candidate cells.
To extract connected structures, any two candidate cells are linked into the same group if they are direct neighbours of each other, that is they share at least a part of a cell face (on the 3D adaptive mesh neighbouring cells may have surface areas which differ by a factor 4).
We note that for the used AMR implementation in the {\sc Flash} code, neighbouring cells can never differ by more than one level of refinement, i.e. a factor of 2 in resolution.

\begin{figure}
  \centering
  \includegraphics[width=.8\columnwidth]{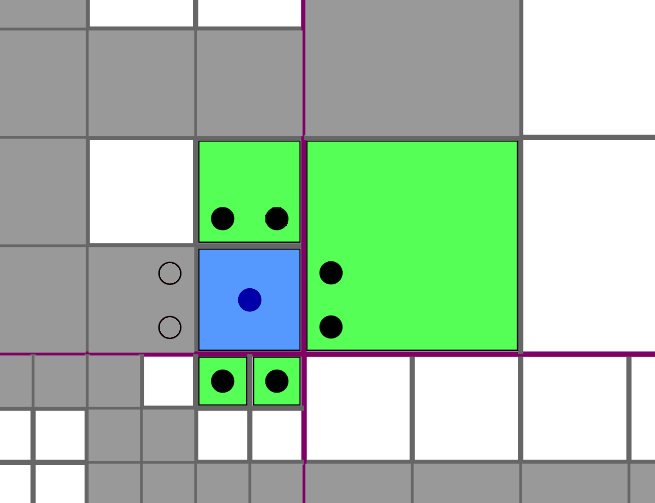}
  \caption[Sketch of Neighbourhood Discovery]{
    2D-sketch visualizing the 3D algorithm to identify neighbouring cells on the adaptive mesh. White and green cells fulfil all selection criteria, while grey cells are not considered further as they do not fulfil one or more criterion.
    The dark grey lines are cell boundaries, while the dark purple lines indicate cell boundaries where the resolution changes. The cell whose neighbours should be found is coloured in blue, its centre is marked by the dark blue circle.
    The other circles represent the spawned neighbour sampling points of the blue cell.
    Their positions are chosen such that also neighbouring cells on a higher refinement level would be properly sampled.
    The neighbour sampling points which are marked by filled circles are accepted, and the green cells are hence marked as valid neighbours.
    The process is then \revb{iterated} for all valid neighbours.
    In this way, connected groups, i.e. subsets of all candidate cells are identified in an iterative process.
    The groups which contain more than \rev{30}~cells are finally accepted as \simcls{} or \simcos{}, respectively.
  }
  \label{fig:cldetect/neighfind}
\end{figure}

An illustration of the corresponding search pattern for finding all neighbouring cells of a selected cell on the adaptive mesh is shown in \reffig{fig:cldetect/neighfind}. An iterative application of this simple algorithm then groups the connected cells into separate groups. A group is a set of cells, which are indirectly linked to each other by any number of neighbour-of-neighbour hops.
If a group contains at least \rev{30}~cells it is saved as a clump. Otherwise it is discarded, as we consider structures smaller than \rev{30}~cells to be insufficiently resolved for further analysis.

To identify \simcos{}, we use a modified version of the clump detection algorithm.
\revb{
As we want to limit ourselves to confined cores, we design our criterion to extract heavily shielded connected objects in order to filter out transient density fluctuations.
We note that this might lead to non-spheroidal, complex-shaped 3D objects in cases where commonly used 2D methods might retain a spheroidal outline or detect multiple spheroidal objects.
}
We again consider all cells with ${\chi_{_\mathrm{CO}}>1\e{-4}}$, as for the \simcls{}.
We then further narrow the selection to the subset of cells with a 3D visual extinction of ${A_{\mathrm{V,3D}}>8\,\mathrm{mag}}$,
above which the dense gas \rev{mass} has been found to scale linearly with the star formation rate \citep{Lada2010, Heiderman2010}. 
The same neighbour-based algorithm as for the \simcls{} is then used to identify \simcos{} from that narrowed selection.
Again, we discard structures smaller than \rev{30}~cells.
The method that we use to calculate ${A_{\mathrm{V,3D}}}$ is discussed in \rev{Appendix~\ref{sec:TreerayOpticalDepth}}.

\rev{We note that we tested different methods to identify our cores.}
One approach is to replace the ${A_{\mathrm{V,3D}}}$ threshold by an gravitational acceleration threshold of $\left|a_{G}\right|>5\e{-8}\,\mathrm{cm}\,\mathrm{s}^{-2}$.
It is motivated by observing the threshold above which the gravitational acceleration is well correlated to the gas velocity, as could be expected for gravitationally dominated substructures with little turbulence.
Obviously, this does not capture the inner part of some cores, as their gravitational field is not sufficiently sloped in vicinity of gravitational field minima.
We remedy this by subsequently removing cavities inside the identified cores.
This method, while vastly more complicated, performs similar to the adopted method described above.
We also tried to detect cores by different density-thresholds, which interestingly failed to identify gravitationally bound structures.

\subsubsection{\revc{Detection of 2D sources}}
\label{subsubsec:detect-2d}
\rev{
To estimate what \simcos{} created by the conditions in our simulations would look like when observed (in 2D),
we apply the \textsc{getsf} \citep{men2021getsf} source detection on our simulations.
As a simple approximation, we first apply a Gaussian convolution with an FWHM of 2.5 pixel to the face-on (projection along the $x$-direction) $\mathrm{H}_\mathrm{2}$ column density map of each simulation.
This corresponds to a beam size of $11.4"$ at a distance of $500\,\mathrm{pc}$.
We then run \textsc{getsf} on the resulting column density map.
The output contains a source column density map as well as the subtracted background column density map.
We extract the detected sources, and calculate their $\mathrm{H}_\mathrm{2}$ mass, $M_{\mathrm{H2}}$, by integrating each objects' background-subtracted column density map.
\revb{The detected \textsc{getsf} sources for the \CF{X}a-runs are shown in \reffig{fig:coldens-x}, marked by gray overlaid circles of corresponding size.}
}

\rev{\begin{table}
  \caption{\rev{
    Number of \textsc{getsf}-sources, \revb{filtered \textsc{getsf}-sources}, and \simcos{} at \T{20}.
  }}
  \label{tab:Detect}
  \centering
  \rev{
  \begin{tabular}{lrrr}
    \hline
    \hline
    Run & \#\textsc{getsf}-src. & \revb{\#\textsc{getsf}-src.} & \#\simcos{} \\
    ~&~& \revb{(filtered)} &~\\
    \hline
    \CF{0.01}a & $340$ & \revb{ $53$} & $13$ \\
    \CF{0.01}b & $357$ & \revb{$109$} & $34$ \\
    \CF{1.25}a & $134$ & \revb{  $2$} &  $2$ \\
    \CF{1.25}b & $131$ & \revb{  $3$} &  $1$ \\
    \CF{2.50}a & $149$ & \revb{ $13$} & $10$ \\
    \CF{2.50}b & $331$ & \revb{ $35$} & $13$ \\
    \CF{5.00}a & $766$ & \revb{ $49$} & $26$ \\
    \CF{5.00}b & $322$ & \revb{  $6$} &  $6$ \\
    \hline
    \hline
  \end{tabular}
  }
\end{table}}

The \textsc{getsf} sources partly coincide with our \simcos{}.
\textsc{getsf} extracts many more (low-mass) sources, which represent small features within the column density map that are located in regions which do not contain any \simcos{}.
\revc{
We note however, that while we put no additional constraint on the \textsc{getsf} sources so far,
our 3D core detection is designed to disregard objects reaching only low visual extinction, thus discriminating against low-density objects that might potentially be transient.
Furthermore, we expect it to not pick up most low-mass \simcos{} below $1\,\mathrm{M}_\odot$,
as we enforce a limit on the minimum number of included grid cells, which in turn sets a lower limit to the core size, thus discriminating against low mass cores.
\revb{
Thus, we additionally consider a filtered subset of the \textsc{getsf} sources with the goal of imposing a comparable constraint on those objects.
We implement this by requiring the selected \textsc{getsf} sources to exceed an average density threshold 
of $\rho_\mathrm{avg} > 10^{-19}\,\mathrm{g}\,\mathrm{cm}^{-3}$,
which we derive from our grid data by finding the 3D density that is typically exceeded where our 3D core criterion of $A_{\mathrm{V,3D}>8\,\mathrm{mag}}$ is met.
The dependency between the 3D density and 3D visual extinction at \T{20} is shown in the appendix, Fig. \ref{fig:cdto-dens}.
We then integrate the source mass from the background-subtracted column density maps
and estimate the 3D-volume of each source by assuming the volume enclosed by a sphere of the same cross section as
\begin{eqnarray}
    V_\mathrm{src} = \frac{4}{3} \sqrt{\frac{A_\mathrm{src}^3}{\pi}} ~\mathrm{,}
    \label{eq:volest}
\end{eqnarray}
where $A_\mathrm{src}$ is the area covered by the source's footprint.
}}

\section{Results}
\label{sec:Results}

\subsection{Evolution of the sheet morphology \& molecular gas fraction}
\label{subsec:sim-morph}

As a function of time, a dense sheet builds up near the collision interface, where the gas is pushed together by the opposing flows of warm neutral medium. The increased density leads to a thermal instability and the gas starts to cool down and become molecular.
Further instabilities arise due to the irregular collision interface and turbulence develops \citep[see][for a discussion of the different instabilities arising in perturbed colliding flows]{heitsch2009gravitational}.
\begin{figure*}
  \centering
  \includegraphics[width=\textwidth]{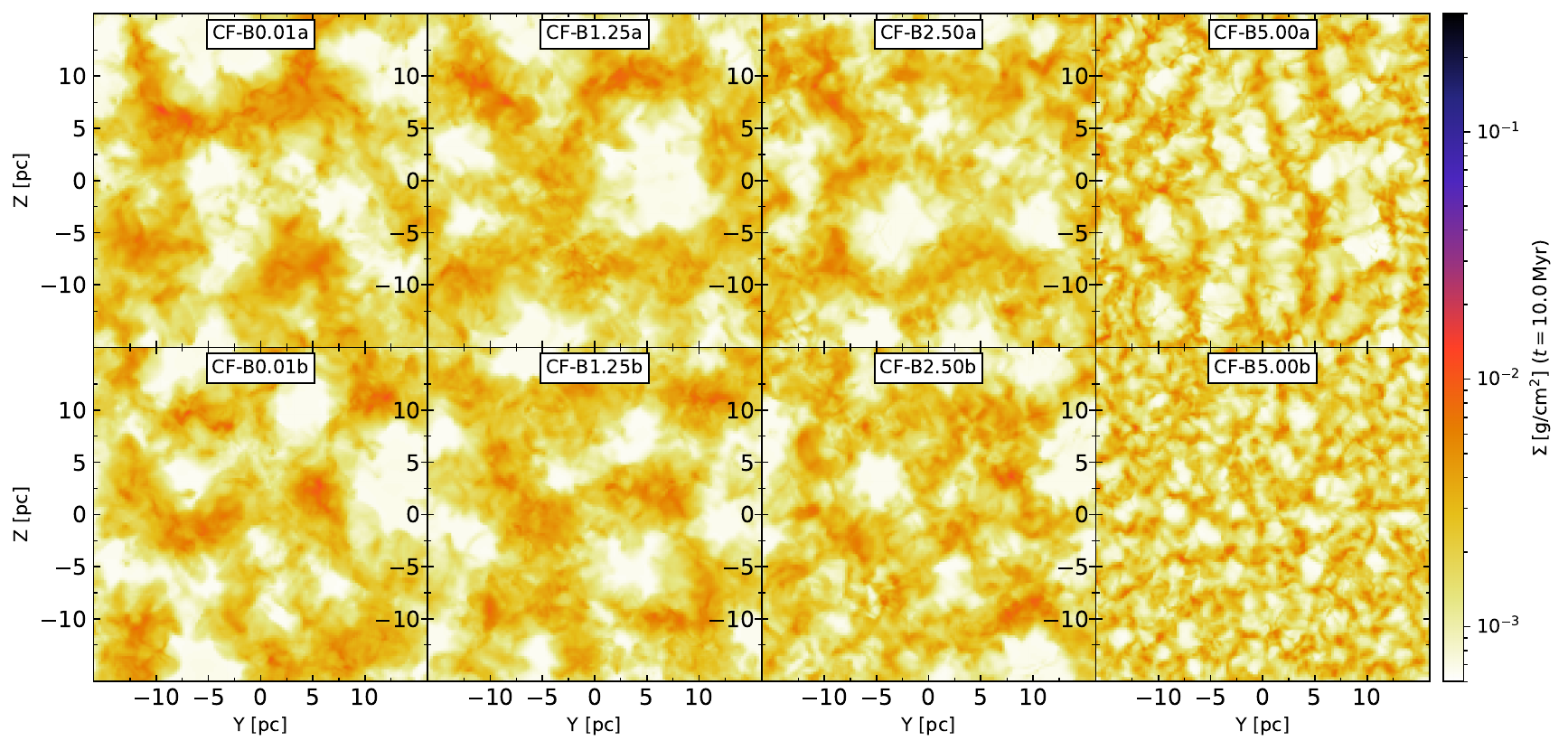}
  \includegraphics[width=\textwidth]{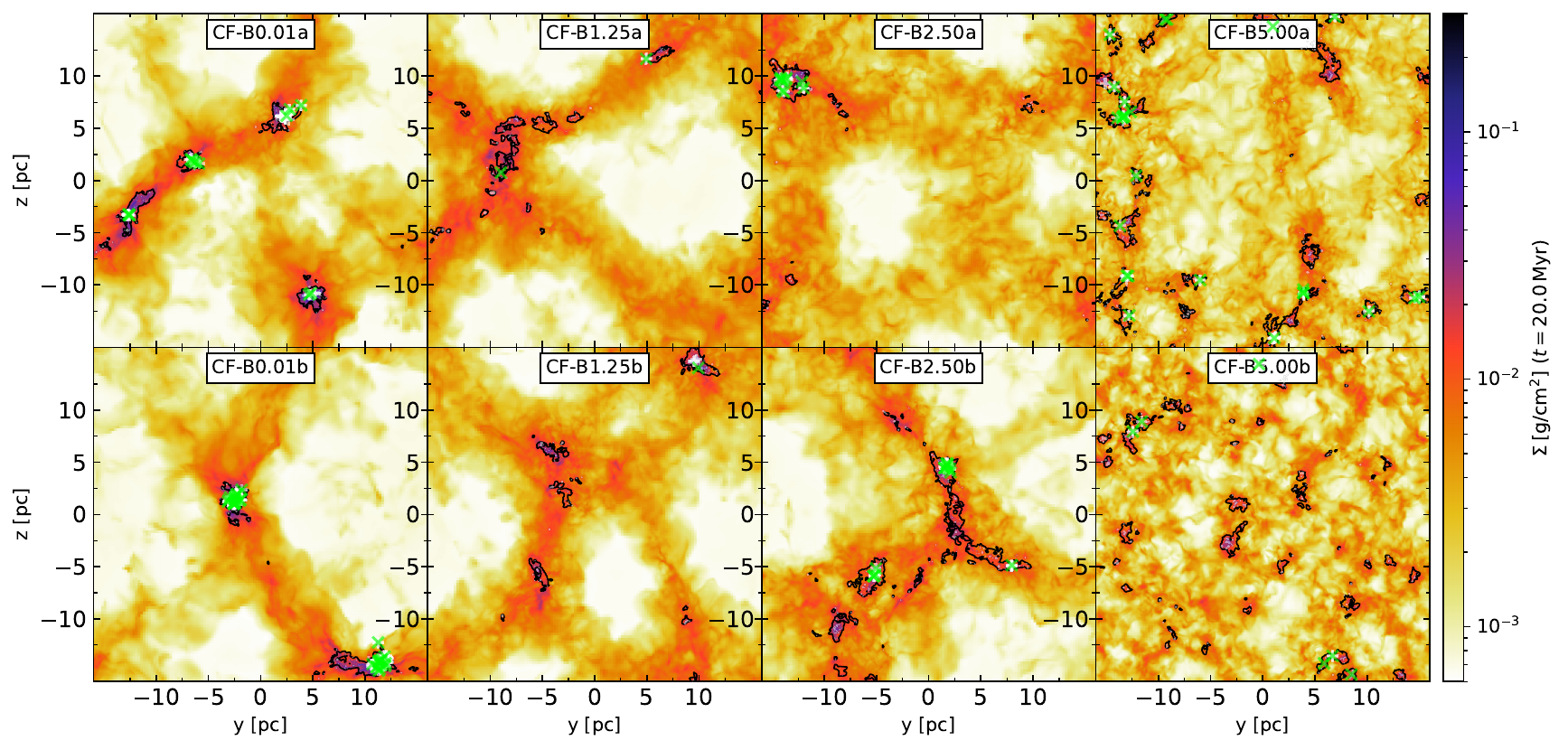}
  \caption{
    \rev{
    Face-on total gas column densities for all simulations at \T{10} (top panel) and \T{20} (bottom panel).
    The top row in each panel represent the simulation runs CFXa with the low spatial frequency interface,
    while the bottom row represent the simulation runs CFXb with the high spatial frequency interface.
    }
    \revb{
    The \simcls{} are indicated by the black contour, the \simcos{} are indicated by the green x markers.
    The \textsc{getsf} sources with an average density of $\rho_\mathrm{avg}>1\e{-19}\,\mathrm{g}\,\mathrm{cm}^{-3}$ are indicated by the white + markers.
    The positions of the filtered \textsc{getsf} sources and the \simcos{} mostly overlap.
    }
  }
  \label{fig:coldens-x}
\end{figure*}
The sheet's thickness and its density contrast, as well as the time of the MC buildup, depend on the initial magnetic field strength.
In \reffig{fig:coldens-x} we show the column density distribution of all simulations,
as integrated along the inflow direction \rev{at 10~Myr (top panel) and 20~Myr (bottom panel)}.
\rev{A supplemental set of $\mathrm{H}_\mathrm{2}$ and $\mathrm{CO}$ column density maps is contained in the online material.}

In the runs with a weak magnetic field (\rev{left}), a network of dense filamentary structures is forming.
Within the \rev{filamentary structures}, molecular clumps are slowly accumulated (indicated by the black contours).
\revc{Depending on the initial conditions, the first \simcls{} appear after $12$~Myr to $14$~Myr.}
With increasing magnetic field strength, the formed \rev{large-scale filamentary structures} are less pronounced \rev{because the magnetic field remains more ordered in this case and gas flows perpendicular to the magnetic field direction are suppressed}.
In particular, for simulations \CF{5.0}a/b (right column) the colliding flow approaches a thick, relatively uniform sheet, modulo smaller-scale density fluctuations (see also Fig.~\ref{fig:coldens-z}).
\revd{
Observations of clouds that at the same time are significantly magnetized and highly filamentary \citep[e.g.][]{andre2014filamentary,Planck2016Magnetic}
might indicate that the inflow pattern of those clouds is not fully aligned with their magnetic field.
This would lead to a less pronounced attenuation of the mass flux perpendicular to the inflow direction, hence on the formation of filamentary structures.
}

The resulting clumps are distributed more evenly within the $\dir{y}-\dir{z}$-plane than in the other runs.
The compact column density features are at least three orders of magnitude denser than the initial flow column density of $\sim 6.6\e{-4}\,\mathrm{g}\,\mathrm{cm}^{-2}$.
At later times, there is a significant fraction of gas above the suggested star formation threshold column density of $\Sigma \gtrsim 2.5 \times 10^{-2} \;\mathrm{g\; cm}^{-2}$
(corresponding to a visual extinction of ${A_{\mathrm{V,2D}} \gtrsim 8}$~mag,
above which the dense gas \rev{mass} has been found to scale linearly with the star formation rate \citep{Lada2010, Heiderman2010}.
Some structures even reach $\Sigma>1\;\mathrm{g\; cm}^{-2}$, which is the suggested column density threshold for massive star formation \citep{Krumholz2008} and which corresponds to the column density where the gas approximately becomes optically thick to its cooling radiation (if the geometry allows for it). 

\begin{figure*}
  \centering
  \includegraphics[width=\textwidth]{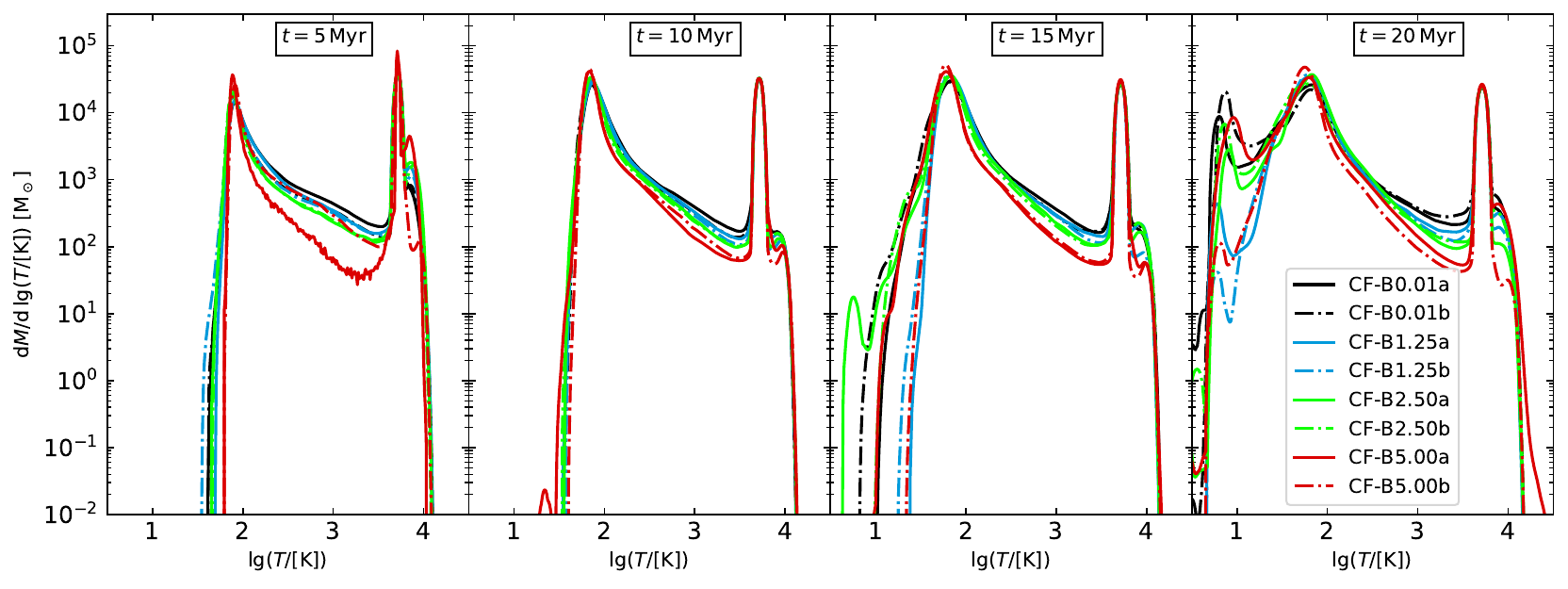}
  \caption{
    Mass-weighted temperature PDF of the simulation domain at 5, 10, 15 and 20~Myr.
    Thermal instability allows for less gas at intermediate temperatures of $300\,\mathrm{K}< T < 4000\,\mathrm{K}$
    when increasing the magnetic field strength.
  }
  \label{fig:pdf-temp-dens}
\end{figure*}
In Fig.~\ref{fig:pdf-temp-dens} we show the mass-weighted temperature probability distribution function (PDF) of all gas for the runs with different magnetic field strength at four different times (5--20~Myr).
The prominent peak at $T\approx 5500$~K shows the inflowing gas, and a second peak of cold gas develops at $T<100$~K.
For increasing magnetic field strength, gas motions perpendicular to the magnetic field direction are increasingly suppressed.
Thus, gas mixing is less efficient for the high magnetization simulations.
This leads to less gas in the thermally unstable regime (between $\sim$ 300--4000~K) \citep{seifried2011forced}.



\begin{figure}
  \centering
  \includegraphics[width=\columnwidth]{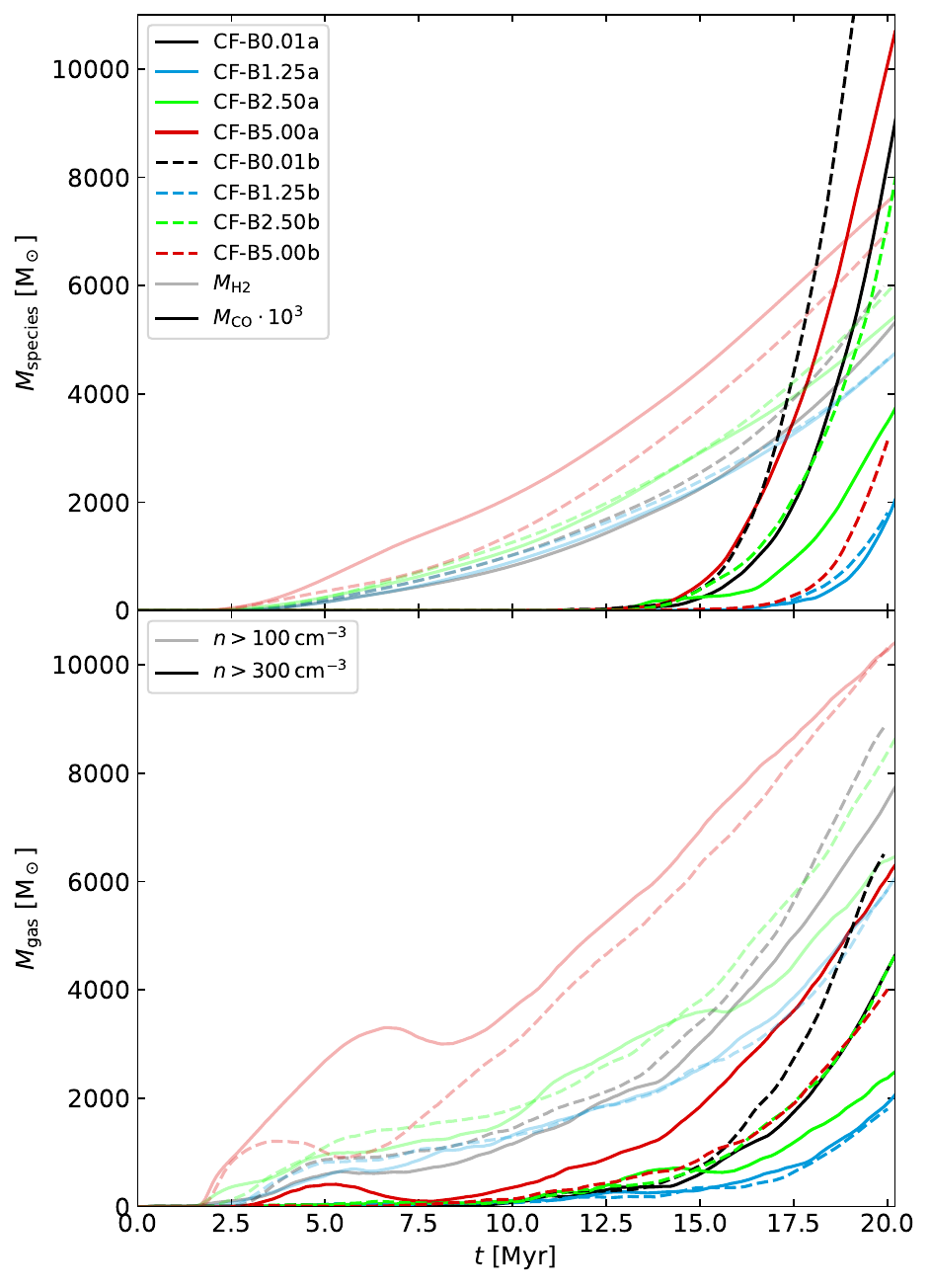}
  \caption{
    Top: Total H$_2$ mass and CO mass contained in the simulation domain versus time.
    H$_2$ formation begins at around \T{2.5}.
    CO formation begins at around \T{14}.
    Bottom: \rev{High-density gas mass} versus time;
    The rate of CO formation closely follows the amount of \rev{high-density gas} in the simulation domain.
  }
  \label{fig:sim-mass}
\end{figure}
The top panel of \reffig{fig:sim-mass} shows the evolution of the total amount of molecular mass in the simulation domain over time.
The formation of H$_2$ \rev{(light-coloured lines)} begins at around \T{2.5} in all simulations.
The amount of molecular gas differs because the sheets substructures are different, leading to different shielding properties
(see \reffig{fig:Av-pdf}, which shows the three-dimensional visual extinction $A_{\mathrm{V,3D}}$-PDF for the computational domain; 
the algorithm that is used to calculate $A_{\mathrm{V,3D}}$ for every grid cell is detailed in the Appendix~\ref{sec:TreerayOpticalDepth}).
\rev{In particular, runs \CF{5.0}a/b have a high molecular gas fraction},
as the almost uniform sheet protects the molecules inside from the dissociating radiation coming from all directions.
The formation of CO \rev{(top panel, dark-coloured lines)} begins later at around \T{14}.
The bottom panel shows the total amount of gas at number densities of $\revm{n_\mathrm{gas}}>100\,\mathrm{cm}^{-3}$ (dash-dotted lines) and $\revm{n_\mathrm{gas}}> 300\,\mathrm{cm}^{-3}$ (solid lines).
While there is no simple correlation between the CO formation and the ambient magnetic field strength, the mass of the formed CO follows the amount of dense gas above $n\gtrsim 300\,\mathrm{cm}^{-3}$,
like already observed by \citet{seifried2017silcc}.
CO formation is slowest for \rev{simulations \CF{1.25}a/b}.

Overall, the formation of CO sets in later than expected from the column density maps, which already show structures with $\Sigma \gtrsim 9\times 10^{-3}\;\mathrm{g\; cm}^{-2}$ at \T{5}.
The reason is that the projected column density always overestimates the local column density experienced by any given cell in the complex, fractal 3D structure \citep[see also][]{haid2019silcc}.
However, it is the latter one, $A_{\mathrm{V,3D}}$, which is relevant for the shielding and hence protection of the molecules from the dissociating interstellar radiation field.

\subsection{Formation of \simcls{} and \simcos{}}
\label{subsec:sim-clmorph}

We apply the clump and core detection algorithms described in Section~\ref{subsec:Detection}
on all simulation data snapshots, which have been recorded in steps of ${0.1\,\mathrm{Myr}}$.
\begin{figure}
  \centering
  \includegraphics[width=.98\columnwidth]{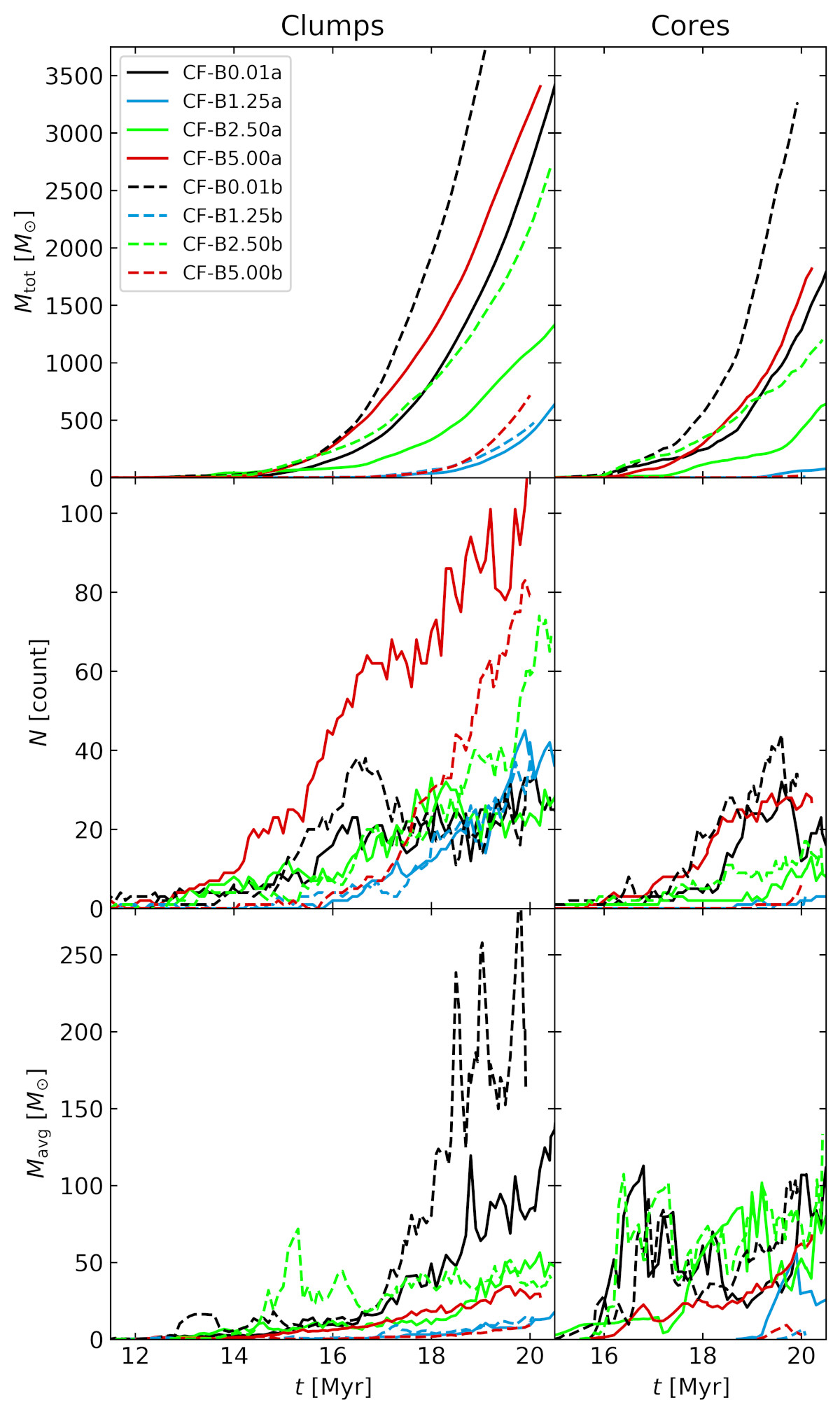}
  \caption{
  \rev{
  Top panel: Total clump mass (left) and core mass (right) vs. time
  2nd panel: Number of \simcls{} (left) and \simcos{} (right) vs. time.
  Bottom panel: Average mass of \simcls{} (left) and \simcos{} (right) vs. time.
  The low magnetization simulations, \CF{0.01}a/b, produce the highest average clump masses.}
  }
  \label{fig:clstat}
\end{figure}
The time evolution of the resulting sets of detected objects is shown in \reffig{fig:clstat}.

The first \simcls{} are detected around \T{12}.
\rev{The total mass in \simcls{} ${M_\mathrm{tot,clumps}}$ (top left panel) increases over time, in agreement with the growing amount of CO.
For the lower magnetization simulations up to \B{1.25} and \CF{2.50}a,
the number of \simcls{} ${N_\mathrm{clumps}}$ (second panel, left) fluctuates between 10 and 45~\simcls{} per simulation, as clumps frequently form and disperse.}
This is in agreement with the findings of \citet[][]{dib2007virial} who discuss the clumps as transient features in the turbulent flow.
\rev{The highest magnetization simulations \CF{5.00}a/b, as well as simulation \CF{2.50}b, form considerably more \simcls{} than the other simulations (2nd panel, left).}
\rev{However, the average mass ${M_\mathrm{avg,clumps}}$ of those \simcls{} (bottom left panel) is below ${50\,\mathrm{M}_\odot}$,
while there is a steep growth of the average clump mass in the low magnetization simulations \CF{0.01}a/b,
which exceeds ${100\,\mathrm{M}_\odot}$ after \T{19}.}

The first \simcos{} are detected around \T{16}.
Generally, the evolution of the core mass ${M_\mathrm{tot,cores}}$ \rev{(\reffig{fig:clstat}, top right panel)} is similar to that of the clump mass, delayed by a few~Myr.
By definition of the detection parameters, the \simcos{} represent sub-domains of the \simcls{}.
Hence, their overall mass is always smaller than that of the containing \simcls{}.
The average core mass ${M_\mathrm{avg,cores}}$ \rev{(second panel, right)} is often higher than the average clump mass because the lighter \simcls{} do not contain any \simcos{} while many of the heavier \simcls{} contain one core that incorporates a large fraction of the parental clump mass.
The number of \simcos{} ${N_\mathrm{cores}}$ fluctuates due to the merging of \simcos{} rather than their dissipation.
Hence, in our simulations \simcos{} are generally not transient features \citep[contrary to][]{dib2007virial} as we will discuss in more detail later on.
\begin{figure}
  \centering
  \includegraphics[width=.98\columnwidth]{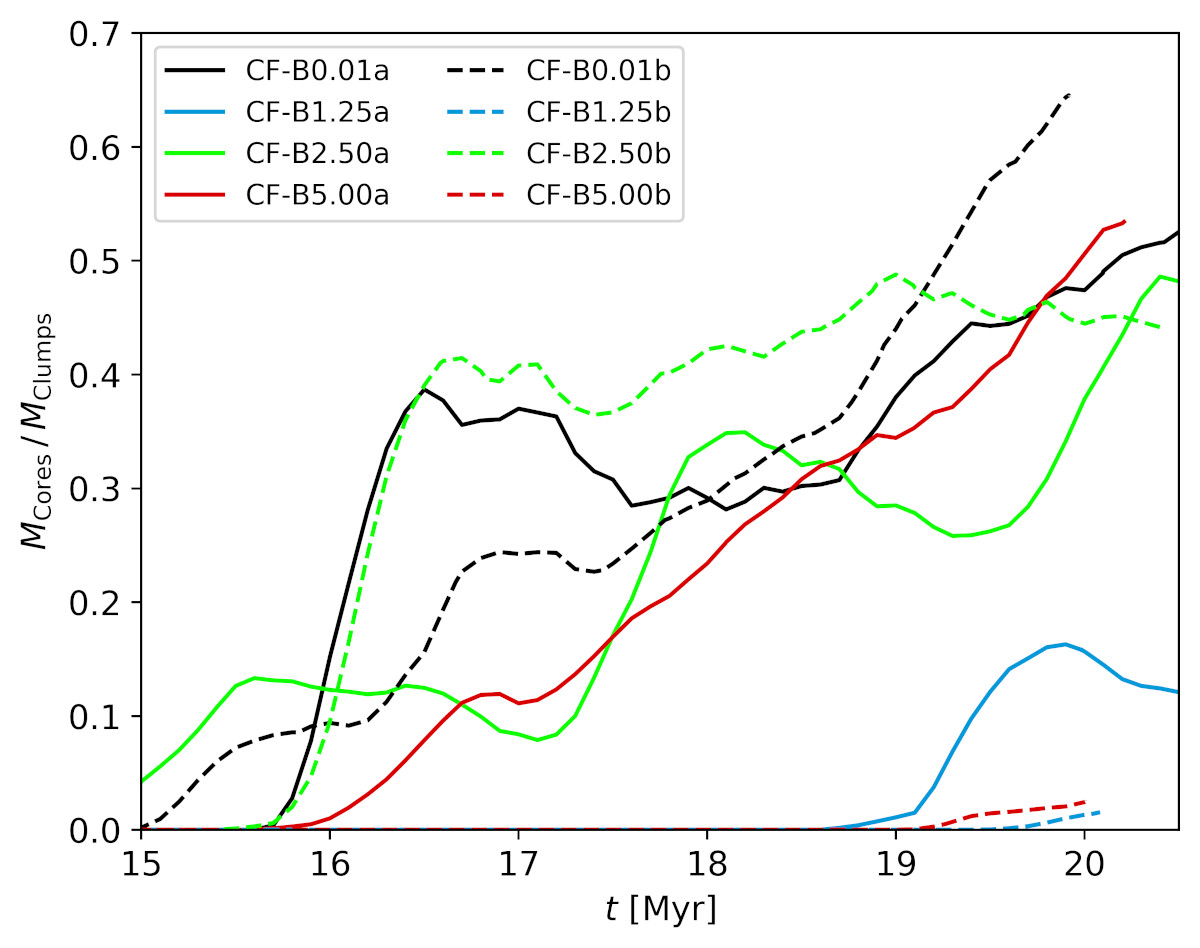}
  \caption{
  Ratio of total core and total clump mass vs. time.
  }
  \label{fig:coremassfrac}
\end{figure}
At \T{20}, close to the end of \rev{most simulations, at least 40\%} of the mass in \simcls{} is concentrated in the \simcos{} (\reffig{fig:coremassfrac}),
with exception of runs \CF{1.25}a/b and \CF{5.00}b.

The structural analysis of the formed objects, Sec.~\ref{subsec:larson}--\ref{subsec:EVTana-compare}, is done at \T{20.0},
where all of the simulations have developed molecular structures.
The \simcls{} at that point in time are marked in \reffig{fig:coldens-x}.
The \simcos{} are usually to small to be reasonably displayed in \reffig{fig:coldens-x}.
Column density maps of the \simcos{} of the \CF{X}a-simulations are shown in the Appendix,
Fig.~\ref{fig:cores-A0}--\ref{fig:cores-D20}; the corresponding maps for the \CF{X}b-simulations are provided with the online material.

\subsection{\revc{Comparison to \textsc{getsf} sources}}
\label{subsec:comp-2d-3d}

\begin{figure}
  \centering
  \includegraphics[width=.98\columnwidth]{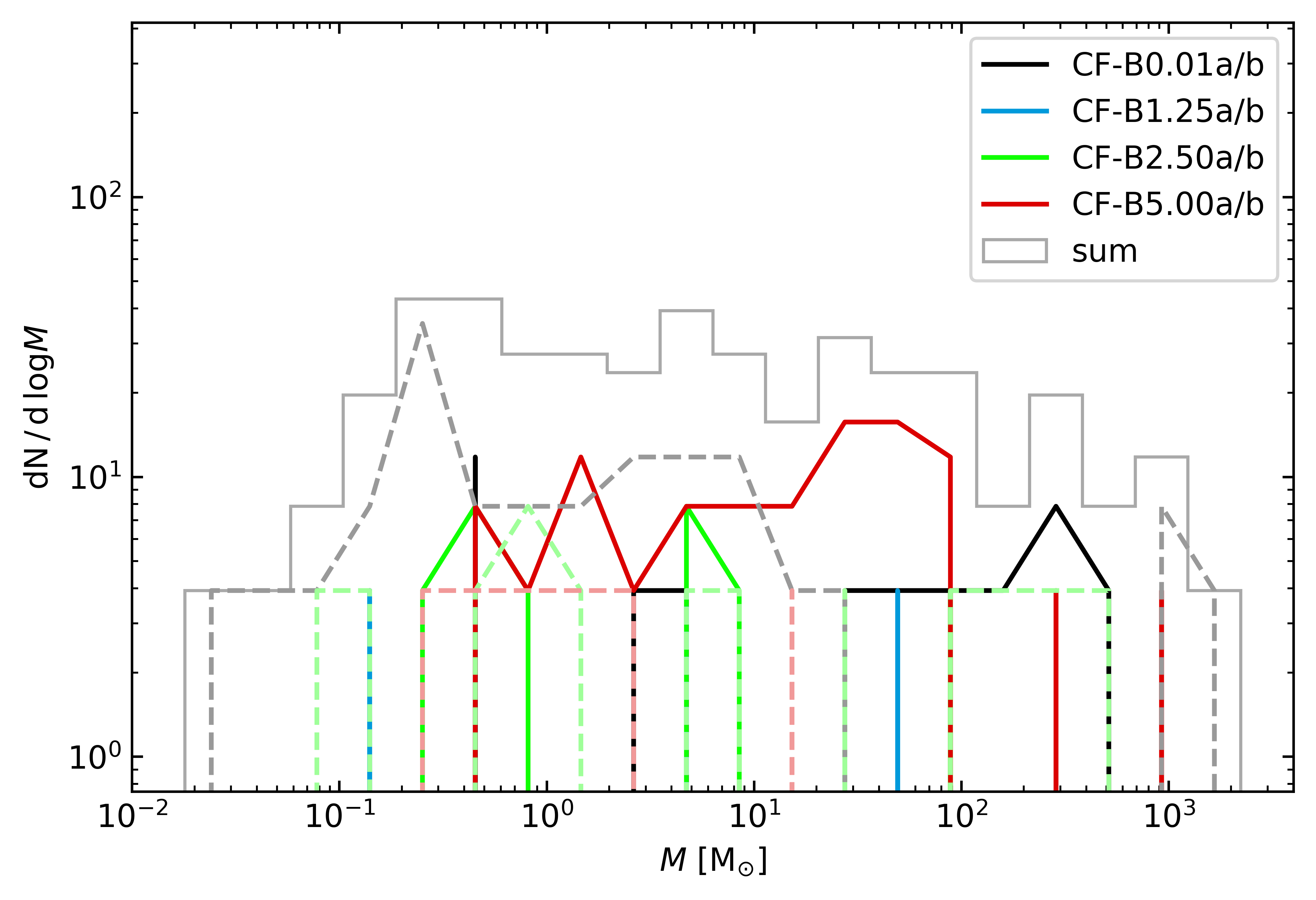}
  \includegraphics[width=.98\columnwidth]{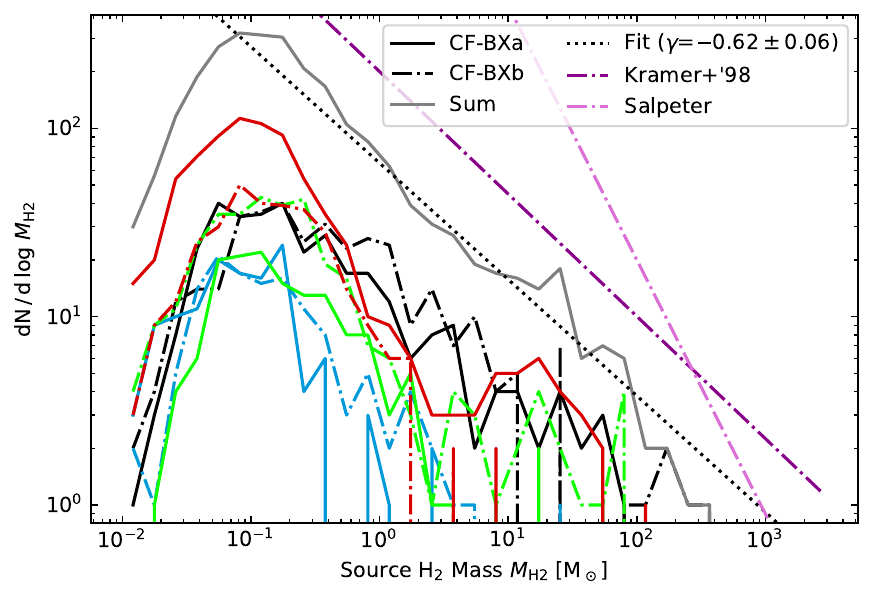}
  \includegraphics[width=.98\columnwidth]{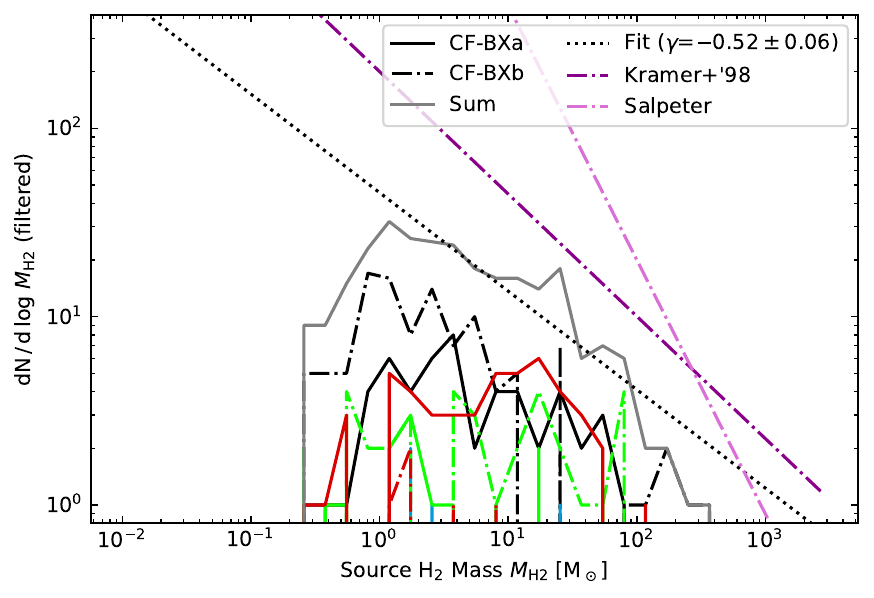}
  \caption{\rev{
  Top: Core mass function of the \simcos{} at \T{20}.
  Centre: Source mass function for the detected \textsc{getsf}-sources at the same time.
  \revb{The dotted line shows our power-law tail fit.}
  \revc{Bottom: Same as centre, but only for sources with estimated densities larger than $10^{-19}\,\mathrm{g}\,\mathrm{cm}^{-3}$.}
  \revc{The orchid and dark magenta lines indicates the slopes of ${\mathrm{d}N/\mathrm{d}\log~M}$ of the Salpeter IMF,
  and of the mass spectrum found by \citet{Kramer1998CO} for CO-clumps, respectively.}
  }}
  \label{fig:cmf}
\end{figure}

\rev{In Fig. \ref{fig:cmf} we display the core mass function of the \simcos{} (top),
as well as of the \revc{unfilterd (center) and filtered (bottom)} \textsc{getsf}-sources for each simulation at \T{20}.
}
\rev{
The mass functions derived using \revb{the sources indicated by} \textsc{getsf} \revb{(center)} peak around a source mass of $0.1\,\mathrm{M}_\odot$.
In contrast, the core mass functions extracted in 3D \revb{(top)} do seem not to feature a peaked distribution.
Furthermore, we count very few \simcos{} in the mass range below $1\,\mathrm{M}_\odot$.
}
\revb{In Fig. \ref{fig:cmf} (bottom), we show the source mass function that results from rejecting those \textsc{getsf} sources with $\rho_\mathrm{avg} < 10^{-19}\,\mathrm{g}\,\mathrm{cm}^{-3}$.}

Comparing the \revb{3D (top) and the unfiltered 2D results (centre)},
one can see that \textsc{getsf} picks up a lot of low-mass sources, also in regions which do not contain any \simcos{}.
On the other hand, \textsc{getsf} splits massive \simcos{} which are connected structures in 3D into smaller pieces.
We also varied the Gaussian smoothing used for \textsc{getsf}, but the overall result of \textsc{getsf} does not change significantly.

Due to the large number of low-mass objects, \textsc{getsf} gives core mass functions with an approximate power-law slope.
\revc{We compute least-squares fits of ${\mathrm{d}N/\mathrm{d}\log M}$ vs. ${\log M}$ for the range ${1<M/M_\odot<200}$.}
\revb{
For the sum over all simulations (grey line in Fig. \ref{fig:cmf}, middle panel), we find a power law slope of
\begin{eqnarray}
    \frac{\mathrm{d}\,N}{\mathrm{d}\log{M}} \propto M^{\substack{-0.62\,\text{(all)}\\-0.52\,\text{(filtered)}}}~\text{,}
    \label{eq:getsf-cmf-slope}
\end{eqnarray}
which is shallower than the Salpeter IMF ($\alpha=-1.35$ with ${\mathrm{d}N/\mathrm{d}\log{M}\propto~M^{\alpha}}$).
A CMF slope of $\alpha=-0.65$ is found by \citet{Kramer1998CO} for CO-clumps.
\revc{
Several studies conducted in the nearby Gould Belt clouds \citep[see e.g.][]{andre2010filamentary,konyves2015census}
lead to the interpretation that the IMF is inherited from the CMF.
}
}
\revd{
While aforementioned studies do not capture high-mass cloud environments which are present in the Milky Way \citep{Motte2022AlmaImfI},
\citet{Nony2023AlmaImfV} still observe a Salpeter-like slope of $\alpha\approx-1.46$ for the prestellar CMF in the high-mass region W43.
While aforementioned observations are not consistent with the CMF of our getsf sources,
\citet{Nony2023AlmaImfV} interestingly find an index of $\alpha\approx-0.6$ for the high-mass end of the protostellar CMF of the same region.
which they attribute to the protostellar cores being clump-fed from their environment \citep[see also][ for further analysis of the top-heavy CMFs]{Pouteau2023}.
}

\revb{
A large part of the total mass of our \textsc{getsf} sources is contained in sources of masses below $2\e{-1}\,\mathrm{M}_{\odot}$,
none of which are retained when filtering out the low-density \textsc{getsf} sources in Fig. \ref{fig:cmf} (bottom), as described above,
while the remaining \textsc{getsf} sources are all located in areas where also \simcos{} were detected (see Fig. \ref{fig:coldens-x}). 
Furthermore, a number of sources with higher masses get rejected.
This results in a distribution that is similarly flat to the one for the \simcos{} (top).
However, the filtered \textsc{getsf} source mass function has a narrower object mass range when compared to the 3D core mass function,
with objects of very low and high masses missing.
For the high mass range, we think this is caused by our larger \simcos{} being covered by multiple separate \textsc{getsf} sources,
while our low-mass \simcos{} might not cause sufficient contrast in the column density distribution to be detected by \textsc{getsf}.
}

\begin{figure}
  \centering
    \includegraphics[width=.98\columnwidth]{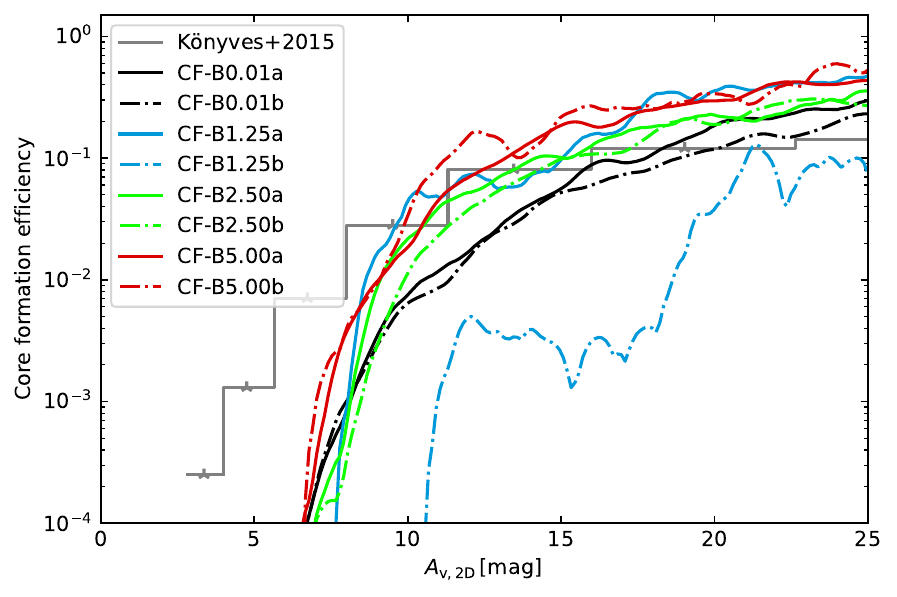}
    \includegraphics[width=.98\columnwidth]{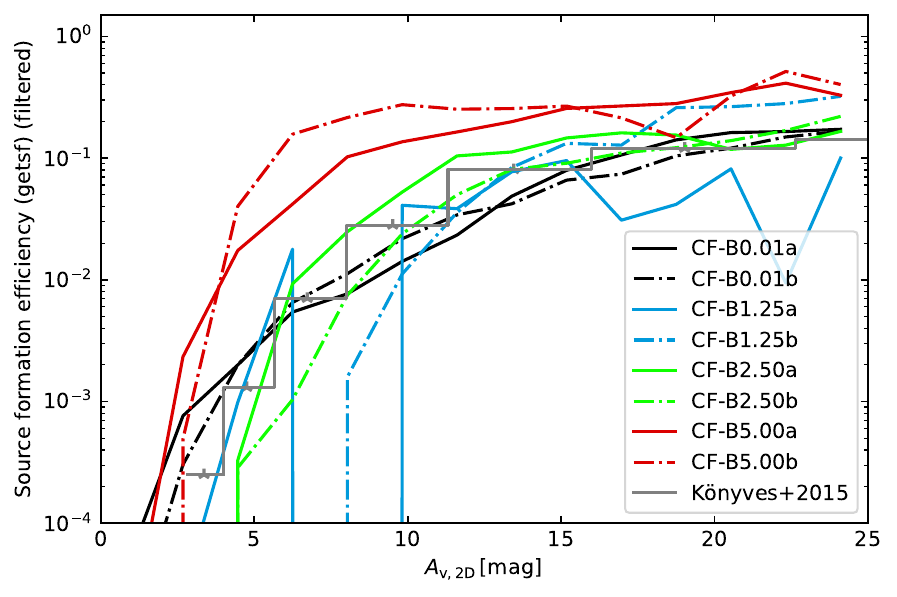}
  \caption{
  Top: \revc{3D} core formation efficiency versus the projected visual extinction $A_{\mathrm{V,2D}}$ at \T{20}.
  \rev{Bottom: Formation efficiency of \revb{filtered \textsc{getsf}-sources} versus the projected visual extinction $A_{\mathrm{V,2D}}$ at \T{20}.}
  The grey line indicates the value found in the observational study from \citet{konyves2015census}.
  We see that our data is in \rev{some} agreement with the observations below 15~mag if we consider the projected column densities.
  }
  \label{fig:effpdf}
\end{figure}
In \reffig{fig:effpdf}, we show the 2D core formation efficiency \citep{konyves2015census}.
\rev{To calculate the 2D core formation efficiency, we calculate the difference between the gas column density map (projection along the $x$-direction as shown in Fig.~\ref{fig:coldens-x}) and the column density map derived for the \simcos{} only.}
Here, $A_{\mathrm{V,2D}}$ is computed as
\begin{equation}
    \frac{A_{\mathrm{V,2D}}}{\mathrm{mag}} = \revm{\frac{N_\mathrm{H}}{1.87\e{21}\,\mathrm{cm}^{-2}}}\,,
\end{equation}
where $N_\mathrm{H}$ is the gas column density \citep{walch2015silcc}.
\rev{
We expect that our simulations will not resolve visual extinctions of above  ${A_{\mathrm{V,2D}}>160\,\mathrm{mag}}$,
as the spatial resolution can only resolve the \revb{Jeans} length for densities up to the order of $n\lesssim 1\e{6}\,\mathrm{parts}\,\mathrm{cm}^{-3}$,
and our densest structures are typically no larger than $0.1\,\mathrm{pc}$.
However, we see only few objects reaching that limit.
}
\rev{
In our projections, the core formation efficiency in gas at intermediate visual extinctions, ${8\,\mathrm{mag}<A_{\mathrm{V,2D}}<20\,\mathrm{mag}}$,
is roughly in agreement with the efficiency found in the observational census of dense cores in the Aquila cloud complex conducted by \citet{konyves2015census} (grey line).
In more detail, the non-magnetized simulation lie slightly below while the magnetized simulations lie slightly above the observational efficiency.
Overall, we basically see no \simcos{} for ${A_{\mathrm{V,2D}}<6\,\mathrm{mag}}$.
At high visual extinctions, ${A_{\mathrm{V,2D}}>20\,\mathrm{mag}}$, we see efficiencies in excess of the observed values,
meaning that the \simcos{} dominate the overall mass budget located in columns with a high $A_{\mathrm{V,2D}}$. However, because the molecular clouds forming in a colliding flow are inherently sheet-like, it can be expected that the core formation efficiency will eventually reach 100\% in the densest regions because there is very little gas in front and behind the sheet, thus limiting the amount of mass which can be contained in foreground and background gas.}
\rev{
For detecting the \simcos{}, we consider the 3D-shielded $A_{\mathrm{V,3D}}$, as obtained from the {\sc TreeRay/OpticalDepth} module.
A direct consequence of our core definition is that almost all the shielded gas mass above ${A_{\mathrm{V,3D}}>8\,\mathrm{mag}}$ is contained in cores.
This demonstrates the substantial difference between the observed and the actual 3D visual extinction experienced by the core forming gas (see also \cite{haid2019silcc}).}

\rev{
In the bottom panel of \reffig{fig:effpdf}, we also show the formation efficiency of the \revb{filtered} \textsc{getsf}-sources as a function of $A_\mathrm{V,2D}$.}
In comparison with the core formation efficiencies (top), \revc{we see higher \revb{source} formation efficiencies already at low visual extinctions.
Overall, the source formation efficiencies are similar to the values observed by \citet{konyves2015census} at up to ${A_\mathrm{V,2D}<15\,\mathrm{mag}}$ for all but the high magnetization simulations.}
The source formation efficiencies at high visual extinctions of ${A_\mathrm{V,2D}>20\,\mathrm{mag}}$
\revb{are overall comparable but slightly more dependent on the simulation's initial conditions than for our \simcos{}.}

\revb{\subsection{Adherence to Larson's relations}}
\label{subsec:larson}
The algorithm used to detect the \simcls{} and \simcos{} results in them to be connected objects.
Other than that, it does not impose additional constraints about their geometry.
In particular, the \simcls{} and \simcos{} are not radially symmetric in general.
Hence, we define an effective object radius
\begin{equation}
  R_I = \sqrt{\frac{5}{6 M}\int_V \rho r^2~\mathrm{d}V} \, ,
  \label{eq:clsize}
\end{equation}
where $R_I$ is the radius of a homogeneous sphere with the object's mass, $M$,
which has the same trace of the moment of inertia tensor. We also calculate its mass-weighted velocity dispersion
\begin{equation}
  \sigma_\mathrm{3D} = \sqrt{\frac{\int_V \rho\vec{v}^2~\mathrm{d}V}{M} -\left(\frac{\int_V \rho\vec{v}~\mathrm{d}V}{M}\right)^2} \, .
  \label{eq:cldisp}
\end{equation}
In order to avoid projection effects, we choose a definition of those quantities that utilises \rev{the full 6D} (PPP,VVV) phase space information which resembles quantities that are also accessible from the observational PPV space. Using those quantities, we test the scaling relations of our substructures against the Larson relations.

\begin{figure*}
  \centering
  \includegraphics[width=\textwidth]{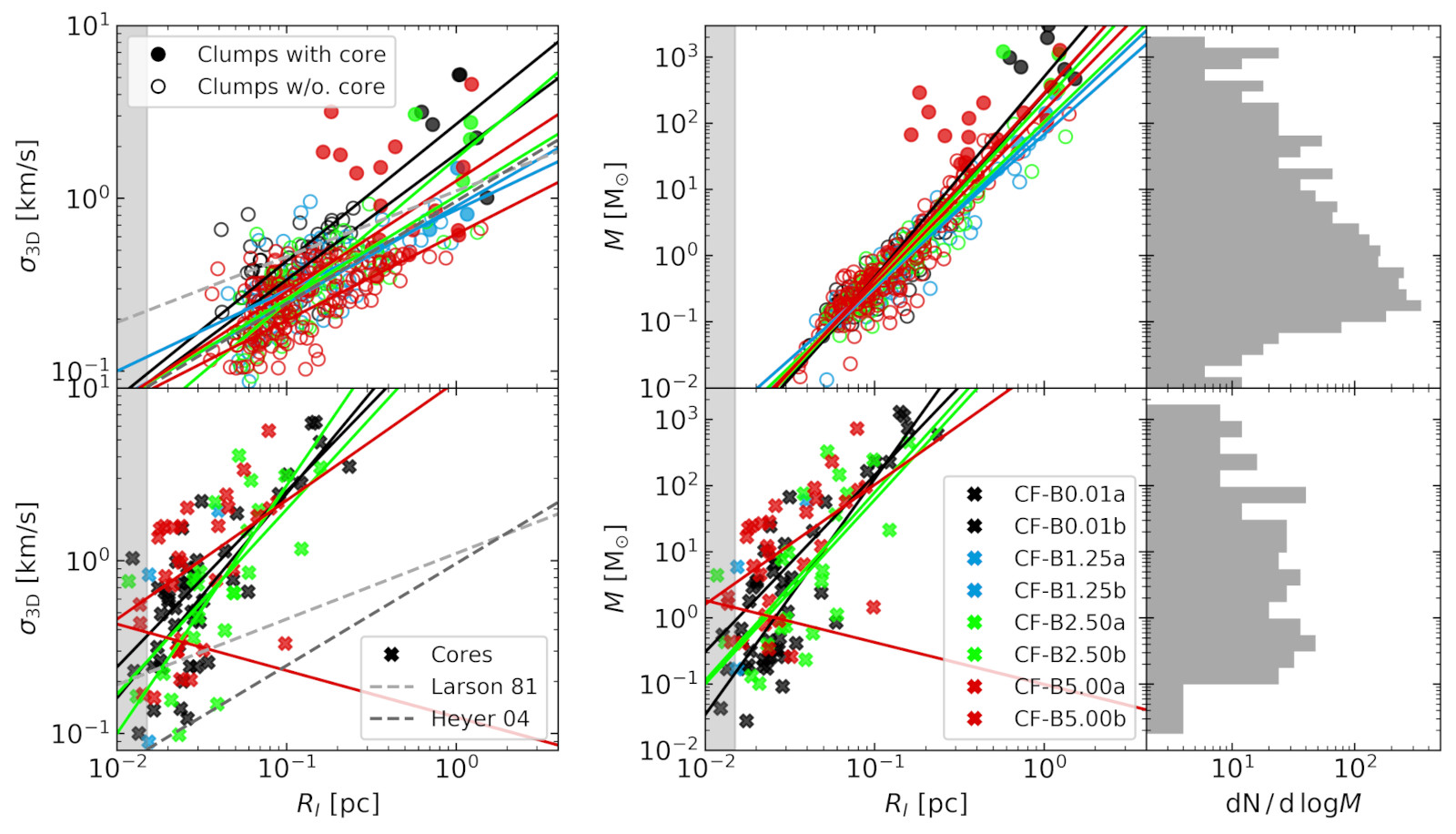}
  \caption{
  \revb{Left}: $\sigma_\mathrm{3D}$-size relation \revb{of our \simcls{} (top) and \simcos{} (bottom)},
  where the dashed grey line represents Larson's first relation \citep{larson1981turbulence}.
  The dashed dark grey line represents the revised relation found by \citet{heyer2004universality}.
  The less massive \simcls{} roughly follow the revised relation.
  \revb{Right}: Mass-size relation \revb{of our \simcls{} (top) and \simcos{} (bottom)}.
  The coloured solid lines indicate least squares linear regressions in logspace for the $\sigma$-size relations and for the mass-size relations of the objects identified in the respective simulations at \T{20}.
  The resultant slopes $\gamma_{\sigma}$ and $\gamma_{M}$ are listed in Tab.~\ref{tab:Rel-Size}.
  \rev{ The grey area marks the size resolution limit, i.e. the range of radii for which a spherical object would be disregarded due to the resolution requirement as enforced by the object detection algorithm.}
  \revb{Rightmost: Object mass histogram of \revb{of our \simcls{} (top) and \simcos{} (bottom)}.}
  }
  \label{fig:Larson}
\end{figure*}
In \reffig{fig:Larson}, we show the Larson linewidth-size relation \rev{(left column)} and the mass-size relation \rev{(center-right column)} for our \simcls{} \rev{(top row)} and \simcos{} \rev{(bottom row)} at \T{20}. 
The corresponding set of \simcls{} and \simcos{} is fitted separately for each simulation using linear regression in log-space, yielding power-laws:
\begin{eqnarray}
  \log{(\sigma_\mathrm{3D}/(\mathrm{km/s}))} &=& \beta_{\sigma}+\gamma_{\sigma}\log{(L/(\mathrm{pc}))} \,,
  \label{eq:sigma-size} \\
  \log{(M/M_{\odot})} &=& \beta_{M}+\gamma_{M}\log{(L/(\mathrm{pc}))} \,,
  \label{eq:mass-size}
\end{eqnarray}
where $\gamma_{\sigma}$ is the power-law slope of the linewidth-size relation, and $\gamma{_M}$ is the slope of the mass-size relation, respectively (see Table~\ref{tab:Rel-Size} for the results).


An observational correlation between the linewidth and size of molecular clouds has been established historically \citep{larson1981turbulence, heyer2004universality}.
Possible origins for a linewidth-size relation are either a turbulent energy cascade or the structures taking on a kinetic-gravitational equipartition (see below).
It is at least questionable if the linewidth-size relation holds for observations of bound clumps \citep{traficante2018testing} or even smaller substructures.
On the top row of \reffig{fig:Larson}, we overplot the original Larson relation \citep[][$\gamma_\sigma=0.38$, dashed light grey line]{larson1981turbulence},
as well as the revised relation described by \citet{heyer2004universality} ($\gamma_{\sigma}=0.59$, dashed, dark grey line). 
\rev{
While simulations \CF{1.25}a/b and simulation \CF{5.00}b have values around $\gamma_{\sigma}\approx 0.5$, the clump ensembles found in the other simulations are around $\gamma_{\sigma}\approx 0.7$, being in excess of Larson's relation.
The steep slope is mainly caused by very turbulent \simcls{} larger than $0.1\,\mathrm{pc}$,
while the ensembles of smaller \simcls{} seem to better agree with the revised relation.
We note that the continuous inflow of our colliding flow model provides a constant amount of kinetic energy.}

A large fraction of the (non-thermal) kinetic energy of the \simcls{} is concentrated in the inner dense regions, defined by the contained \simcos:
\revb{
The velocity dispersions of some \simcos{} are similar to the velocity dispersion of the containing clump (see Fig.~\ref{fig:Larson}, bottom).
These \simcos{}, which inhabit the upper left area of the linewidth-size diagram \citep[as seen in][]{Ballesteros2018Turbulence},
also contain a considerable fraction of the total mass of their parent clumps.
However, the corresponding radii of the \simcos{} are far smaller than those of the \simcls{} that contain them.
Furthermore, the \simcls{} which contain a low mass core often contain multiple \simcos{} and/or a high mass core.
The lower-mass \simcos{} also show lower velocity dispersions.
In Appendix~\ref{sec:magdir}, Fig.~\ref{fig:vdisp-core-parent}, we show a direct comparison between the respective velocity dispersions of each core and its parent clump.
}

The third Larson relation \citep{larson1981turbulence} states that GMCs are observed to have comparable surface densities, independent of their size, thus following a density-size relation of $\rho\propto R^{-1.1}$.
If the MC substructures in our simulations were to follow that relation, we would expect a mass-size relation of $M\propto R^{\gamma_{M}}$ with $\gamma_{M}=1.9$.
However, for our per-simulation clump ensembles at \T{20}, 
we find slightly to moderately steeper slopes $\gamma_{M}\geq 2.1$ (\reffig{fig:Larson} bottom panels).
This is caused by the larger \simcls{} which host \simcos{} and hence show an increase in surface density.

Larson's second relation states that GMCs assume kinetic-gravitational equipartition \citep{myers1983dense},
with according ensembles of objects following the relation ${\sigma\propto\rho^{0.5} R}$.
To be consistent with the linewidth-size relation ${\sigma\propto R^{\gamma_{\sigma}}}$ and mass-size relation ${\rho\propto R^{\gamma{_M}-3}}$,
this requires the slope coefficients to follow the relation
\begin{equation}
    \gamma_{\sigma} = \frac{1}{2}\left(\gamma_{M}-3\right) + 1\, .
    \label{eq:Larson2Fit}
\end{equation}
A test of that relation against our object ensembles at \T{20} is shown in Table~\ref{tab:Rel-Size}.
\begin{table}
  \caption{
    Linear regression slopes $\gamma_{\sigma}$ of the $\sigma$-size relation (Eq.~\ref{eq:sigma-size}) and $\gamma_{M}$ of the mass-size relation (Eq.~\ref{eq:mass-size})
    of \simcls{} and \simcos{}, respectively.
    The term ${\gamma_{\sigma}-\frac{1}{2}\left(\gamma_{M}-1\right)}$ vanishes if the slope coefficients are consistent with kinetic-gravitational equipartition (see Eq.~\ref{eq:Larson2Fit}).
    The given errors are the standard error, as calculated from the square root of the diagonal elements of the covariance matrix.
  }
  \label{tab:Rel-Size}
  \centering
  \rev{
  \begin{tabular}{l|cc|c}
    \hline
    \hline
    \simcls{} & $\gamma_{\sigma}$ & $\gamma_{M}$ & \revc{${\gamma_{\sigma}-\frac{1}{2}\left(\gamma_{M}-1\right)}$}\\
    \hline
    \CF{0.01}a & $ 0.73\pm 0.09$ & $2.86\pm 0.19$ & $-0.20\pm 0.16$\\
    \CF{0.01}b & $ 0.79\pm 0.11$ & $3.03\pm 0.21$ & $-0.22\pm 0.18$\\
    \CF{1.25}a & $ 0.47\pm 0.06$ & $2.27\pm 0.11$ & $-0.17\pm 0.10$\\
    \CF{1.25}b & $ 0.56\pm 0.09$ & $2.42\pm 0.14$ & $-0.15\pm 0.13$\\
    \CF{2.50}a & $ 0.83\pm 0.09$ & $2.74\pm 0.17$ & $-0.04\pm 0.15$\\
    \CF{2.50}b & $ 0.60\pm 0.07$ & $2.46\pm 0.12$ & $-0.13\pm 0.11$\\
    \CF{5.00}a & $ 0.64\pm 0.06$ & $2.81\pm 0.13$ & $-0.26\pm 0.11$\\
    \CF{5.00}b & $ 0.50\pm 0.05$ & $2.62\pm 0.09$ & $-0.31\pm 0.08$\\
    \hline
    \simcos{} & $\gamma_{\sigma}$ & $\gamma_{M}$ & \revc{$\gamma_{\sigma}-\frac{1}{2}\left(\gamma_{M}-1\right)$}\\
    \hline
    \CF{0.01}a & $ 1.02\pm 0.22$ & $ 2.68\pm 0.49$ & $+0.18\pm 0.41$\\
    \CF{0.01}b & $ 1.18\pm 0.18$ & $ 3.57\pm 0.44$ & $-0.10\pm 0.37$\\
    \CF{2.50}a & $ 1.43\pm 0.70$ & $ 2.84\pm 1.70$ & $+0.51\pm 1.39$\\
    \CF{2.50}b & $ 1.07\pm 0.29$ & $ 2.79\pm 0.70$ & $+0.18\pm 0.57$\\
    \CF{5.00}a & $ 0.69\pm 0.28$ & $ 1.80\pm 0.57$ & $+0.28\pm 0.49$\\
    \CF{5.00}b & $-0.27\pm 0.64$ & $-0.64\pm 1.27$ & $+0.55\pm 1.10$\\
    \hline
    \hline
  \end{tabular}
  }
\end{table}
For our per-simulation ensembles of \simcls{} at \T{20}, above relation is violated significantly,
while for our per-simulation ensembles of \simcos{} at \T{20} the above relation is roughly followed within the given uncertainties.
This is consistent with a picture where the linewidth-size relation of the clumps is regulated by turbulence,
while the cores are approaching kinetic-gravitational equipartition.

\begin{figure}
  \centering
  \includegraphics[width=\columnwidth]{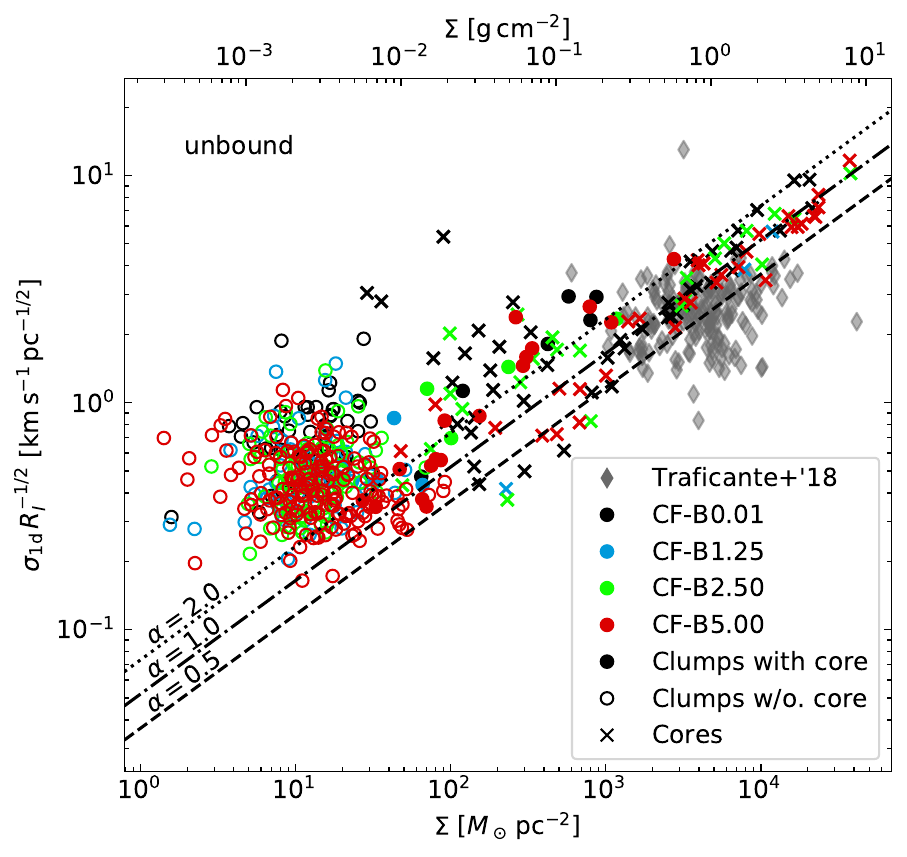}
  \caption{
  Larson scaling coefficient ${\nu_{\circ,G}\,\alpha^{1/2}=\sigma_\mathrm{1D} R_I^{-1/2}}$ vs. clump (circles) and core (crosses)
  surface density ${\Sigma=M/(\pi R_I^2)}$ at \T{20}.
  The dashed, dash-dotted and dotted lines represent
  ${\alpha_\mathrm{vir}=0.5}$, $1.0$, and $2.0$;
  objects represented by points on the dash-dotted line are in gravitational-kinetic equipartition.
  Clumps are represented by a filled marker if they contain at least one core.
  }
  \label{fig:Heyer}
\end{figure}

Slightly extending on \citet{heyer2009re} by explicitly including the object's virial parameter $\alpha$,
the mass of a self-gravitating object that is not in kinetic-gravitational equipartition can be expressed as
\begin{equation}
    M = M_\mathrm{vir}/\alpha = 5\sigma_\mathrm{1D}^2 R/\left(G\alpha\right) \,,
    \label{eq:VirialMass}
\end{equation}
with $\sqrt{3}\sigma_\mathrm{1D}\,=\,\sigma_\mathrm{3D}$.
When using the column density estimate
\begin{equation}
    \Sigma = \frac{M}{\pi R_I^2} \,,
    \label{eq:Coldens}
\end{equation}
and the velocity scaling coefficient
\begin{equation}
    \nu_{\circ,G}\left(\Sigma\right) = \left(\pi G\Sigma /5\right)^\frac{1}{2} \,,
    \label{eq:VelScaling}
\end{equation}
this recovers a modified version of Larson's linewidth-size relation:
\begin{equation}
    \sigma_\mathrm{1D} = \nu_{\circ,G}\,\alpha^\frac{1}{2} R_I^\frac{1}{2} \,,
    \label{eq:LarsonScaling}
\end{equation}
where the Larson scaling coefficient ${\nu_{\circ,G}\,\alpha^{1/2}=\sigma_\mathrm{1D} R_I^{-1/2}}$ quantifies the effect on the linewidth size relation, if variations of the surface density and deviations from the kinetic-gravitational equipartition are present.
We show a comparison of the Larson scaling coefficients ${\sigma_\mathrm{1D} R_I^{-1/2}}$ of our \simcls{} and \simcos{} to their approximate surface density in \reffig{fig:Heyer}.
There, a clear distinction between those \simcls{} which contain a core and those which do not is visible.
This leads to a combination of two effects:

The \simcls{} which do not contain a core (open circles) sit at the lower end of the surface density spectrum.
Most of them are far from being in gravitational-kinetic equipartition.
Thus, their ensemble is scattered over a broad range of virial parameters, centred around a large average $\alpha$ of about $6$.
This is not equally the case for the \simcls{}, which do contain at least one core (filled circles).
These clumps, which are typically the largest clumps, are closer to $\alpha=1$. This can also be seen from \reffig{fig:Larson} (bottom panels) because, according to our definition, the clump mass must always be larger than the core mass.
For \simcls{} smaller than $\sim 0.3$~pc, there is no corresponding core with smaller or comparable mass. 
Hence, the coreless \simcls{} constitute this population of small to medium sized \simcls{}.
This implies that $\alpha$ is particularly high for small and medium sized \simcls{}.
As a result, the linewidth-size relation is skewed towards more shallow slopes by the coreless \simcls{}.

On the other hand, the surface density of the \simcls{} which do contain at least one core is larger than that of the coreless population.
Therefore, the Larson scaling coefficient ${\nu_{\circ,G}\,\alpha^{1/2}}$ (Eq.~\ref{eq:LarsonScaling}) increases considerably for these clumps.
This leads to an increase in linewidth for these large clumps, despite them being close to $\alpha=1$.
This effect leads to a steepening of the linewidth-size relation, which can outweigh the flattening introduced by the coreless \simcls{}. 

The combination of both effects yields a linewidth-size pattern, where the low- and medium-sized \simcls{} are distributed around the same linewidth-size relation characterised by a slope of $\gamma_{\sigma}=0.5$,
while the largest \simcls{} tend to sit above the linewidth predicted by that relation,
thus leading to our observed slope coefficients in excess of Larson's linewidth-size relation.
Finally, for the \simcos{}, we find surface densities which are considerably higher than that of most \simcls{},
and in excess of ${2\e{3}\,\mathrm{M}_\odot\,\mathrm{pc}^{-2}}$.
This is in agreement with the recent observations by \citet{traficante2018testing}, who study the Larson relation for massive clumps identified in HiGal data.
For almost all \simcos{}, the object mass is within a factor~of~2 of their virial mass,
close to gravitational-kinetic equipartition.


\subsection{Specific angular momentum profiles of \simcos{}}
\label{subsec:cores-jana}
\rev{ The specific angular momentum profile $j(r)$ of cores is important for constraining models of star formation \citep{Bodenheimer1995},
especially with regard to the formation of a circumstellar disk \citep[see][for a discussion on the net turbulent angular momentum]{Walch2012}.} 
\rev{ We analyse the rotational profile of our \simcos{} at \T{20}.}

\begin{figure}
  \centering
	\includegraphics[width=1.0\columnwidth]{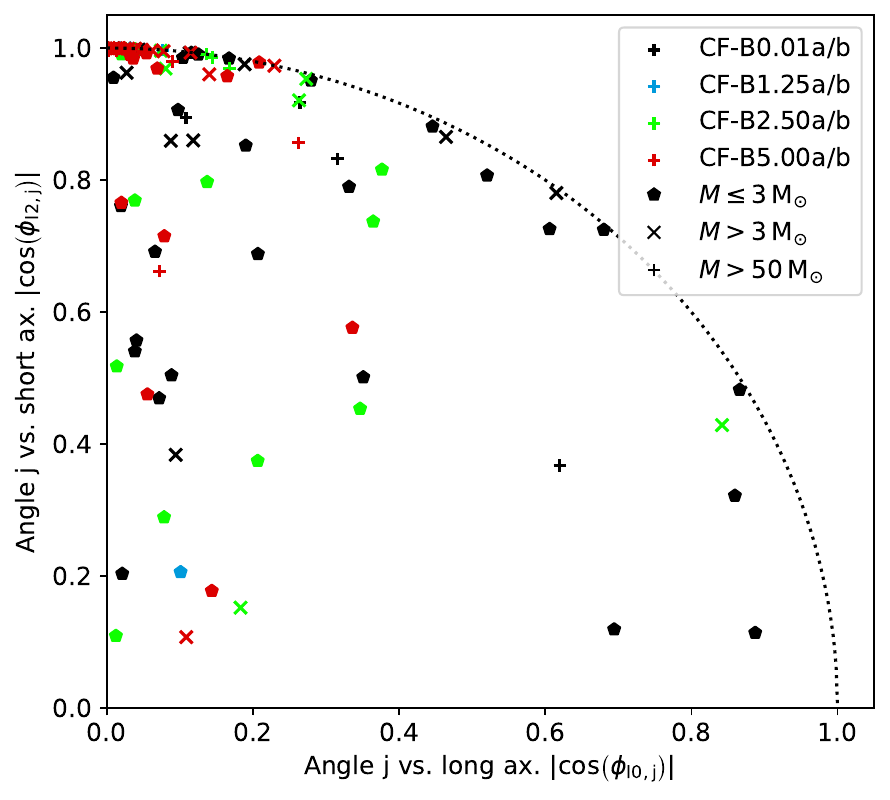}
  \caption{\rev{
    Direction of the rotational axis of each core in relation to the direction of the objects' principal component axis.
    \simcos{} with masses $M\le 3\,\mathrm{M}_\odot$, $M>3\,\mathrm{M}_\odot$, and $M>50\,\mathrm{M}_\odot$ are indicated by a pentagon/x/+ marker, respectively.
    The rotational axis of most \simcos{} is aligned with the shortest principal component axis.
    This is consistent with the rotation of oblate objects and indicates rotational flattening.}}
  \label{fig:rotdir}
\end{figure}
\revb{
We check the overall direction of each core's rotation by calculating the angles $\Phi_{\mathrm{I}_{i},\mathrm{j}}$ between its overall specific angular momentum vector $\vec{j}$ and its principal axis $\vec{I}_{i}$, \revc{where $\vec{I}_{2}$ is the shortest principal axis, i.e. the axis with the largest principal moment of inertia.}}
\revb{
Figure \ref{fig:rotdir} shows this quantity for the stable axes $\vec{I}_{0}$ and $\vec{I}_{2}$ for each core.
Most of the \simcos{} are approximately aligned with the shortest principal axis, $\cos(\Phi_{\mathrm{I}_{2},\mathrm{j}}) \approx 1$,
which is consistent with the rotation of oblate objects.}
It is likely that the objects are flattened by rotation.

\begin{figure}
  \centering
	\includegraphics[width=1.0\columnwidth]{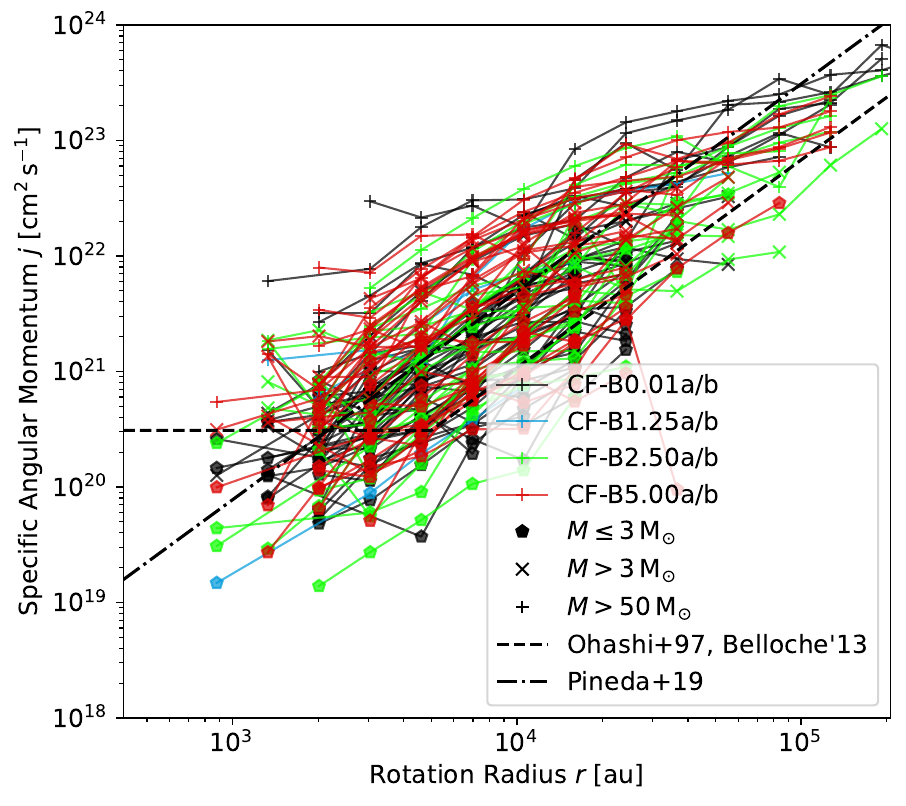}
  \caption{ \rev{
    Specific angular momentum profile of the \simcos{}.
    Each line corresponds to the profile of one core.
    \simcos{} with masses $M\le 3\,\mathrm{M}_\odot$, $M>3\,\mathrm{M}_\odot$, and $M>50\,\mathrm{M}_\odot$ are indicated by pentagon/x/+ markers for each radial bin, respectively.
    The dash-dotted line shows the best-fit power-law given by \citet{Pineda2019Specific} for their study of three sources,
    the dashed line shows the profile as proposed by \citet{Ohashi1997Kinematics} and \citet{Belloche2013Observation}.}}
  \label{fig:angmom}
\end{figure}
\revb{ 
We analyse the specific angular momentum profile of our \simcos{} in the following way.
First, we compute the total angular momentum $\vec{L}_{\rm tot} = \sum_{i=1}^{N_{\rm cells}}\vec{r}_{_{\rm COM,i}} \times \vec{p}_{_{\rm COM,i}}$ for each core with respect to its center of mass (COM).
Here, $N_{\rm cells}$ is the number of cells associated with the respective core.
Next, we rotate the coordinate system such that $\vec{L}_{\rm tot}$ points in the $z$-direction. The position and velocity vectors in the rotated coordinate system are denoted with a $\prime$.
We calculate $\vec{j^{\prime}}_{_{\rm COM,i}} = \vec{r^{\prime}}_{_{\rm COM,i}} \times \vec{v^{\prime}}_{_{\rm COM,i}}$ and $\|\vec{j^{\prime}}_{_{\rm COM,i}}\|$ for all cells $i$.
We bin the radial positions equally in log-space. The bin centers are plotted as the so-called rotation radius, $r$, on the x-axis of Fig.~\ref{fig:angmom}.
For each bin~$b$, we compute the mass-weighted mean of each component of the specific angular momentum, $\vec{j^{\prime}}_b$, as well as the respective absolute value, $j^{\prime}=\|\vec{j^{\prime}}_b\|$.
The resulting angular momentum profiles for all \simcos{} are shown in Figure \ref{fig:angmom}.}

\revb{
For our lower mass \simcos{} ($M<3\,\mathrm{M}_{\odot}$, pentagon markers), we recover power-law slopes $j^{\prime}\propto r^{\alpha}$ in a range of $\alpha=0.45\text{--}2.77$ and an average slope of $\alpha_\mathrm{avg}=1.55$.
For rigidly rotating bodies, we would expect a slope of $\alpha=2$.
Observations of actual low-mass cores yield slopes in the range of ${\alpha=1.60\pm 0.20}$ \citep{Goodman1993Velocity} to ${\alpha=1.80\pm 0.04}$ \citet{Pineda2019Specific}.
As only the line-of-sight velocity is accessible in observations, a possibility to estimate the specific angular momentum \revc{of a mapped core is} to calculate the velocity gradient across the map \citep{Goodman1993}.
\revc{
A linear gradient is expected for solid body rotation.
Moreover, \citet{Burkert2000Rotational} initialise a range of turbulent cloud cores.
Considering only two of the three angular frequency components to compare with observations, they compute a slope of $\alpha=1.5$.}
Here, we calculate $\vec{j^{\prime}}$ directly from the 3D velocities without taking into account any projection effects.
Therefore, the inferred radial distributions of $j^{\prime}$ might not be directly comparable with observations.}

\revb{
For most of our \simcos{}, we observe that the $\vec{j^{\prime}}$-components perpendicular to the core's main rotation axis are typically an order of magnitude smaller than the parallel component. We believe that this is a good indicator for a quite established net rotation in our cores, and it is consistent with above findings (see Fig. \ref{fig:rotdir}), which point toward a rotational flattening of the cores.
}

\rev{
The actual values of $j^{\prime}$ at any given $r$ have a scatter of about two orders of magnitude.
This is somewhat related to the mass of our \simcos{},
in that those \simcos{} with masses greater than $50\,\mathrm{M}_\odot$ tend to have the largest specific angular momenta at any given radius,
while those \simcos{} with masses below $3\,\mathrm{M}_\odot$ generally show the lowest values of $j^{\prime}$ at any given $r$.
This is caused by the difference in geometric complexity of those \simcos{}.
The \simcos{} with a clearly defined radial density distribution, which are mostly found in the intermediate mass range between $3\,\mathrm{M}_\odot$ and $50\,\mathrm{M}_\odot$ show similar or slightly higher specific angular momenta compared to the observational fit.
\revd{
Our lower mass \simcos{} often do not have a clearly defined radial density distribution.
Most of them have a subsonic velocity dispersion,
while the respective embedding parent clump is rather turbulent (compare Fig. \ref{fig:vdisp-core-parent} in the appendix).
}
In contrast, some of our high-mass \simcos{} with masses larger than $50\,\mathrm{M}_\odot$ contain more than one density peak with a somewhat distinct density profile around each of those peaks
and tend to extend to larger radii.}
\rev{
Here, we would like to remind the reader that the detection of our \simcos{} depends on the CO abundances and local visual extinction of the gas,
but is not immediately tied to the shape of the gravitational potential (see Section  or of the density field \ref{subsec:Detection}).
\simcos{} that were detected despite not containing any gravitational potential minimum tend to carry comparatively little specific angular momentum.
}

\subsection{Time-independent virial analysis}
\label{subsec:clumps-virana}
In this Section, we apply the EVT (see Section~\ref{sec:VirTh}) to the \simcls{} and \simcos{} at \T{20}.
Therefore, we calculate the volume energy terms (see Eqs.~\ref{eq:EV_W}-\ref{eq:EV_mag}) for each object.
We also to calculate the the surface terms (see Eqs.~\ref{eq:gs-SPhi}--\ref{eq:gs-Smag}), as well as the gravity term (see Eq.~\ref{eq:EV_W}),
by first calculating the corresponding differentials for each cell.
We then evaluate all terms by finite summation over the corresponding cells of each object.
The differentials are computed using a central finite difference scheme on guarded cell data.

\subsubsection{Time-independent EVT terms}
\label{subsec:EVTana-static}
This subsection focuses on the evaluation and discussion of the time-independent terms containing the volume energies $\varepsilon$,
the surface terms $\tau$ and the gravitational binding energy $W$ on the \rhs~of \refeq{eq:EVT};
the time-dependent term containing $\dot{\Phi}_I$ (see Eq.~\ref{eq:ES_Phi}) is discussed in the subsequent Section~\ref{subsec:EVTana-timedep}.

\begin{figure*}
  \centering
  \includegraphics[width=\textwidth]{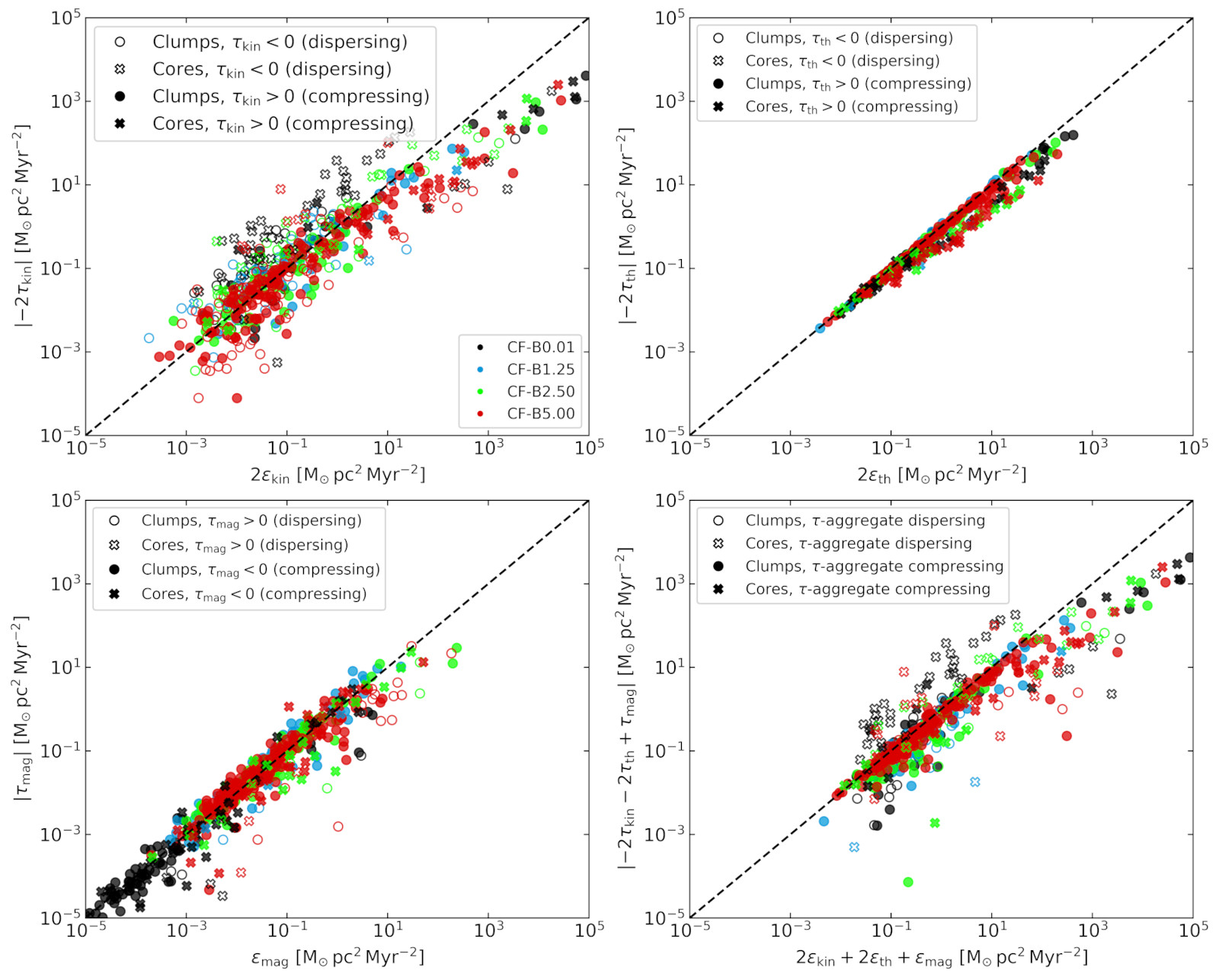}
  \caption{
  Surface vs. volume terms for \simcls{} (circles) and \simcos{} (crosses) at \T{20}.
  We show the kinetic EVT terms (top left), the thermal EVT terms (top right), the magnetic EVT terms (bottom left) and the aggregate of
  kinetic, thermal and magnetic terms (bottom right) for \rev{the} simulations with different initial magnetic field strength (colours).
  A solid marker indicates a surface term that provides a negative contribution to the \rhs~of the EVT, hence opposing the corresponding volume term and confining, respectively compressing, the structure.
  \rev{
  In contrast, an empty marker indicates  surface term that provides a positive contribution to the \rhs~of the EVT, hence expressing a dispersing effect on the structure.
  }
  Overall, the surface and volume terms correlate with each other over five or more orders of magnitude.
  There is a trend that the volume energy terms are more important,
  in particular for the \simcos{}, but the surface terms \rev{are not negligible}.
  Apparently, the initial magnetic field strength does not influence the general trends. 
  }
  \label{fig:VolSurf}
\end{figure*}
\reffig{fig:VolSurf} shows a comparison of the values of the kinetic (top left panel), thermal (top right panel) and magnetic surface terms (bottom left panel), as well as their aggregate (bottom right panel),
to the corresponding volume energy terms in the EVT as of \refeq{eq:EVT}.
Simulations with different initial magnetic field strength are distinguished by colour.
\simcls{} are plotted with circles, while \simcos{} are shown using crosses.
Filled symbols denote surface terms which are negative (when including their sign as appearing on the \rhs of Eq.\ref{eq:EVT}),
implying that the object is confined or compressed by the surface term.
All three MHD volume energy terms act to stabilise the object and are positive by construction. 

Overall, the surface and volume terms correlate with each other over five orders of magnitude or more for our \simcls{} and \simcos{} at \T{20}.
For many objects, the highest energies are contributed by the kinetic terms.
There is a trend that the volume energy terms are more important, in particular for the \simcos{}, but the surface terms are certainly not negligible.
Independent of the initial magnetic field strength, all populations of objects show similar trends with no clear systematic difference between the different runs. 

Considering the kinetic terms (top left), the surface energy term is mostly smaller than the volume energy term by up to $2$~orders of magnitude for objects with high kinetic energy.
Further, the kinetic surface term is negative for some of the \simcls{} and \simcos{}, while being positive for others.
This indicates that the velocity fields affecting the objects' surfaces promote the dispersion of some objects (empty markers), while others are confined by ram pressure (filled markers).
The thermal surface term (top right) of all detected objects contributes a negative value towards the \rhs~of the EVT, implying thermal pessure confinement.
For a majority of \simcls{}, the magnitude of the thermal surface term is comparable to that of the corresponding volume term.
However, for almost all \simcos{}, the thermal surface term is about ten times smaller than the thermal energy inside the object.
The comparison of the magnetic energy terms (bottom left)
again shows that the surface and volume terms are mostly of the same magnitude.
However, compared to kinetic energies, the magnetic energies are overall smaller.
For most of the \simcls{}, the effect described by the magnetic surface term $\tau_\mathrm{mag}$ is of confining nature (magnetic surface tension) and of similar order than the counteracting support from the internal magnetic pressure.
Exceptions to this can mainly be attributed to objects from the high magnetization simulations (runs \CF{5.00}a/b shown in red).
Interestingly, \revb{some} of the \simcos{} have a positive magnetic surface term, which has a dispersing rather than confining effect.
\revb{In these cases, the magnetic field does not facilitate core formation.
However, since $\tau_\mathrm{mag} >0$ is usually small in comparison with the other $\tau,\varepsilon$-terms, we do not expect it to have a major impact on the development of those \simcos{}.}
In total (bottom right panel), the surface energies of most \simcls{} are comparable to the corresponding volume energies. 
However, most \simcos{} have smaller, non-confining surface terms.

\subsubsection{Clumps and cores: Which Energy is important?}
\label{subsec:EVTana-compare}
\begin{figure}
  \centering
  \includegraphics[width=1.0\columnwidth]{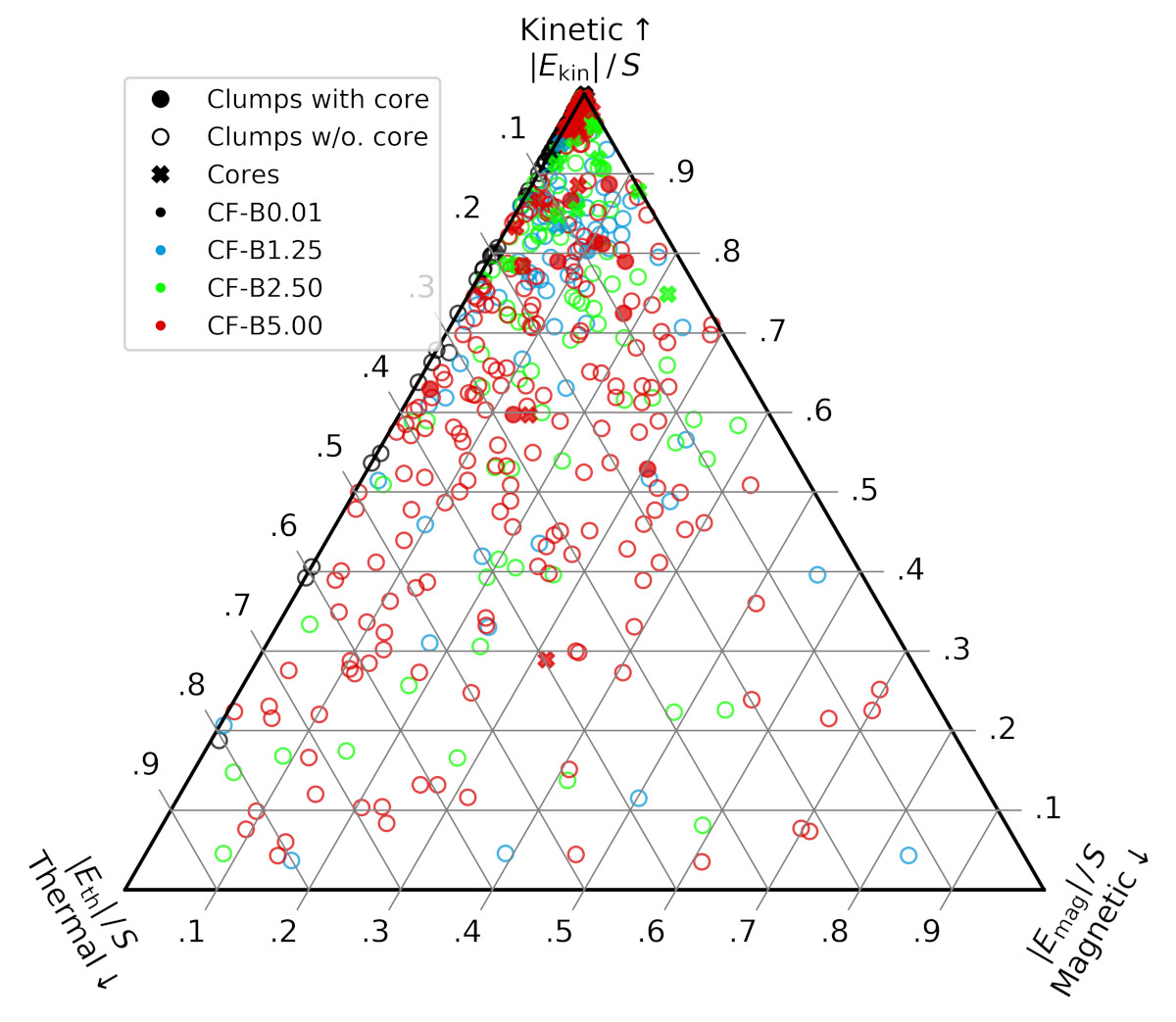}
  \caption{
  Ternary diagram showing a comparison of the magnitude absolute magnetic vs. thermal vs. kinetic terms using the combination of volume and surface terms (see Eqs.~\ref{eq:EVT-kin}--\ref{eq:EVT-mag})
  of \simcls{} (circles) and \simcos{} (crosses) at \T{20}.
  The \simcos{} are dominated kinetically and are all located in the top corner.
  In contrast to the volume- and surface terms themselves, the aggregated magnetic term is small compared to the thermal and kinetic terms for most objects.
  }
  \label{fig:CO-ter-dyn}
\end{figure}
In \reffig{fig:CO-ter-dyn}, we show the relative significance of the magnetic, kinetic and thermal energy for each object (\simcls{} and \simcos) at \T{20}.
Here, we introduce the combinations of surface and volume terms contributing to $E_\mathrm{MHD}$ in Eq.~\ref{eq:EMHD} as
\begin{eqnarray}
    E_\mathrm{kin} &=& 2(\varepsilon_\mathrm{kin}-\tau_\mathrm{kin})\\
    \label{eq:EVT-kin}
    E_\mathrm{th} &=& 2(\varepsilon_\mathrm{th}-\tau_\mathrm{th})\\
    \label{eq:EVT-th}
    E_\mathrm{mag} &=& (\varepsilon_\mathrm{mag}+\tau_\mathrm{mag}).
    \label{eq:EVT-mag}
\end{eqnarray}
For normalization we define 
\begin{equation}
    S_\mathrm{MHD} = \left|E_\mathrm{kin}\right| + \left|E_\mathrm{th}\right| + \left|E_\mathrm{mag}\right|\,.
\end{equation}
In the appendix, Fig.~\ref{fig:CO-ter-VS}, we show the individual contributions of volume and surface terms in ternary diagrams.

First, we note that all \simcos{} (crosses) are clustered in the upper corner of the diagram, which means that $E_\mathrm{kin}$ dominates.
Importantly, this implies that the \simcos' energetic state can be well estimated from the velocity field (kinetic energy) and gravity (see later) alone.
Furthermore, the kinetic surface term is inferior for many \simcos{} (see the upper left panel of \reffig{fig:VolSurf});
yet, a few \simcos{} have high, outwards-pointing surface terms, which means that these \simcos{} could be ablated or shredded \citep[see e.g.][]{Scannapieco2015Launching} by the surrounding flow.

For most of the larger-scale \simcls{} (circles), $E_\mathrm{kin}$ is still very important, but the contributions of $E_\mathrm{th}$ in particular, and $E_\mathrm{mag}$ occasionally, cannot be neglected.
As expected, for the \simcls{} identified in the low-magnetization simulations \CF{0.01}a/b,  the contribution of $E_{\mathrm{mag}}$ is minuscule in comparison to the thermal and kinetic energy terms. 
Yet, also for most of the objects from all other simulations, including \CF{5.00}a/b, the contribution of $E_{\mathrm{mag}}$ is minor.
This is surprising, because the mass-to-flux ratios that can be expected for the central region (as of Eq.~\ref{eq:MFRrate}) suggest that at least some of the clumps should be magnetically subcritical, hence the magnetic component in the EVT should be relevant.
Possible explanations for this are numerical diffusion of the magnetic field and (numerical) reconnection diffusion \citep{lazarian2012magnetization}.

Moreover, it seems that the thermal energy term can be important for the \simcls{}, in particular for runs \CF{5.00}a/b.
This could indicate the dissipation of magnetic and kinetic energy into internal energy, thus heating the clumps.

Summarised, magnetic fields seem to be of little importance for our MC substructures, i.e. our \simcls{} as well as \simcos{}.
This is in good in agreement with the flip of the magnetic field orientation with respect to dense structures to a perpendicular configuration which occurs in our simulations \citep{seifried2020parallel}. This happens above 100--1000~$\mathrm{cm}^{-3}$ and generally indicates a subdominant magnetic field.
It also matches the results of other authors finding magnetic fields to play the least role in dense interstellar regions:
A virial analysis from \citet{ganguly2022} finds little importance of the magnetic field for even larger structures.
\citet{ibanez2022gravity} find magnetic fields not to be dynamically important for dense clouds with $n\;>\;100\,\mathrm{cm}^{-3}$.

\begin{figure*}
  \centering
  \includegraphics[width=\textwidth]{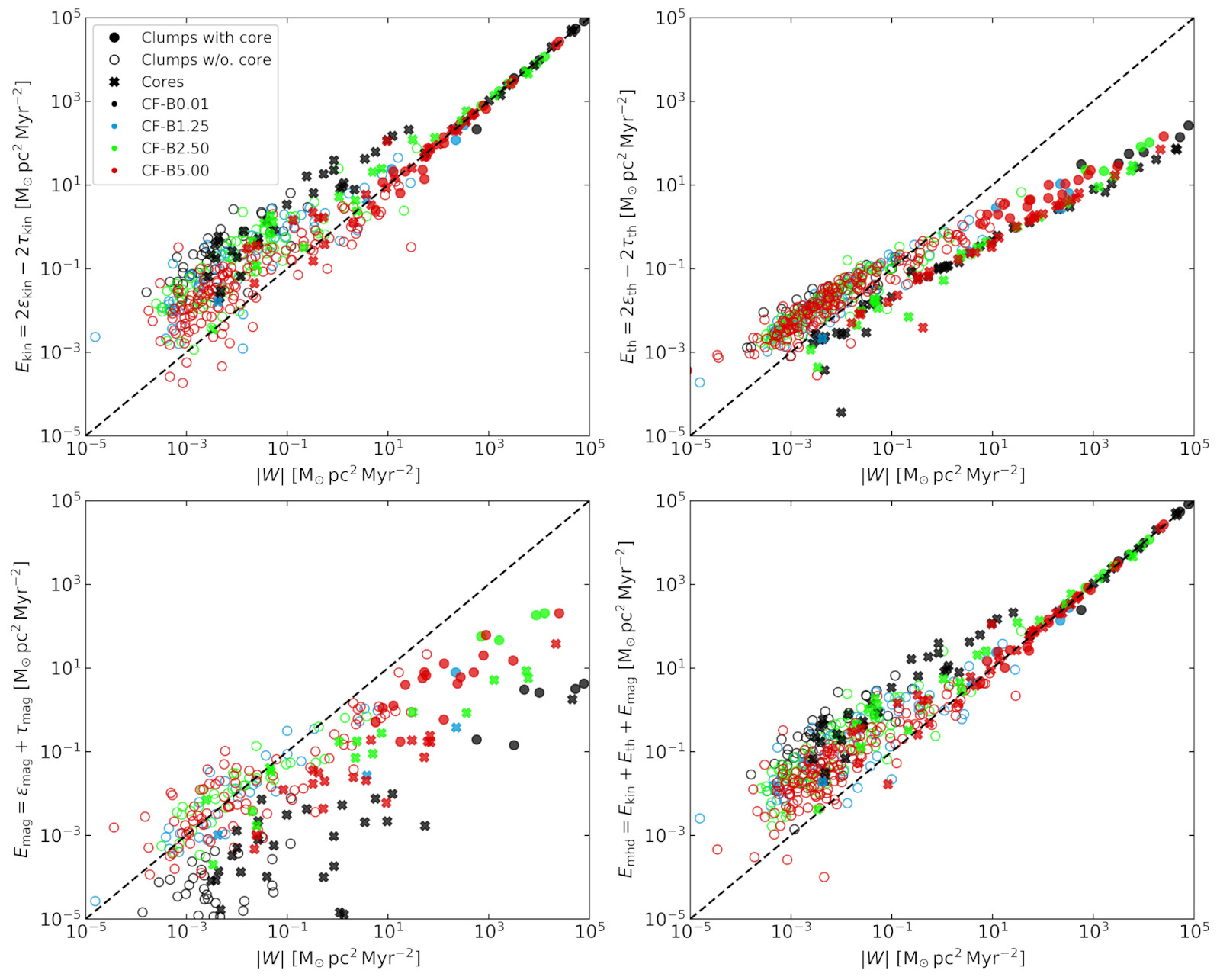}
  \caption{
    Clump (circles) and core (crosses) EVT dynamic components vs. the gravitational binding energy ${W}$ at \T{20}.
    We plot the kinetic (top left), thermal (top right), magnetic terms (bottom left) EVT terms as well as their aggregate (bottom right).
  }
  \label{fig:DynW}
\end{figure*}
The virial balance of the considered \simcls{} and \simcos{} is given by the interplay of the MHD contributions, discussed above, with the contribution of the gravitational binding energy $W$, towards the \rhs~of the EVT, \refeq{eq:EVT}.
\rev{In \reffig{fig:DynW}, a comparison of the kinetic (top left), thermal (top right), and magnetic (bottom left) terms as well as their aggregate (bottom right) as a function of the gravitational binding energy is shown for all \simcls{} and \simcos{} at \T{20}.}
We identify two important thresholds:
At gravitational binding energies below ${\sim 1\,\mathrm{M}_\odot\,\mathrm{pc}^2\,\mathrm{Myr}^{-2}}$,
the thermal contribution for a majority of \simcls{} as well as the magnetic contribution for some of the \simcls{} exceeds the gravitational binding energy.
Beyond that threshold, thermal and magnetic contributions do appear to only weakly scale with $W$, meaning their impact on the gravitational boundedness of those objects diminishes.
For most objects of gravitational binding energies below ${\sim 30\,\mathrm{M}_\odot\,\mathrm{pc}^2\,\mathrm{Myr}^{-2}}$,
the kinetic EVT contribution as well as the resulting aggregate of all MHD contributions (lower right panel) exceeds the gravitational binding energy,
marking those objects as being gravitationally unbound.
This regime of ${|W|\lesssim 30\,\mathrm{M}_\odot\,\mathrm{pc}^2\,\mathrm{Myr}^{-2}}$ contains most of the \simcls{} and a small number of \simcos{}.

For almost all objects with ${|W|\gtrsim 30\,\mathrm{M}_\odot\,\mathrm{pc}^2\,\mathrm{Myr}^{-2}}$,
the kinetic contribution approximates the gravitational binding energy. Also, the thermal and magnetic contributions are minuscule beyond that threshold.
This implies that objects above this threshold are "virialized", and their dynamics is mainly of kinetic nature.
The gravitational binding energy of those objects tightly correlates with the object mass and type.
For the \simcls{}, those thresholds correspond to an object mass of 9--60~$\mathrm{M}_\odot$, respectively;
for the \simcos{}, they correspond to an object mass of 3--20~$\mathrm{M}_\odot$ (see \reffig{fig:mass-W} in the appendix).
Note that the \simcls{} which do contain \simcos{}, most of the clump mass is actually in the core (see Section~\ref{subsec:larson}).

\subsubsection{The importance of surface terms}
\label{subsec:EVTana-surf}
In general, the effect of the gravitational field on some object can either be confining due to self-gravity or dispersing due to tidal forces caused by the external mass distribution.
However, we find only two such tidally dispersed objects for which the gravitational binding energy term $W$ would be non-negative (see \reffig{fig:mass-W} in the appendix).
Moreover, we find the thermal surface effects, expressed by the term ${-2\tau_\mathrm{th}}$, to be of confining nature for all objects (see Fig. \ref{fig:VolSurf}).
The magnetic and kinetic effects at the objects' surface, expressed by the surface terms $\tau_\mathrm{mag}$ and ${-\tau_\mathrm{kin}}$ in the EVT can be of confining or dispersing nature, which we have shown to both occur for different objects.
This poses the question, whether virialized objects are confined simply by being gravitationally bound or rather by being confined by MHD forces.
To explore this, we define two quantities.
The virialization ratio
\begin{equation}
    Q_\mathrm{vir}=\frac{(-W)-E_\mathrm{MHD}}{(-W)+E_\mathrm{MHD}}
    \label{eq:Qvir}
\end{equation}
considers the time-independent terms of the EVT to give an indication of the virialization of an object, where ${Q_\mathrm{vir}<0}$ indicates an unbound object, while ${Q_\mathrm{vir}=0}$ indicates a virialized object and ${Q_\mathrm{vir}>0}$ indicates an object with excess binding energy, which should be gravitationally collapsing.
The confinement balance ratio, $Q_\mathrm{cob}$, defined as
\begin{equation}
    Q_\mathrm{cob}=\frac{W-E_\tau}{\min(E_\tau,0)+\min(W,0)}\mathrm{ ,}
    \label{eq:Qcob}
\end{equation}
with
\begin{equation}
    E_\tau = -2\tau_\mathrm{kin} -2\tau_\mathrm{th} +\tau_\mathrm{mag} \,,
\end{equation}
provides information on all the terms which contribute negatively, i.e. in a confining/compressive way, towards the \rhs~of the EVT. It can be interpreted as follows:
\begin{itemize}
    \item {${Q_\mathrm{cob}=-1}$: Surface terms compressing, negligible gravitational energy,}
    \item {${Q_\mathrm{cob}=0}$: Surface terms and gravity equally compressing,}
    \item {${Q_\mathrm{cob}=1}$: Net zero surface terms,}
    \item {${Q_\mathrm{cob}>1}$: Dispersive surface terms,}
    \item {${Q_\mathrm{cob}>2}$: Surface term dispersion outweighs gravity.}
\end{itemize}

\begin{figure}
  \centering
  \includegraphics[width=\columnwidth]{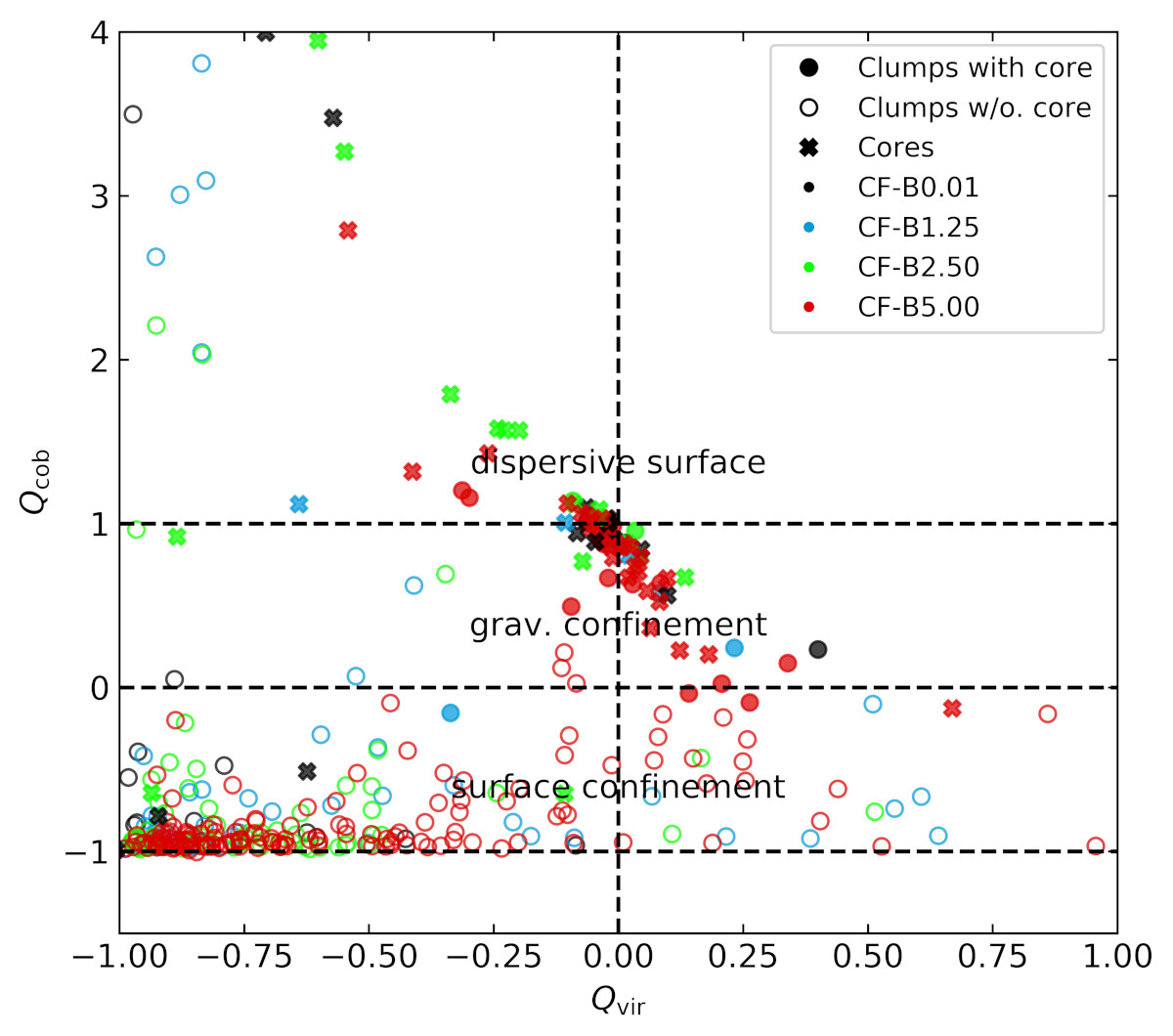}
  \caption{
    Confinement balance ratio vs. virialization.
    The virialization $Q_\mathrm{vir}$ (Eq.~\ref{eq:Qvir}) on the \dir{x}-axis gives an indication of the
    degree of object virialization ($-1$ being unbound, $0$ being virialized);
    The confinement balance ratio $Q_\mathrm{cob}$ (Eq.~\ref{eq:Qcob}) on the \dir{y}-axis gives an indication of the balance of the gravitational binding energy and the surface MHD terms
    ($<-1$: $E_\tau$ inwards, $W$ outwards;
    $-1$: $E_\tau$ inwards, $W=0$;
    $1$: $W$ inwards, $E_\tau=0$;
    $>1$: $W$ inwards, $E_\tau$ outwards).
   Most \simcls{} are either unbound or confined by the MHD surface terms.
   Most \simcos{} are gravitationally bound rather than bound by the surface terms.
  }
  \label{fig:SurfWH}
\end{figure}

We show a comparison of both ratios in \reffig{fig:SurfWH}. A large portion of the \simcls{} experiences weak confinement ($Q_\mathrm{vir} \lesssim -0.5$), most often caused by MHD surface terms ($Q_\mathrm{cob} \approx -1$).
Five \simcls{} are clearly compressed beyond virialization (${Q_\mathrm{vir}>0.25}$) by the surface terms (at least for four out of the five cases: $Q_\mathrm{cob}<0 $).
Few \simcls{} are approximately virialized ($Q_\mathrm{vir}~0$) with varying contributions of the MHD surface terms.
Most of the \simcos{} are either virialized or close to being virialized ($Q_\mathrm{vir}~0$), with their confinement being dominated by gravity.
For most of the not fully virialized \simcos{} ($Q_\mathrm{vir}<0$) the deficit of confinement can be explained by an according amount of dispersive MHD surface terms ($Q_\mathrm{cob}>1$).

\subsection{Time-dependent EVT term}
\label{subsec:EVTana-timedep}
\begin{figure}
  \centering
  \includegraphics[width=1.0\columnwidth]{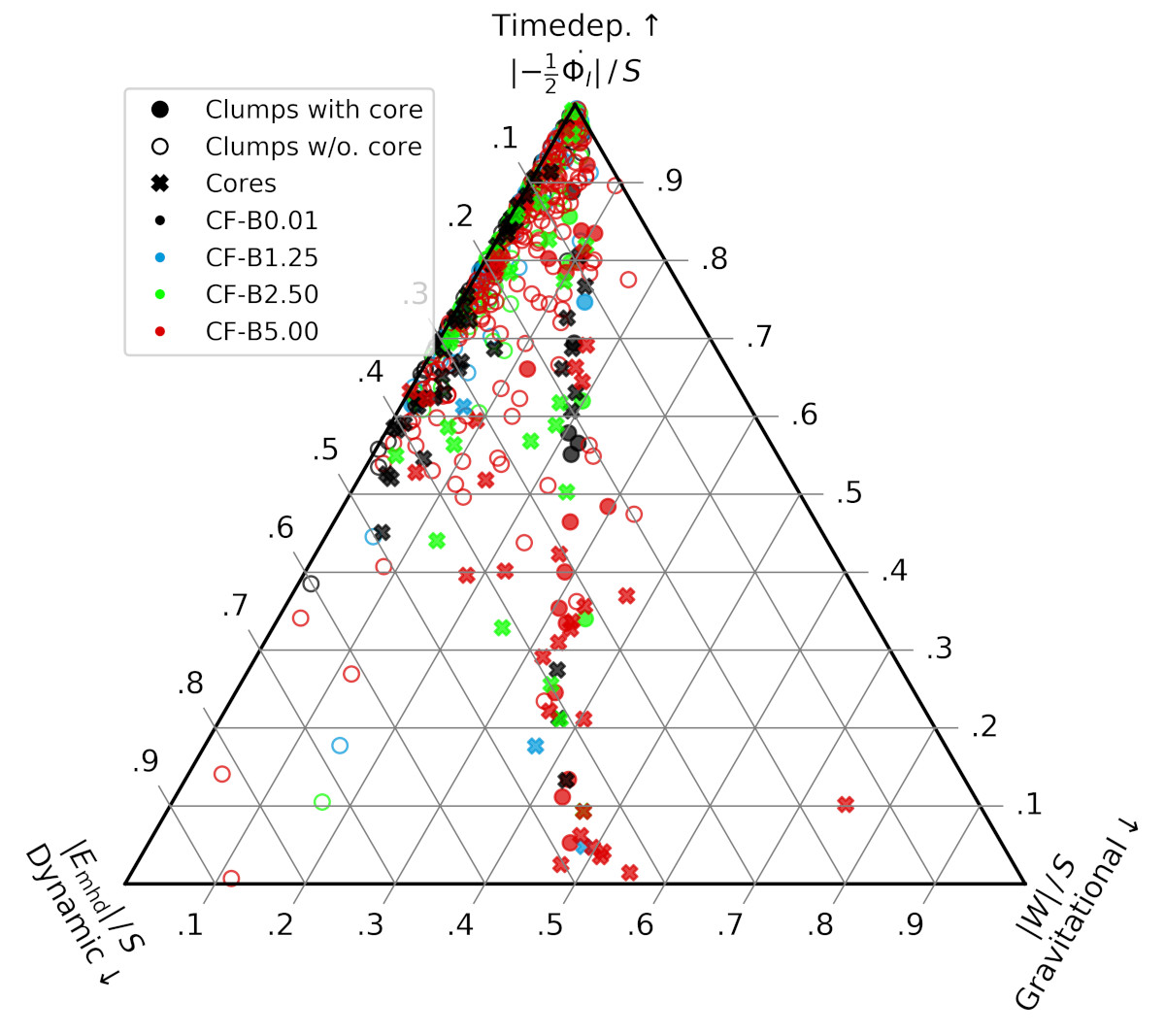}
  \caption{
  Ternary clump (circles) and core (crosses) balance of absolute
  gravitational vs. temporal vs. dynamic EVT components at \T{20}.
  }
  \label{fig:EVT-balance}
\end{figure}

The analysis presented in the previous Section addresses the static terms present in the virial theorem.
In addition to those, the virial theorem in its Eulerian formulation (see \refeq{eq:EVT})
contains the time-dependent component ${-\frac{1}{2}\dot{\Phi}_I}$,
where $\Phi_I$ is given in \refeq{eq:ES_Phi}. 
A comparison of the absolute contributions towards the EVT
of the aggregated MHD component $E_\mathrm{MHD}$ (see \refeq{eq:EMHD}),
the gravitational binding energy, $W$, and the time-dependent term ${\frac{1}{2}\dot{\Phi}_I}$
is shown in \reffig{fig:EVT-balance}. The values are normalized with
\begin{equation}
    S_\mathrm{EVT} = \left|E_\mathrm{MHD}\right| +\left|W\right| +\left|-\frac{1}{2}\dot{\Phi}_I \right|.
\end{equation}
In this way, the ternary is an isosceles triangle, displaying the relative contributions of the different terms, where a value right in the centre of the triangle implies that all three terms are equal. We show all \simcls{} (circles) and \simcos{} (crosses) at \T{20}. 

In the Lagrange frame, ${-\frac{1}{2}\dot{\Phi}_I}$ would not exist \citep{mckee1992virial} as an object would not have a fixed surface and hence the volume of the object would be allowed to change according to the flow properties. More or less any object deformation in the Lagrange frame should cause a non-zero ${-\frac{1}{2}\dot{\Phi}_I}$ in the Eulerian frame. In addition, accretion at a non-uniform rate also causes a contribution to ${-\frac{1}{2}\dot{\Phi}_I}$ (see below).

In \reffig{fig:EVT-balance}, we see that ${-\frac{1}{2}\dot{\Phi}_I}$ is basically never zero, but is rather significant for the majority of the \simcls{},
whose gravitational binding energy is small compared to the MHD contribution towards the EVT (left of the vertical center line).
Some of the \simcls{} fall onto the vertical center line implying comparable contributions of gravity and MHD terms,
while ${-\frac{1}{2}\dot{\Phi}_I}$ can still be significant (depending on their vertical position).
In contrast, a majority of the \simcos{} (crosses) have comparable contributions of gravity and MHD terms and exhibit only a small ${-\frac{1}{2}\dot{\Phi}_I}$ contribution towards the EVT.

We conclude that, while a majority of the \simcls{} is gravitationally unbound (see also Fig. \ref{fig:DynW}),
${-\frac{1}{2}\dot{\Phi}_I}$ is the dominant term for most of the unbound and some of the bound \simcls{}.
This implies that the EVT is overall not controlled by the object properties $E_\mathrm{MHD}$ and $W$, but mainly through the flux of $I_E$ at the statically defined object surface.
Hence, the choice of an object with a fixed surface is extremely difficult.
Most clump-like objects seem to be random fluctuations in the turbulent flow rather than well defined entities.
Only \simcls{} which actually contain \simcos{} develop in such a way that their dynamics "decouples" from the surrounding turbulent flow.
We further conclude that the \simcos{} are either gravitationally bound or at least gravitationally supported,
with the quantity $I_E$ mainly evolving along $E_\mathrm{MHD}$ and $W$.

From \refeq{eq:ES_Phi}, $\Phi_I$ can be viewed as contributing towards $\dot{I}$ by means of a mass flux density ${\rho\vec{v}\cdot\vec{n}_\mathrm{S}}$ accreting mass through the static surface elements $\mathrm{dS}$ at a distance $r$ from the centre of mass,
henceforth adding to the inertia quantity $I_\mathrm{E}$ (\refeq{eq:I_E}).
We now consider a simplified model for $\Phi_I$ by assuming mass flux onto a sphere of radius $R_I$
as of \refeq{eq:clsize},
\begin{equation}
  \Phi_{R} = -R_{I}^2\dot{M}\,\mathrm{,}
  \label{eq:PhiEst}
\end{equation}
and its time derivative
\begin{equation}
  \dot{\Phi}_{R} = -R_{I}^2\ddot{M} -2R_{I}\dot{R}_{I}\dot{M}\,\mathrm{,}
  \label{eq:PhiDotEst}
\end{equation}
where $\dot{M}$ is the object's mass accretion rate.
We note that while, due to the Eulerian reference frame, a static outer surface is assumed for the \simcls{} and \simcos{},
the radius approximation $R_I$ as of \refeq{eq:clsize} is still time-dependent,
as it is dependent on the distribution of the mass inside the volume that is bounded by that surface.

\begin{figure}
    \centering
    \includegraphics[width=\columnwidth]{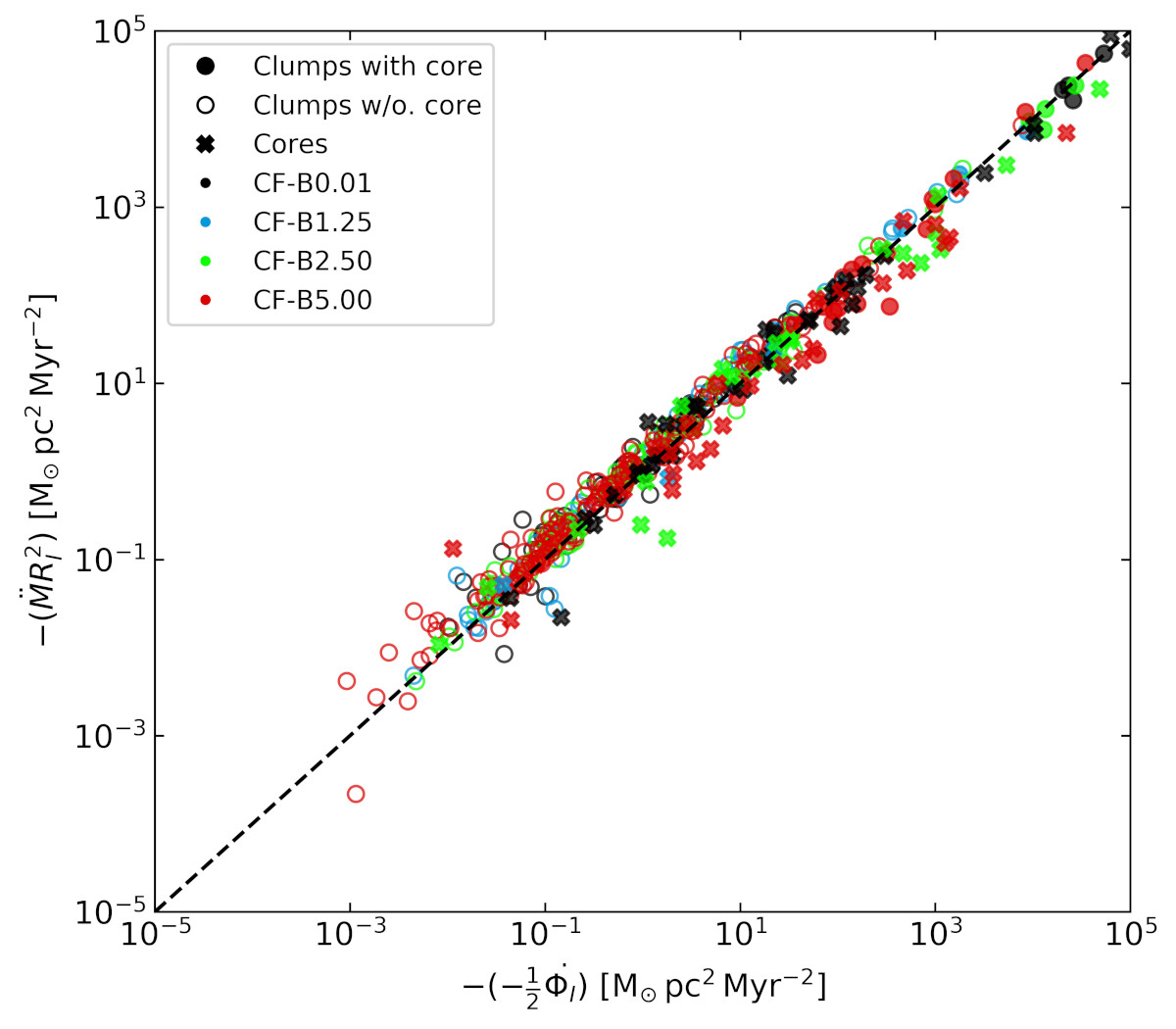}
    \caption{
    Comparison of the time derivative of the flux of moment of inertia through the surface,
    as modelled by the change of the mass accretion rate at fixed estimated radius $R_I$ (abscissa; see \refeq{eq:clsize}),
    vs. ${-\frac{1}{2}\dot{\Phi}_I}$ as derived from the Eulerian virial theorem (ordinate)
    of the detected \simcls{} (circles) and \simcos{} (crosses) at \T{20}.
    Independent of the initial magnetic field strength, the value of ${-R_I^2\ddot{M}}$ from the fixed radius model
    approximates the value of ${\frac{1}{2}\dot{\Phi}_I}$ very well over several orders of magnitude.
    }
    \label{fig:Timedep/AccrDot}
\end{figure}

\reffig{fig:Timedep/AccrDot} shows a comparison of the ${R_{I}^2\ddot{M}}$-component from \refeq{eq:PhiDotEst}
against the value of ${\dot{\Phi}_I}$ as obtained by numerical differentiation of the value obtained by \refeq{eq:gs-SPhi}. We note that we apply a central differencing scheme in time on the basis of single MHD time steps. 
Both quantities are correlated for \simcls{} and \simcos{}.
For many of the \simcls{} and independent of the initial magnetic field strength,
both quantities agree within factor 3 or better, often within a factor of 2,
while for most of the \simcos{}, the quantity ${-R_I^2\ddot{M}}$ is lower than $\dot{\Phi}_I$.
We conclude that for the \simcls{}, the quantity $\dot{\Phi}_I$ is closely related to changes $\ddot{M}$
of the objects mass accretion rate, as calculated in the Eulerian reference frame.
All \simcos{} seem to lie slightly below the one-to-one relation.
This seems to be counter-intuitive as we would expect the \simcos{} rather than the \simcls{} to be dominated by accretion.
However, we only plot the first term of \refeq{eq:PhiDotEst} in \reffig{fig:Timedep/AccrDot} and neglect the change of $R_I$,
which occurs when adding mass to a self-gravitating object.

\section{Discussion}
\label{sec:Discussion}

\subsection{Time evolution of cores in the Euler vs. Lagrange frame}
\label{subsec:timedep-meaning}
The time dependencies as given hitherto are related to the Eulerian reference frame, but they are not suited to trace the identified objects over time.

To capture the time evolution of our \simcos{},
we first run the object identification for a number of consecutive simulation snapshots, separated by ${\delta t=100\,\mathrm{kyr}}$ each. 
We then re-identify the \simcos{} from each of the snapshots and link them in order to determine their evolutionary tracks.
We use the following algorithm in order to determine if any two identified structures from consecutive snapshots should be
identified as the same object:
We first calculate the relative bulk velocity $\vec{v}_\mathrm{rel}$ of both objects.
The cell-centre coordinates of the earlier object are then propagated forward by ${\vec{v}_\mathrm{rel}\delta t}$.
For each structure, we define a "forward inclusion value $Q_{+,ij}$, by calculating the mass weighted fraction of forward propagated cell centres of the structure $i$ from the earlier snapshot that are located inside the structure $j$ found in the later snapshot.
The set of forward inclusion values for each of the earlier structures is then scaled according to:
\begin{equation}
    \widetilde{Q}_{+,ij} = \frac{Q_{+,ij}}{\mathrm{max}(1,\sum_j Q_{+,ij})}\,\mathrm{.}
\end{equation}
In an analogous way, we calculate the "backwards inclusion value", $\widetilde{Q}_{-,ij}$, by tracing the structure found in the later snapshot backwards in time.
The earlier structure is then considered to be a part of the later structure if ${\widetilde{Q}_{+,ij}>0.5}$,
and vice versa using $\widetilde{Q}_{-,ij}$.
Both structures are then considered to represent the same object if they are a part of each other by above definition; the objects' track at that time step is considered "undisturbed".

We only apply the tracking algorithm to the \simcos{} (starting at \T{20} in both directions in time),
as they are found to have high inclusion values, while the \simcls{} seem to be more transient and are hence harder to track over time.

\begin{figure}
  \centering
  \includegraphics[width=\columnwidth]{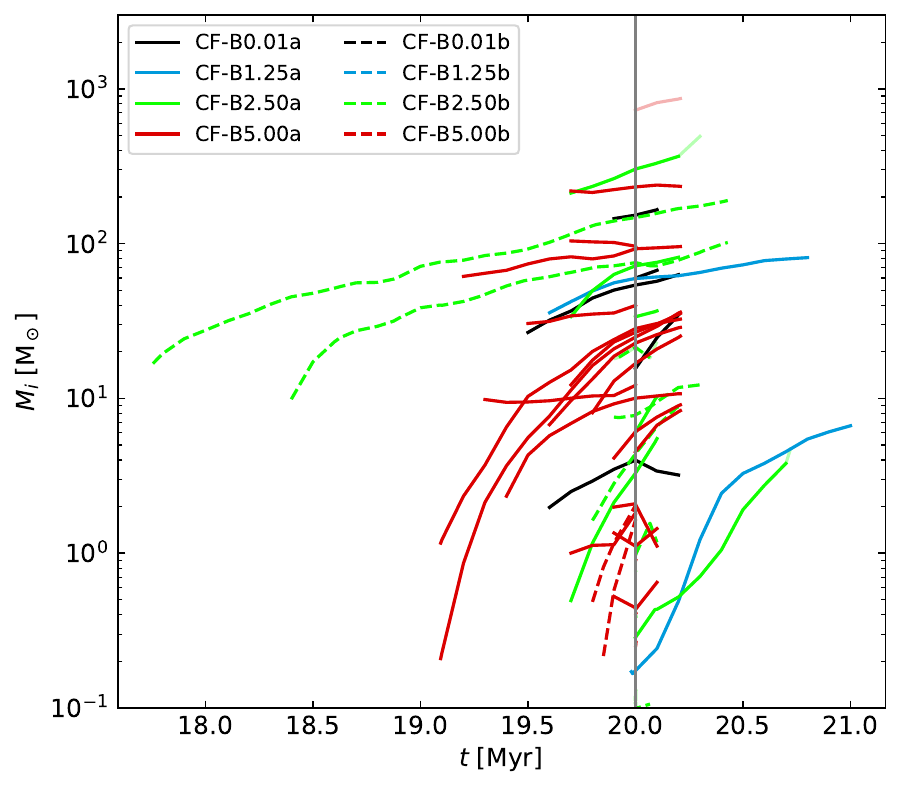}
  \caption[Mass Evolution Traces]{
  Time evolution of the mass of individual \simcos{} as connected via our tracking algorithm. 
  The solid lines show tracks of uniquely \simcos{}, while the dotted lines show tracks of \simcos{} which either fragmented or merged with other \simcos{} while having less than half of the original core's mass.
  }
  \label{fig:TimeDep/Traces}
\end{figure}
In \reffig{fig:TimeDep/Traces}, we show the resulting mass as a function of time for the \simcos{} in different simulations.
Most \simcos{} seem to be accreting, with typical accretion rates in the order of ${10--100\,\mathrm{M}_\odot\,\mathrm{Myr}^{-1}}$ and the trend of having decelerating accretion rates.
A few of the \simcos{} do not stay in isolation but merge with others or fragment, respectively.
These are depicted using dotted lines.

\begin{figure}
  \centering
  \includegraphics[width=\columnwidth]{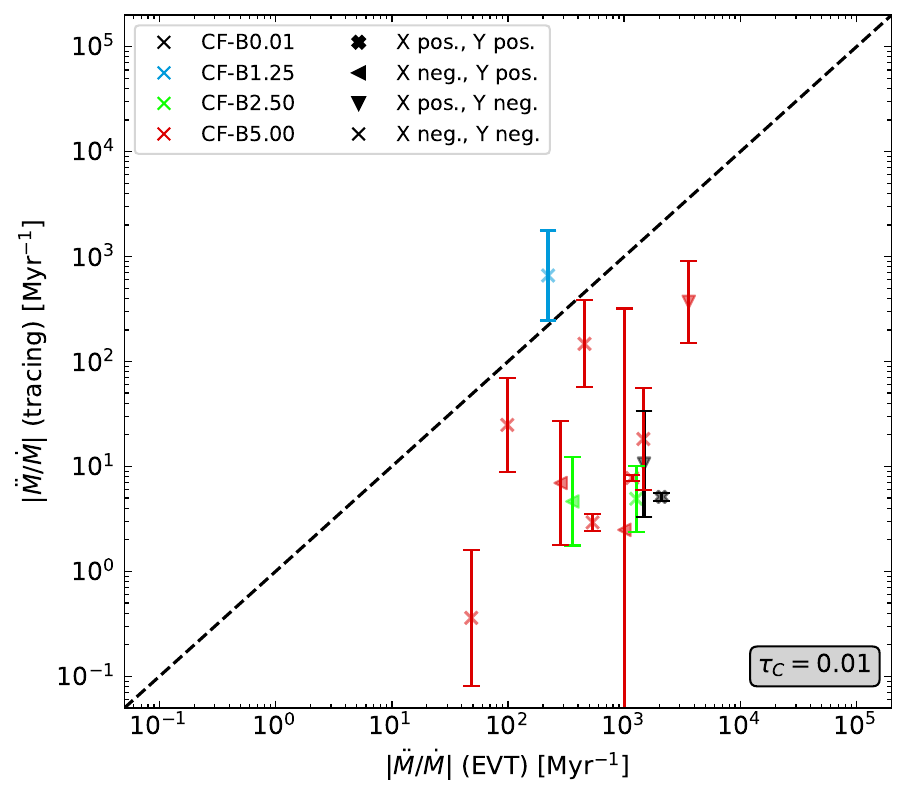}
  \caption{
  Comparison of relative rate of change of the mass accretion rate, ${\ddot{M}/\dot{M}}$, of trackable \simcos{} at \T{20},
  as obtained from object tracing (see Fig.~\ref{fig:TimeDep/Traces}) vs. from the EVT derivation.
  The error bars display the difference in value between calculating $\ddot{M}$ from ${M}$ using intervals of two plotfiles versus one plotfile.
  We include a one-to-one line (dashed line) to guide the eye.
  Both quantities are largely uncorrelated.
  }
  \label{fig:TimeDep/MddMdComparison}
\end{figure}
From the mass evolution, we calculate the relative rate of change of the mass accretion rate, ${\ddot{M}/\dot{M}}$.
We may also compute the same quantity from the mass flux across the \simcos' surface at a given time in the Eulerian reference frame. 
The two approaches are compared in \reffig{fig:TimeDep/MddMdComparison}.
Note that we show the absolute values $\left|{\ddot{M}/\dot{M}}\right|$, while positive and negative values are indicated with different symbols. 
A separate comparison for the accretion rate $\dot{M}$ and its rate of change $\ddot{M}$ is shown in the appendix, Figures \ref{fig:md-comparison}, \ref{fig:mdd-comparison}.
Most \simcos{} have a positive ${\ddot{M}/\dot{M}}$ as derived from the mass tracks, while the respective quantity for the same objects obtained from the Eulerian frame are roughly equally positive or negative.
Also, the typical values of $\|{\ddot{M}/\dot{M}}\|$ are different between the different methods: from the mass tracks, we find typical values of $\sim {10\,\mathrm{Myr}^{-1}}$, while we obtain a mean of $\sim {10^3\,\mathrm{Myr}^{-1}}$ from the Eulerian approach.
A Kendall rank correlation test measuring the correlation between both quantities gives a coefficient as low as $\tau_C=0.01$.
We conclude that ${\ddot{M}/\dot{M}}$, as measured by tracking the actual \simcos{}, is very weakly correlated to the same quantity measured in the Eulerian reference frame.
This implies that an object's surface area as well as its volume are not very well defined and change as a function of time.
Hence the time-evolution of individual objects seems to be not very well captured by the Eulerian analysis.
This result is supported by the fact that we find large ${-\frac{1}{2}\dot{\Phi}_I}$-terms (see Section~\ref{subsec:EVTana-timedep}).

\subsection{Analysis of \simcos{}}
\label{subsec:Discussion/CoreCand}
\begin{figure}
    \centering
    \includegraphics[width=\columnwidth]{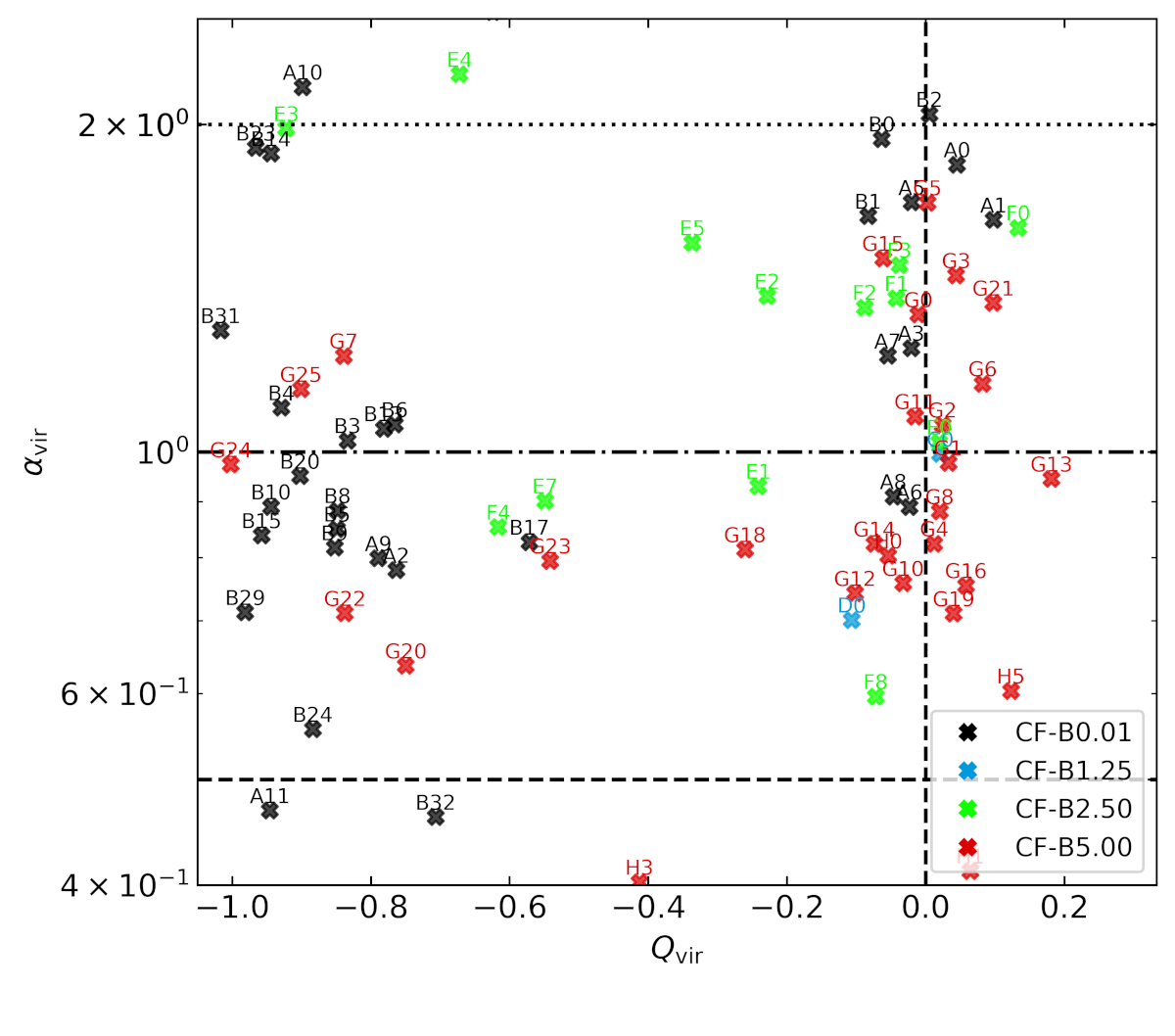}
    \caption{
    Observed virial state vs. full virial analysis of all \simcos{}.
    The virialization ratio $Q_\mathrm{vir}$ on the \dir{x}-axis, as of \refeq{eq:Qvir},
    corresponds to the fully evaluated virial balance (see \reffig{fig:EVT-balance}).
    A virialized state as of the EVT definition is represented by ${Q_\mathrm{vir}=0}$ (dashed vertical line),
    while an unbound object would have ${Q_\mathrm{vir}<0}$. On the \dir{y}-axis, we show $\alpha_\mathrm{vir}$,
    the usually used virial parameter, as obtained
    by ${\alpha_\mathrm{vir}=5\sigma_\mathrm{1D}^2/(R\Sigma\pi G)}$ analogous to the Heyer-plot shown in \reffig{fig:Heyer}.
    }
    \label{fig:CoreCand/Examples}
\end{figure}

An indication of the virial state of a molecular cloud structure that can be derived from observational data \rev{was} given by the Heyer plot \rev{above}, \reffig{fig:Heyer},
while a more comprehensive assessment of the virial state \rev{was} obtained from the evaluation of the EVT for the same object, \reffig{fig:EVT-balance}.
A comparison of both results is shown in \reffig{fig:CoreCand/Examples}.
\rev{
We note that some of our detected \revc{3D} objects are far from being spherical.
Clumps with elongated and irregular shapes are also seen in observations \citep[see eg., Fig. 7 in][]{Clarke2023Filament}.
In contrast, observed \revc{dense} cores are often recorded as being \revb{spheroidal}.
}
\revc{
This might at least partially be a consequence of their shape being affected by observational effects \citep[see e.g. Fig.~3 in][]{Hsu2023Cmf}.
However, we note that the mass, velocity dispersion \revd{(Fig. \ref{fig:vdisp-core-parent}), and infall velocity (Fig. \ref{fig:vrad}) \citep[compare e.g.][]{Keown2016Infall} of many of our \simcos{}
are considerably higher than observed for low-mass dense cores.}
They are rather similar to those of high-mass protostellar cores, for which morphological features have been observed \citep[see e.g. Fig.~1 in][]{Beuther2023Sf}.
This could be due to our search algorithm, which requires 3D-cores to be highly shielded objects in 3D. We note that the $A_{\mathrm{V,3D}}$ is generally significantly lower than the observed visual extinction in 2D \citep[see][]{haid2019silcc}.
Also, we require at least 30 cells per object in order to be able to properly carry out the virial analysis.
As shown in Fig.~\ref{fig:Larson}, spherical objects with radii less than ${1.506\times 10^{-2}}$~pc are unresolved.}

\revc{Column density projections} of the \simcos{} can be found in the appendix, Fig. ~\ref{fig:cores-A0}--\ref{fig:cores-D20}.

The comparison in \reffig{fig:CoreCand/Examples} shows that the two methods of assessing the virial state of a core are not fully consistent with each other:
All \simcos{} occupy the range \rev{${0.4<\alpha_\mathrm{vir}<2.1}$}.
In contrast, some of them exhibit small virialization ratios $Q_\mathrm{vir}$, as defined in \refeq{eq:Qvir}, in a range of ${Q_\mathrm{vir}<-\frac{1}{3}}$,
which implies they are actually far from being virialized when considering all relevant terms.
We stress that for determining $\alpha_\mathrm{vir}$,
only information on the velocity and mass distribution inside the object volume is incorporated,
while $Q_\mathrm{vir}$ from the full virial analysis is constructed using knowledge on the kinetic, thermal and magnetic state
of the gas inside the objects volume and on its surface as well as the gravitational field configuration caused by the global mass distribution.
The difference in assessment between both methods can be attributed to lacking information on any of those particular fields in the construction of $\alpha_\mathrm{vir}$.

This inconsistency is not caused by tidal interactions arising from the respective external mass distribution
because the \simcos{} masses are well correlated with their gravitational binding energies $W$ (see \reffig{fig:mass-W}).
Kinetic interactions dominate the virial state of the \simcos{},
while thermal as well as magnetic interactions are of little importance (see \reffig{fig:CO-ter-dyn}).
For all \simcos{}, the thermal and magnetic surface terms are considerably smaller than the corresponding volume energy terms in the EVT (see \reffig{fig:VolSurf}),
whereas it can be different for the kinetic surface term.
Furthermore, considering the confinement balance ratio $Q_\mathrm{cob}$ (\reffig{fig:SurfWH}),
the kinetic as well as total surface term for most of the \simcos{} is positive and clearly correlated with the binding energy deficit,
as compared with a virialized state of those \simcos{}.
This suggests that the kinetic surface term is the main source of the inconsistency between $Q_\mathrm{vir}$ and $\alpha_\mathrm{vir}$.

\section{Conclusions}
\label{sec:Conclusions}

We use 3D ideal MHD simulations, which include gravity and a chemical model that considers radiative cooling and heating
to simulate a scenario of colliding magnetized gas flows in order to model the formation of molecular clouds
\revb{and their substructures (\simcls{} and \simcos).}
We analyse four simulations with varying initial magnetic field strength between 0.01--5~$\mu$G.

As a function of the initial magnetic field strength the structure of the molecular cloud changes gradually.
\revb{
With increasing field strength\revb{, large scale} filamentary structures become less prominent since gas motions perpendicular to the magnetic field are suppressed.
Since the magnetic field is initially directed parallel to the flow and perpendicular to the sheet that forms in the centre,
the entire central sheet starts to fragment in our high magnetization runs, leading to a higher number of fragments with a smaller average mass each.}

Based on the local CO abundance, we detect molecular \simcls{}, i.e. connected volumes with at least \rev{30} cells, in the simulations.
\revb{Additionally, we detect \simcos{} which we define as connected objects exhibiting a local 3D visual extinction of at least 8~mag throughout,
while being located inside a clump. 
We reject \simcls{} and \simcos{} below a minimum resolution threshold of 30 cells.
While this creates a bias against low-mass dense cores, this allows us to fully analyse the 3D structure of the remaining objects.}
We find that after $20\,\mathrm{Myr}$, 30\%-50\% of the total molecular mass in our simulations is found in \simcls{},
20\%-60\% of which are concentrated in the contained \simcos{}.
We find projection effects to be important:
Core formation efficiencies around 10\%, as derived from the 2D projected column density maps, seemingly in agreement with observations \citep[][]{konyves2015census}.
However, like \citet{haid2019silcc}, we find that the column-density-based visual extinction $A_{\mathrm{V,2D}}$ overestimates the local shielding $A_{\mathrm{V,3D}}$.
\revb{
As a consequence of our core detection criterion the \simcos{} comprise almost the entire shielded gas above a 3D visual extinction of $A_{\mathrm{V,3D}}>8$.
}

For each of our simulations, at \T{20}, the ensemble of smaller \simcls{} roughly follows the $\sigma$-size relation found by \citet{heyer2004universality},
which is steeper than the original Larson relation.
However, \simcls{} that are more massive than approximately ${10\,\mathrm{M}_\odot}$ deviate towards higher velocity dispersions.
This is \rev{partly} explained by the increased average surface density of those \simcls{} caused by the occurrence of contained \simcos{}.
\rev{The dynamical state of these \simcls{} is mostly consistent with the \citet{heyer2009re}-relation assuming virial parameters $\alpha_\mathrm{vir}>2$.}
Overall, we observe $\sigma$-size and mass-size relations for our \simcls{} that are rather consistent with them being transient objects that are formed by turbulence.
Most of our \simcls{} exhibit a virial parameter that is not consistent with objects in gravitational-kinetic equipartition,
populating a low surface density tail in the Heyer relation.
The \simcos' surface densities are $\sim 1.5\,\mathrm{dex}$ larger than those of most \simcls{} and populate the ${0.5<\alpha_\mathrm{vir}<2.0}$ range of virial parameters. 
Interestingly, there is no clear correlation of the clump and core ensembles' $\sigma$-size- and mass-size-slopes, as well as of the scaling coefficient in the Heyer relation, with the initial magnetic field strength $B_0$.

Our full virial analysis reveals that most of the detected \simcls{} are unbound objects far from virial equilibrium.
For those, the virial surface and volume quantities are often of comparable magnitude, suggesting that they are transient objects.
Thermal and magnetic energy are only significant for a few of the less massive, coreless \simcls{}. 
Compression (or confinement), if present, is mostly driven by ram pressure or thermal pressure rather than gravity (considering self-gravity and the weight of the surrounding medium), except for the largest \simcls{}.
On the other hand, more massive \simcls{} (containing \simcos) tend to be closer to virial equilibrium ($\alpha_\mathrm{vir}\sim 1.5 - 4$).
They are dominated by the interplay of volume kinetic energy and gravity and can be found toward higher surface densities in the Heyer relation.


All \simcos{} are kinetically dominated while thermal and magnetic contributions towards the virial theorem are small.
In fact, most \simcos{} are close to virial equilibrium, yet some are slightly affected by dispersing kinetic surface effects. 
Hence, the cores are overall well described by the normal virial parameter, which explains why they are located close to ${\alpha_\mathrm{vir}=1}$ in the Heyer relation,
valid for objects in gravitational-kinetic equipartition.

The quantitative analysis of the EVT's time-dependent term ($-\frac{1}{2}\dot{\Phi}_I$) shows that the flux of the moment of inertia through an objects' surface is typically significant for the unbound \simcls{}.
This term becomes less important for the denser \simcls{}, which host \simcos{}, and for the \simcos{} themselves.
We show that this term is mostly caused by the rate of change of the mass accretion rate through the objects' static surface in the Eulerian reference frame.
However, this mass flow does not correspond to a real accretion rate following the common interpretation of self-gravitating, well-defined cores which gain mass by accretion. 
This can be seen when we compare the rate of change of the mass accretion rate derived from the EVT to the same value derived for specific \simcos{} which can be tracked in time. 

\revb{To summarise, the \simcls{} and \simcos{} in our simulations form and evolve dynamically.
They may be transient objects, they may evolve in isolation, or merge with other objects.
Often, our \simcos{} are flattened structures.
The fate of a particular core is difficult to track, but our virial analysis gives insights into the stability of the identified objects, in particular into the role of gravity and their boundedness. 
In observations, the identification of a core depends on the projection angle, the beam size, etc..
For example, flattened objects would appear quite distinct from different points of view. 
We do not claim that the \simcls{} and \simcos{} discussed in this paper are directly comparable to observed clumps and cores.
In fact, they are certainly not directly comparable.
\revc{Firstly,} synthetic observations would be required for a direct comparison.
\revc{Secondly, the resolution of our simulations restricts us from analysing low-mass dense cores,
whereas \revd{most of the remaining \simcos{} resemble massive cloud cores rather than low-mass dense-cores.}}
Of course, ultimately, a detailed comparison of simulations and observations is key.
We plan to follow up on this point in subsequent work.}


\section*{Data availability}
The data underlying this article will be shared on reasonable request to the corresponding author.

\section*{Acknowledgements}
\revb{We thank the reviewer for the detailed and helpful input.}
\revb{We also thank Masato I. N. Kobayashi for the insightful discussions during the revision process.}
We thank the Deutsche Forschungsgemeinschaft (DFG) for funding through the SFB~956 ''The conditions and impact of star formation'' (sub-projects C5 and C6) \revb{ as well as via Collaborative Research Center 1601 (SFB~1601; sub-project B6).}
The software used in this work was developed in part by the DOE NNSA- and DOE Office of Science-supported Flash Center for Computational Science at the University of Chicago and the University of Rochester. We particularly thank the Regional Computing Center Cologne for providing the computational facilities for this project by hosting our supercomputing cluster "Odin" and the Leibniz-Computing Center in Garching for providing computational resources on SuperMuc via project pr62ni.
\revb{We thank the MPI for Astrophysics in Garching for the cooperation which provided us with the compute cluster Odin and thus, a substantial fraction of the necessary computing resources for this project.}
The authors are also thankful for VisIt \citep{HPV:VisIt}, which allowed us to produce volume-rendered visualizations.
VisIt is supported by the Department of Energy with funding from the Advanced Simulation and Computing Program and the Scientific Discovery through Advanced Computing Program.



\bibliographystyle{mnras}
\bibliography{virpaper}


\clearpage
\appendix

\section{TreeRay optical depth}
\label{sec:TreerayOpticalDepth}
The \codename{TreeRay/OpticalDepth} module is used to calculate the local extinction $A_{\mathrm{V,3D}}$ of the external UV field.
This Section provides a short overview of the algorithm; for a complete description of the algorithm and its implementation, please refer to \cite{wunsch2018}.

The module works by evaluating the solution to
\begin{equation}
    I_{v} = I_{v,0}\exp{\left(-\tau_{v}\right)}\, ,
    \label{eq:RadTrans}
\end{equation}
where $I_{\nu}$ is the specific intensity of the UV field at the target location and at frequency $\nu$,
$I_{\nu,0}$ is the specific intensity of the external field source, and $\tau_{\nu}$ is the optical depth, which is proportional to the total hydrogen column density along the path connecting both as well as its associated absorption cross-section.

For any target cell in the simulation domain, a \codename{healpix} sphere \citep{gorski2005healpix} using $N_\mathrm{PIX}=48$~pixels is constructed,
each pixel representing a beam radiating towards the pixel from a specific direction.
Those pixels then comprise a temporary $N_{\mathrm{H},i_\mathrm{PIX}}$ column density map, with each pixel mapped to the AMR octree-nodes on the line of sight along the beam.
The \codename{TreeRay} mechanism then walks the entire octree at each timestep, in turn filling those maps.
The mean visual extinction of a given cell is then determined by
\begin{equation}
    A_{\mathrm{V,3D}} = -\frac{1}{2.5}\ln{\left[ \frac{1}{N_\mathrm{PIX}} \sum_{i_\mathrm{PIX}=1}^{N_\mathrm{PIX}} \exp{\left(-2.5\frac{N_{\mathrm{H},i_\mathrm{PIX}}}{1.87\e{21}\,\mathrm{cm}^{-2}}\right)} \right]}\, ,
\end{equation}
where $1.87\e{21}\,\mathrm{cm}^{-2}$ is the ratio between the hydrogen column density and the visual extinction given by \citet{draine1996structure}.

\begin{figure}
  \centering
  \includegraphics[width=.98\columnwidth]{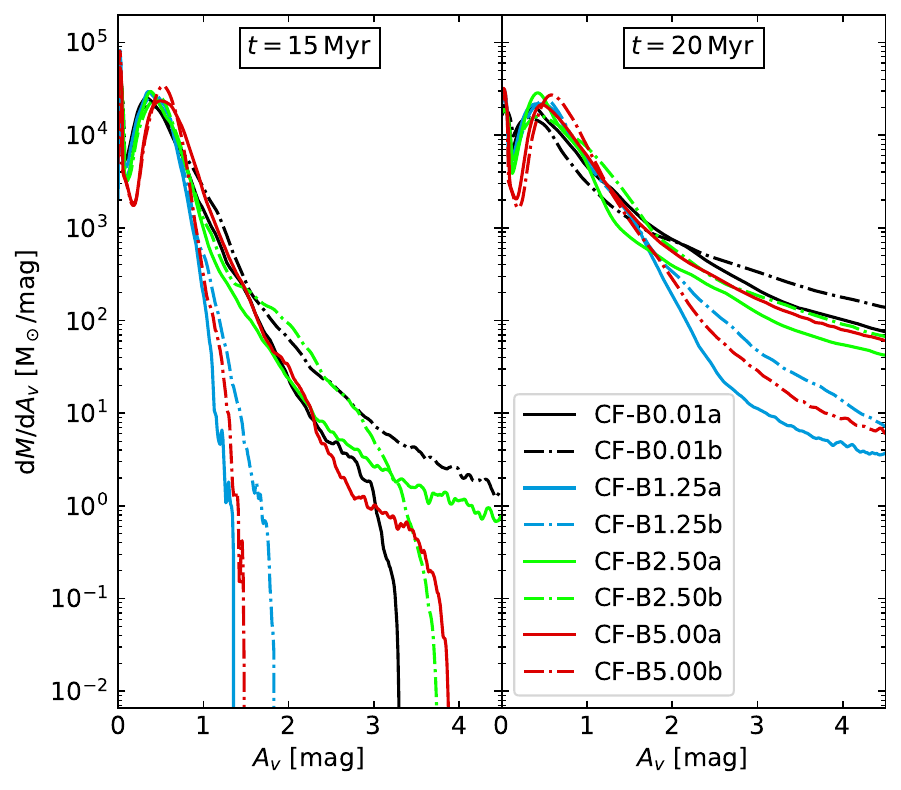}
  \caption{
  Mass-weighted PDF of the 3D visual extinction $A_{\mathrm{V,3D}}$.
  \rev{
  At \T{15}, a large amount of gas is shielded at extinction levels around $A_{\mathrm{V,3D}}=0.35$~mag (up to runs \CF{2.5}a/b),
  while bulk gas shielding is increased to around $A_{\mathrm{V,3D}}=0.6$~mag for runs \CF{5.0}a/b.
  Almost the entire gas mass sits at $A_{\mathrm{V,3D}}<3$~mag.
  At \T{20} in most runs (all but \CF{1.25}a/b and \CF{5.00}b),
  several $100\,\mathrm{M}_\odot$ of gas are located at large extinction levels $A_{\mathrm{V,3D}}>4$~mag.
  We note that the peak at $A_{\mathrm{V,3D}}=0$ is attributable to the inflow area.
  }
  }
  \label{fig:Av-pdf}
\end{figure}

Fig. \ref{fig:Av-pdf} shows a mass-weighted PDF of the resulting 3D visual extinction $A_{\mathrm{V,3D}}$.

\section{Morphology}
\begin{figure*}
\centering
  \includegraphics[width=\textwidth]{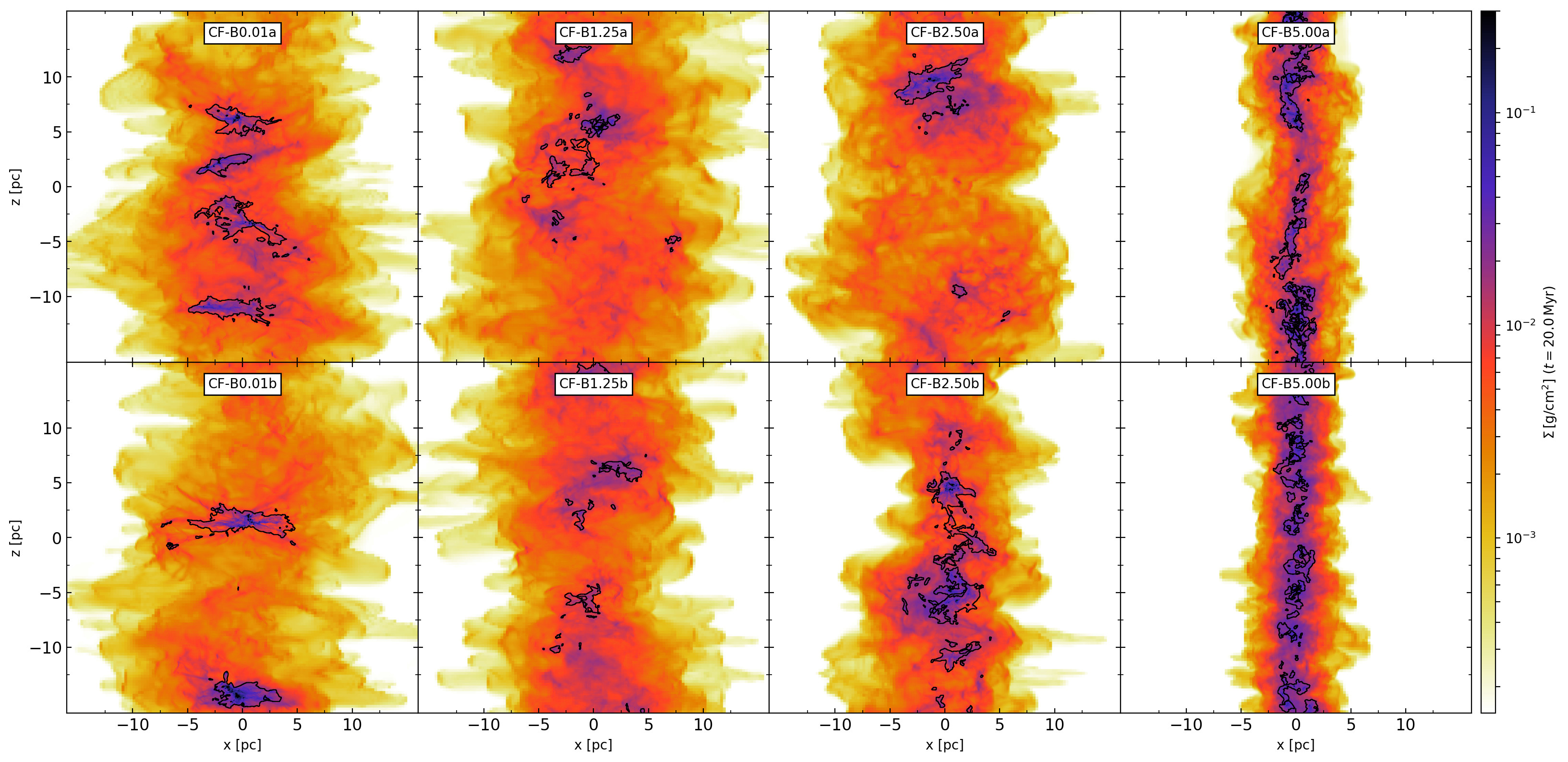}
\caption{
    \rev{
    Edge-on total gas column densities for all simulations at \T{20}.
    The top row represents the simulation runs CFXa with the low spatial frequency interface,
    while the bottom row represents the simulation runs CFXb with the high spatial frequency interface.
    }
    The detected \simcls{} are indicated by the black contour.
    }
    \label{fig:coldens-z}
\end{figure*}

Supplementing Fig. \ref{fig:coldens-x} in Section \ref{subsec:sim-morph},
Fig. \ref{fig:coldens-z} shows the edge-on column density maps for all simulations at \rev{\T{20}.}
The line of sight is along the \dir{z}-direction and perpendicular to the flow direction.

\label{sec:magdir}
\revb{
While the magnetic field is initially homogeneous and aligned with the inflow direction, $\vec{B}=B_0\vec{e_x}$,
we allow the magnetic field to dynamically evolve through the runtime of the simulation,
as described by Eqs. \ref{eq:EMHD}.
}

\begin{figure}
  \centering
  \includegraphics[width=\columnwidth]{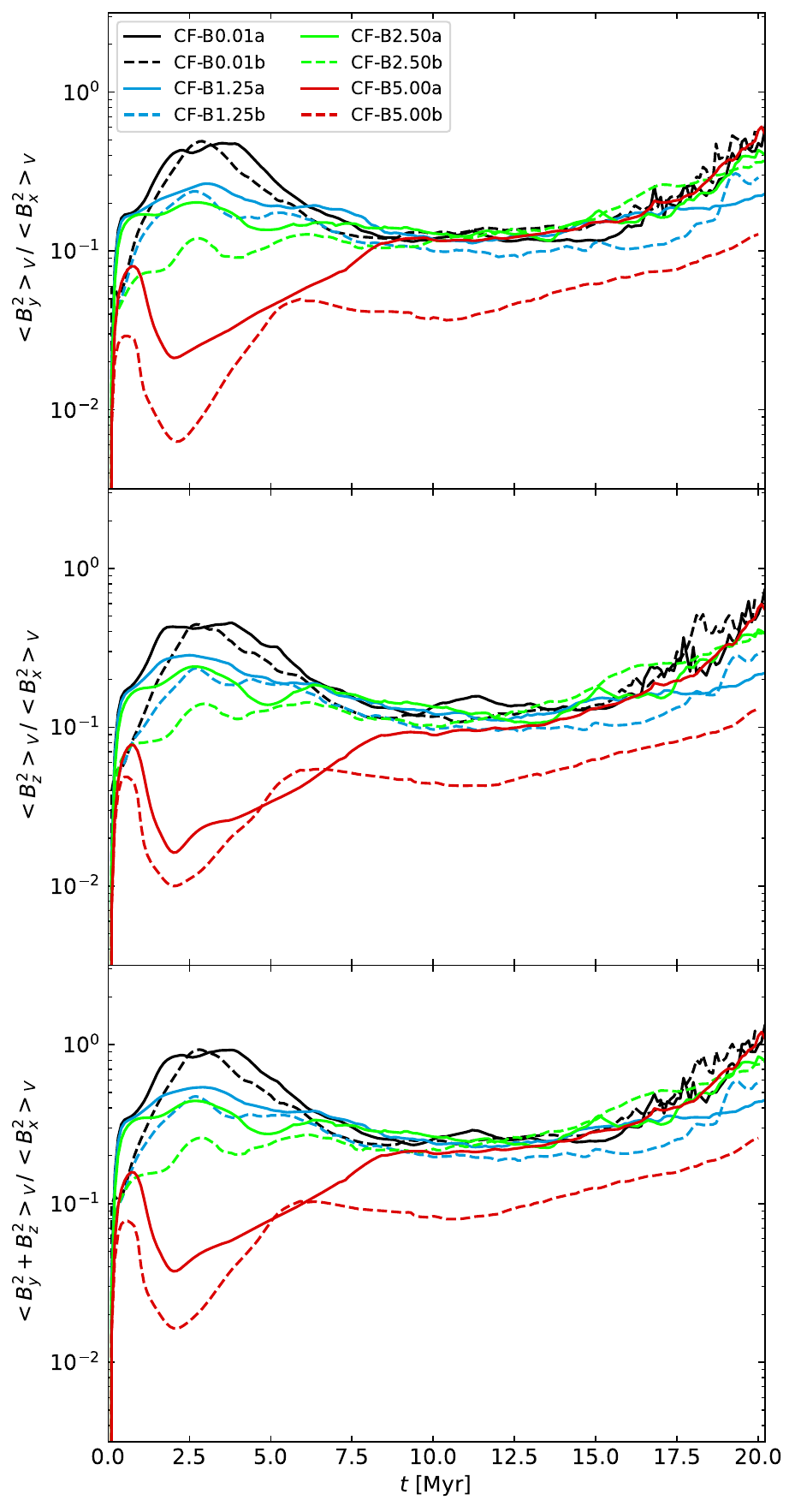}
  \caption{\revb{
      Ratios of volume weighted averages $\langle B_y^2\rangle_V/\langle B_x^2\rangle_V$ (top), $\langle B_z^2\rangle_V/\langle B_x^2\rangle_V$ (centre),
      and $\langle B_y^2+B_z^2\rangle_V/\langle B_x^2\rangle_V$ (bottom) vs. $T$ for the central region $|X|<4\,\mathrm{pc}$.
      }}
  \label{fig:B-ratios}
\end{figure}
\revb{
Figure \ref{fig:B-ratios} shows the ratios of different magnetic energy components $E_{B_i}\propto \langle B_i^2\rangle_V$ to $E_{B_x}\propto \langle B_x^2\rangle_V$ for the central region of $|X|<4\,\mathrm{pc}$ around the central region.
The top panel shows $i =y$, the middle panel $i=z$, and the bottom panel shows the sum of the $y-$ and $z-$components. 
Initially, $B_y = B_z = 0$, but the magnetic field builds up quickly.
For all simulations except \CF{5.0}a/b, the ratios peak around \T{3}, then drop slightly, but continue to grow after \T{10}.
At \T{20}, the combined magnetic energy contained in the \dir{y} and \dir{z}-components is between 30\% and more than 100\% of the \dir{x}-component.
}

\begin{figure*}
  \centering
  \includegraphics[width=\textwidth]{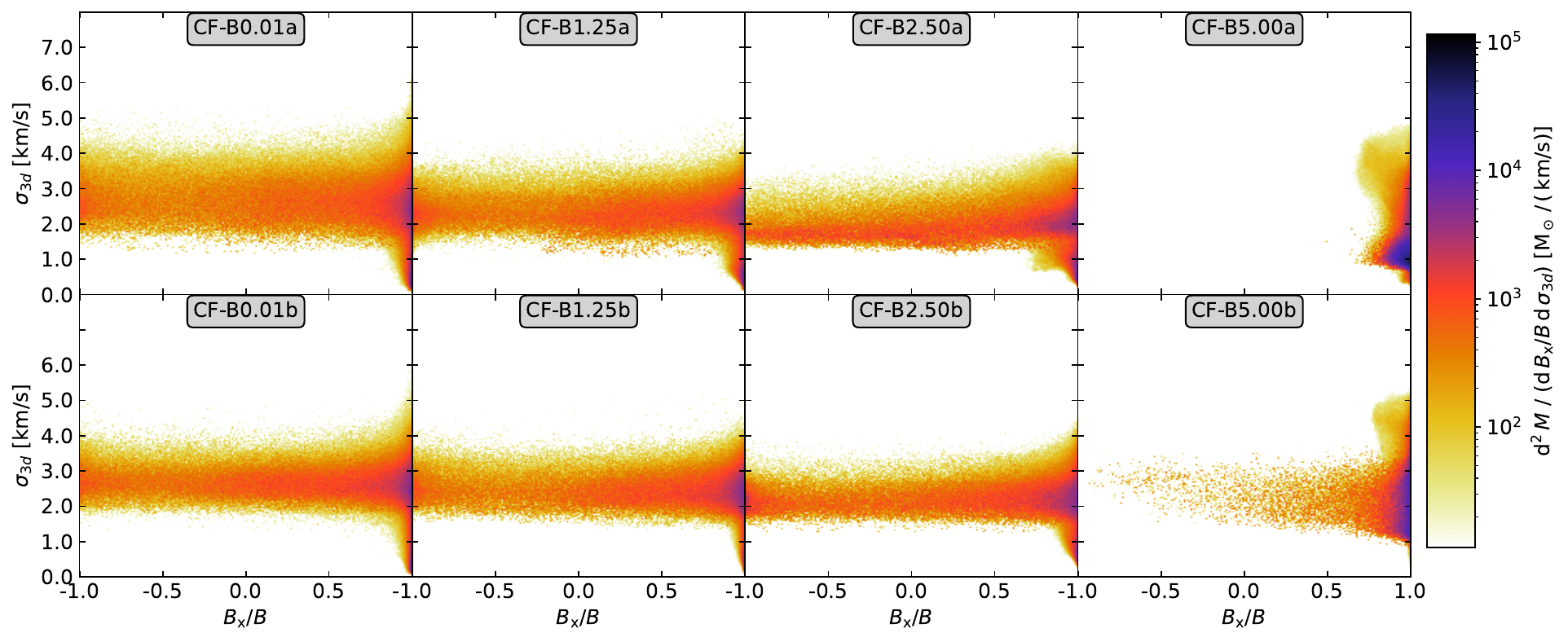}
  \includegraphics[width=\textwidth]{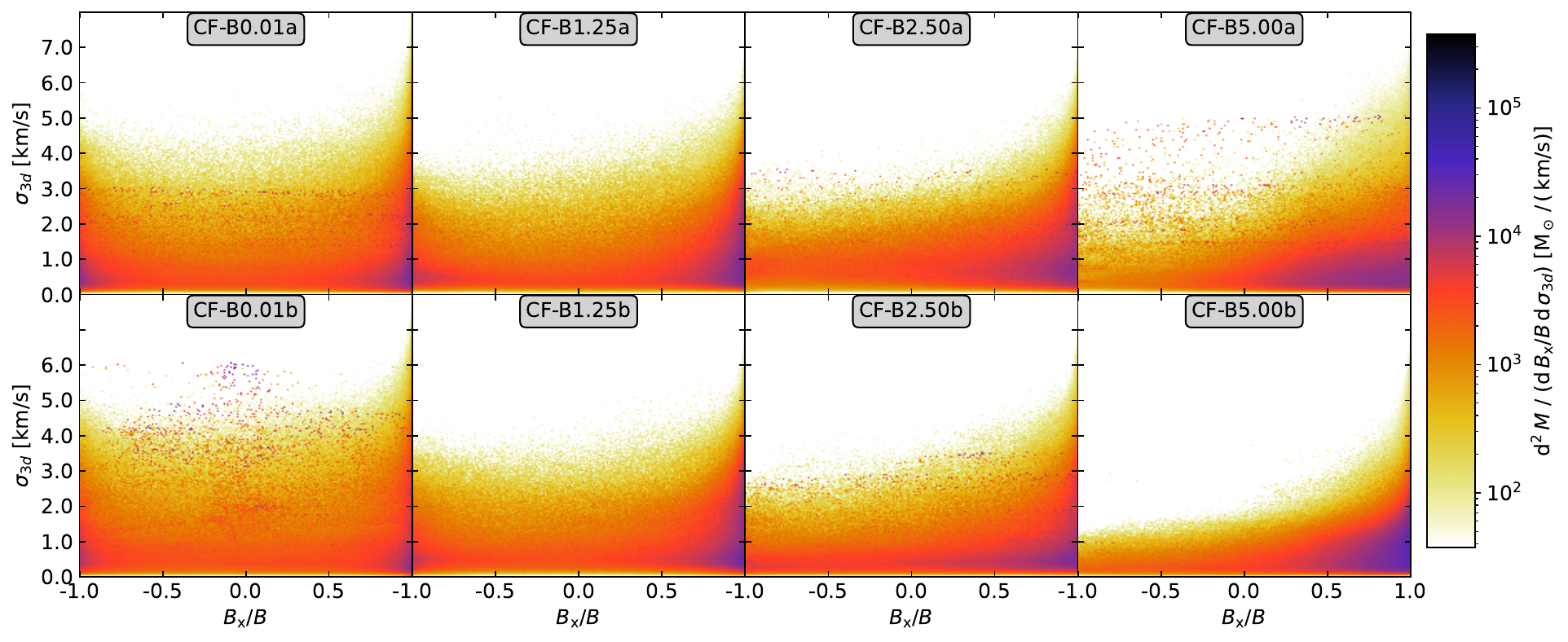}
  \caption{\revb{
      Mass-weighted 2D-Histogram of $B_x/|B|$ vs. the mass weighted local velocity dispersion at \T{3} (top) and \T{20} (bottom).
      Here, $B_x/|B|$ equates to the cosine of the angle between the direction of the magnetic field and its initial direction.
      The local velocity dispersion is calculated using a Epanechnikov Kernel on a support radius covering a 4-cell radius.
      The colour indicates the normalised mass load of each bin.
      }}
  \label{fig:B-dir}
\end{figure*}
\revb{
Fig. \ref{fig:B-dir} shows the relation between $B_x/|B|$, indicating the angle of the magnetic field vs. the initial field direction,
and the local gas velocity dispersion $\sigma_{3D}$, at \T{3} (top panel) and \T{20} (bottom panel), respectively.
The locality of the velocity dispersion in reference to each point is ensured by applying weights according to the Epanechnikov Kernel on a support radius covering a 4-cell radius around that point.
We expect $\sigma_{3D}$ to be close to zero for the mass located in the inflow area,
while having a magnetic field vector $\vec{B}$ close to the initial field direction,
hence populating the lower right corner of each plot in Fig. \ref{fig:B-dir}.
}

\revb{
At \T{3} (top), for a large part of the mass with intermediate $\sigma_{3D}$ in the range of $1.5\text{--}3\,\mathrm{km}\,\mathrm{s}^{-1}$,
the magnetic field directions are distributed over a wide range of $B_x/|B|$.
We interpret this as an indication for the reordering of the magnetic field by the turbulent motions of the gas in the shocked sheet.
At \T{20} (bottom), the properties of the gas are distributed over a wide range in $B_x/|B|$ and $\sigma_{3D}$. 
In particular for low initial field strengths, a notable amount of mass is located near $B_x/|B|=0$, indicating that the magnetic field is perpendicular to the initial field in those locations.
}

\revb{
\section{Derivation of the \textsc{getsf} filter criterion}
For the sources detected by \textsc{getsf}, we consider a filter criterion in the form of a minimum threshold for each source's average density.
The criterion is designed to exclude those sources that are unlikely to have a counterpart among our \simcos{}, which, in turn, are detected using a threshold for the 3D visual extinction $A_{\mathrm{V,3D}}$.
\begin{figure*}
  \centering
  \includegraphics[width=\textwidth]{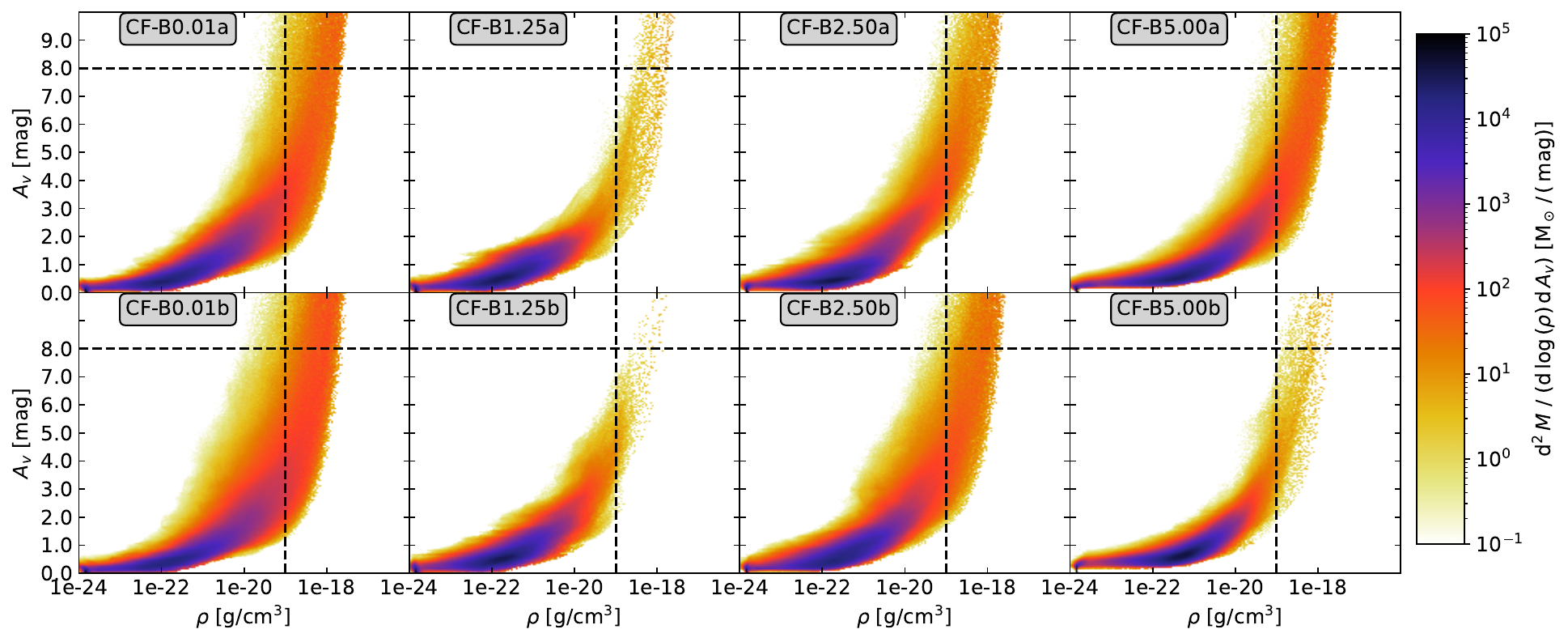}
  \caption{\revb{
    Mass-weighted histogram of the 3D visual extinction $A_{\mathrm{V,3D}}$ vs. the gas density $\rho$ at \T{20},
    with $A_{\mathrm{V,3D}}$ being calculated using the Treeray method.
    The horizontal dashed lines indicate the $A_{\mathrm{V,3D}}>8\,\mathrm{mag}$ threshold of our 3D core detection.
    The vertical dashed lines indicate the density threshold $\rho_\mathrm{avg}>1\e{-19}\,\mathrm{g}\,\mathrm{cm}^{-3}$ that is used to filter our \textsc{getsf} sources.
  }}
  \label{fig:cdto-dens}
\end{figure*}
The relation between both thresholds with respect to the mass (colour-code) present in the simulation domain of each run at \T{20} is displayed in Fig. \ref{fig:cdto-dens}.
To filter our \textsc{getsf} sources, we select a density threshold of $\rho_\mathrm{avg}>1\e{-19}\,\mathrm{g}\,\mathrm{cm}^{-3}$.
This is the highest average density for which a major fraction of the cell masses that adhere to our 3D core criterion ($A_{\mathrm{V,3D}}>8\,\mathrm{mag}$) are not ruled out.}

\section{Properties of detected \simcls{} and \simcos{}}
\begin{figure}
  \centering
  \includegraphics[width=\columnwidth]{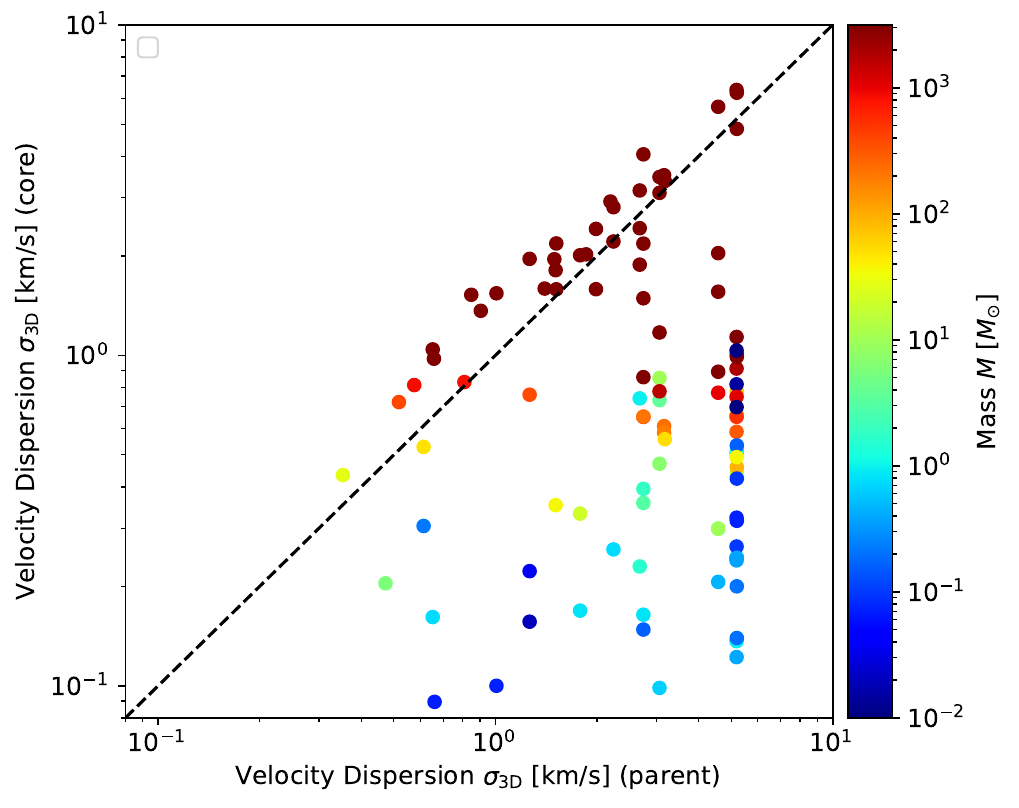}
  \caption{
    Velocity dispersions $\sigma_\mathrm{3D}$ of all \simcos{} vs. their containing clumps.
    The dashed diagonal line marks equality.
    The marker colours indicate the core mass.
    It is obvious that high-mass \simcos{} have much higher velocity dispersions than low-mass \simcos{}.
    In some cases, the core dispersion even exceeds the parent's velocity dispersion.
    The typical local sound speed is ${\sim 0.22\,{\rm km\, s}^{-1}}$.
    Thus, low-mass \simcos{} often have subsonic velocity dispersions.
  }
  \label{fig:vdisp-core-parent}
\end{figure}
\revb{
Figure \ref{fig:vdisp-core-parent} shows the 3D velocity dispersion $\sigma_\mathrm{3D}$ of each core vs. the corresponding value for its containing clump.
For the highest mass \simcos{} (${M>1000\,\mathrm{M}_\odot}$), $\sigma_\mathrm{3D}$ is in the same order than for the containing clump.
In contrast, for most low-mass \simcos{} (${M<3\,\mathrm{M}_\odot}$) $\sigma_\mathrm{3D}$ is considerably smaller than for the containing clump,
and often below the sound speed (typically ${\sim 0.22\,{\rm km\, s}^{-1}}$) inside the \simcos{}.}

\begin{figure}
  \centering
  \includegraphics[width=\columnwidth]{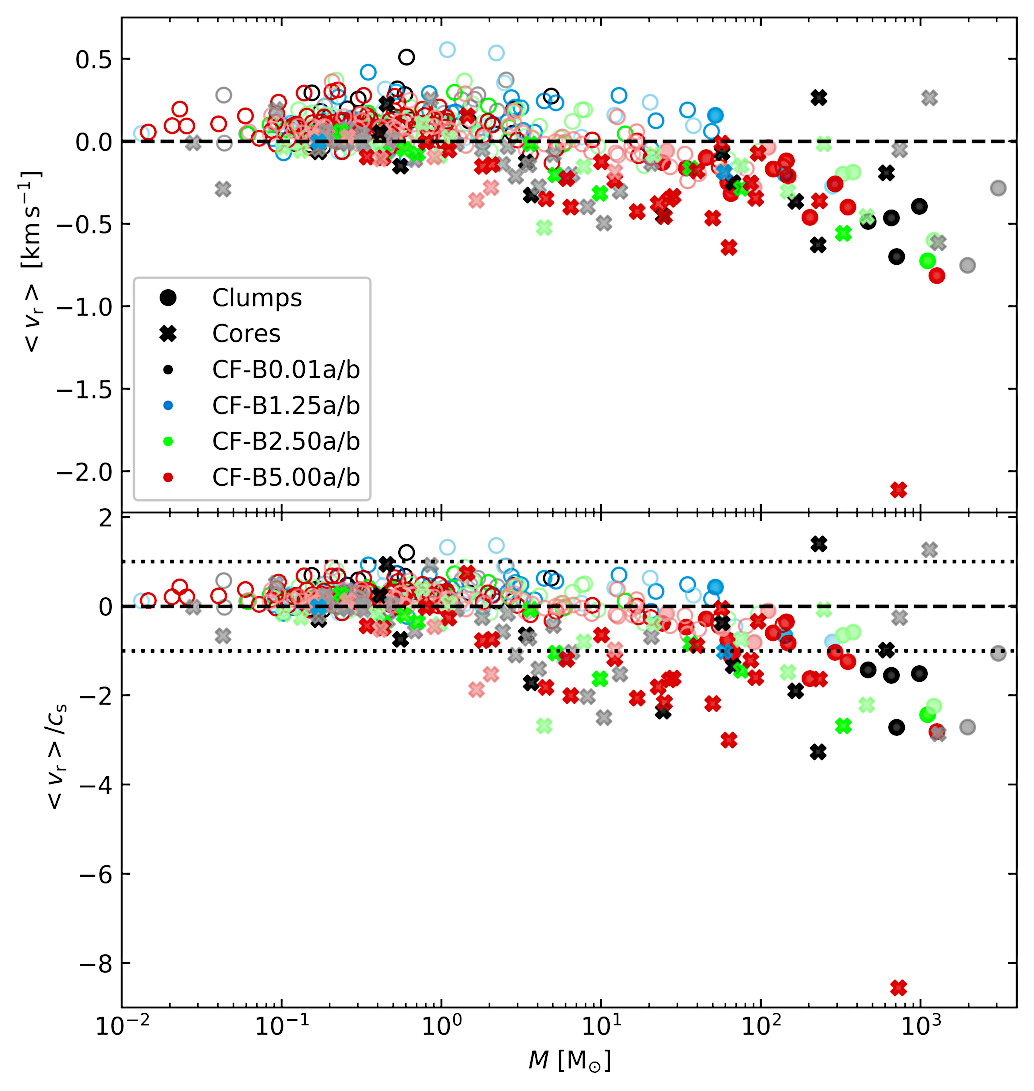}
  \caption{
  Top: Mass weighted average radial velocity $<v_\mathrm{r}>$ vs. object mass $M$.
  \revc{Bottom: Radial mach number $<v_\mathrm{r}>\,c_s^{-1}$ vs. object mass $M$.}
  Positive numbers mean velocities pointed away from the objects centre of mass, v.v..
  Most of the \simcos{} have negative radial velocities, indicating collapse.
  }
  \label{fig:vrad}
\end{figure}
Figure \ref{fig:vrad} (top) shows the mass weighted average radial velocity $<v_\mathrm{r}>$ vs. object mass $M$.
Most of the \simcos{} have high negative radial velocities, indicating collapse.
\revc{
Figure \ref{fig:vrad} (bottom) shows the radial mach number $<v_\mathrm{r}>\,c_s^{-1}$ vs. object mass $M$.
The local sound speed is calculated as ${c_s=\sqrt{\frac{5}{3}\frac{\left<p\right>_V}{\left<\rho\right>_V}}}$,
where ${\left<p\right>_V}$ is the volume weighted average pressure and ${\left<\rho\right>_V}$ is the volume weighted average density inside each object.
The high-mass \simcos{} have infall velocities of up to mach $3.5$.
}

\section{Virial Analysis}
\begin{figure}
  \centering
  \includegraphics[width=\columnwidth]{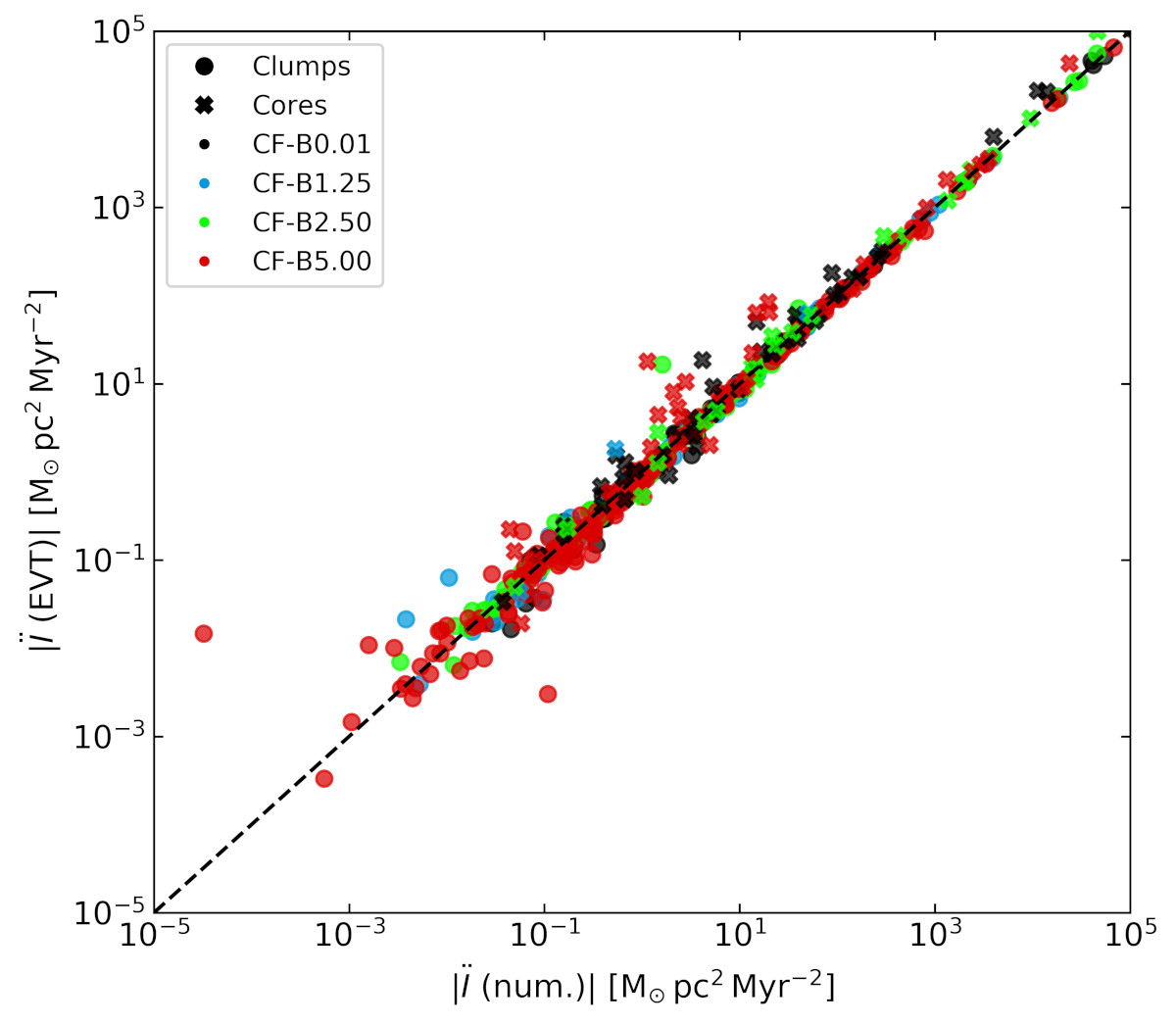}
  \caption{
  Comparison of clump (left) and \simcos{} (right) sum of virial terms, as calculated by \refeq{eq:EVT},
  vs. direct numerical derivation of $\ddot{I}$, each at \T{20}.
  }
  \label{fig:ddI-Check}
\end{figure}

To test the numerical validity of our virial analysis, we compare the temporal evolution of the moment of inertia $I_E$ of
the selected spaces containing the \simcls{} and \simcos{} in our simulations,
to that indicated by evaluation of the Eulerian virial theorem.
\reffig{fig:ddI-Check} shows a comparison of the value of $\ddot{I_E}$ determined by evaluation
of the aggregate of all virial terms represented by the \rhs of \refeq{eq:EVT} at \T{20}
to the analogous value determined by numerical double differentiation of $I$
via application of a finite difference scheme to three consecutive time-steps of simulation data
encompassing that point in simulation time.
For the vast majority of \simcls{}, it shows the virial analysis to predict the numerical value of $\ddot{I}$
with little deviation over several orders of magnitude.

\section{Balance of volume- and surface terms}
\begin{figure*}
  \centering
  \includegraphics[width=\textwidth]{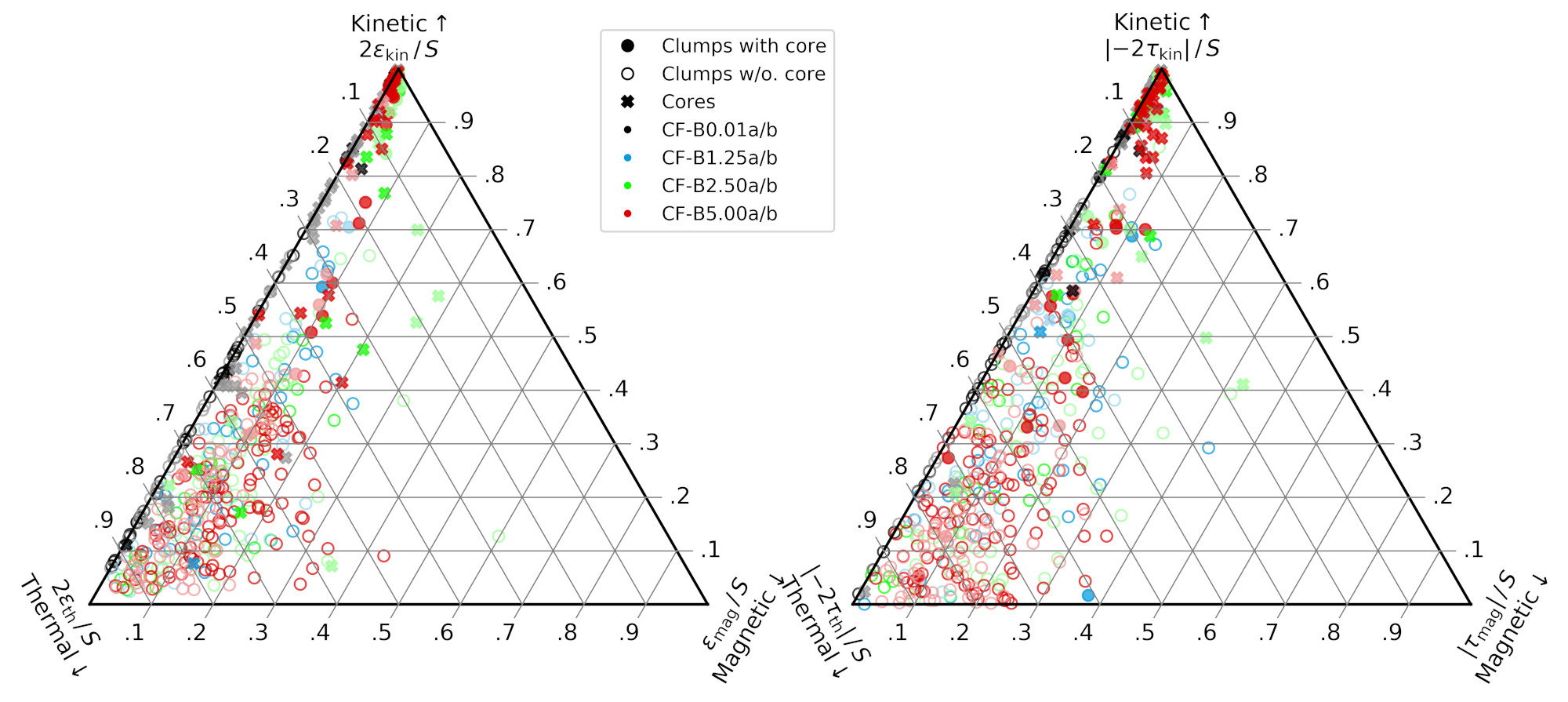}
  \caption{\rev{
  Ternary clump (circles) and core object (crosses) balance of absolute volume energy terms (top)
  and surface terms (bottom) at \T{20}.
  }}
  \label{fig:CO-ter-VS}
\end{figure*}
While all of the components of the volume energy terms are strictly positive (\reffig{fig:CO-ter-VS}),
the surface terms can be both negative or positive, depending on the configuration of the respective clump.
Furthermore, the volume energy of some of the \simcos{} to be largely dominated by kinetic energy,
whereas the other \simcos{} are distributed over all sectors of the ternary, indicating none of the terms to be negligible.
The surface term composition of the \simcos{} is distinct to that of the volume energy,
in that the thermal surface term is only an inferior part of the surface term for most of the \simcls{},
hinting at most of the \simcos{} not being pressure confined by their surroundings.

\begin{figure}
  \centering
  \includegraphics[width=\columnwidth]{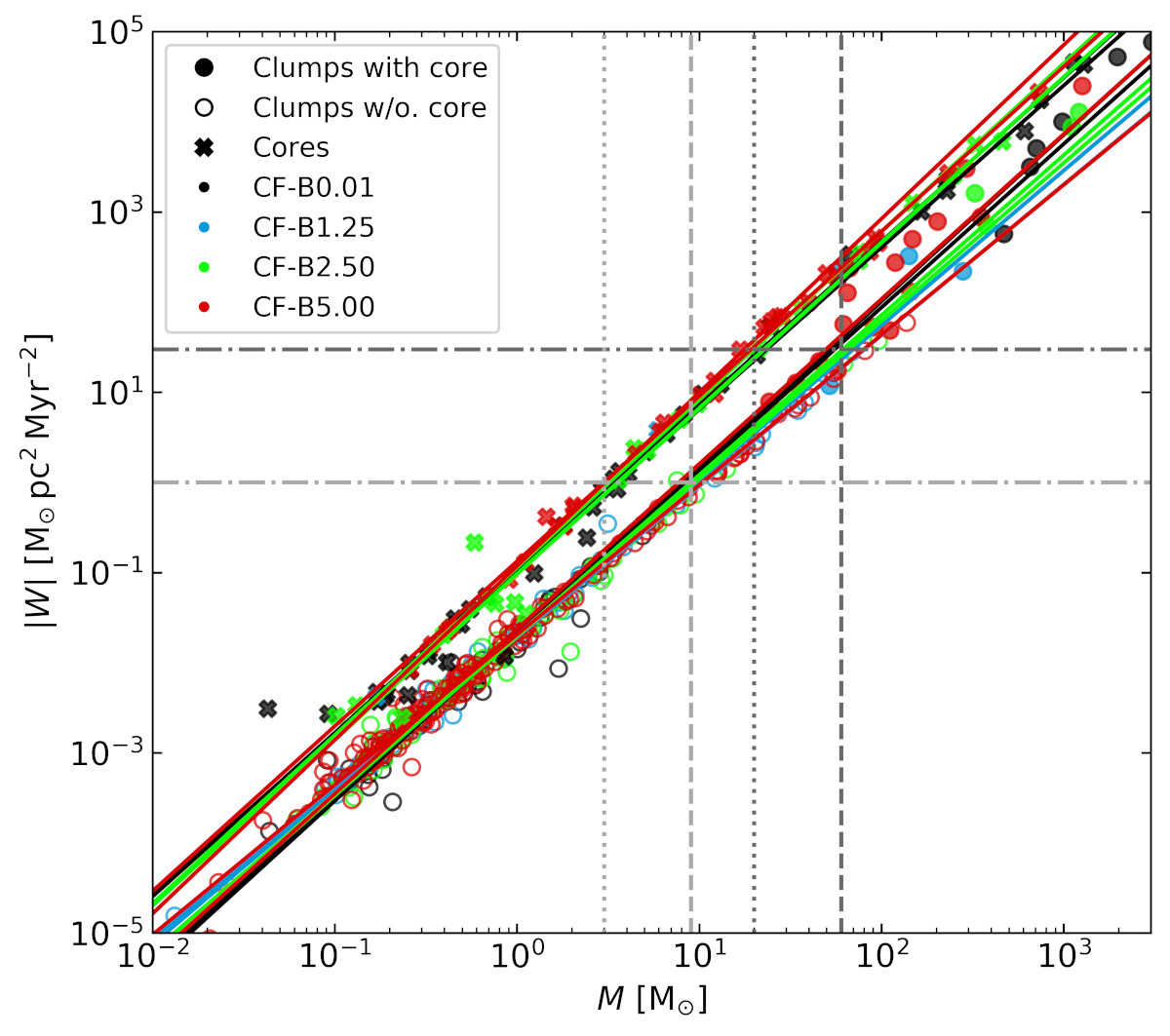}
  \caption{
  Gravitational binding energy $W$ vs. object mass $M$.
  A small number of objects satisfy $W>0$ because of tidal forces overwhelming the gravitational potential caused by the considered objects mass,
  and are not represented in the diagram.
  }
  \label{fig:mass-W}
\end{figure}
Fig. \ref{fig:mass-W} shows the relation between the mass $M$ and the gravitational binding energy $W$ of the \simcos{}.
For the \simcos{} and the coreless \simcls{}, there is a power-law like correlation between $W$ and $M$, respectively, albeit differing in coefficient.

\section{Mass accretion of tracked \simcos{}}
\begin{figure}
  \centering
  \includegraphics[width=\columnwidth]{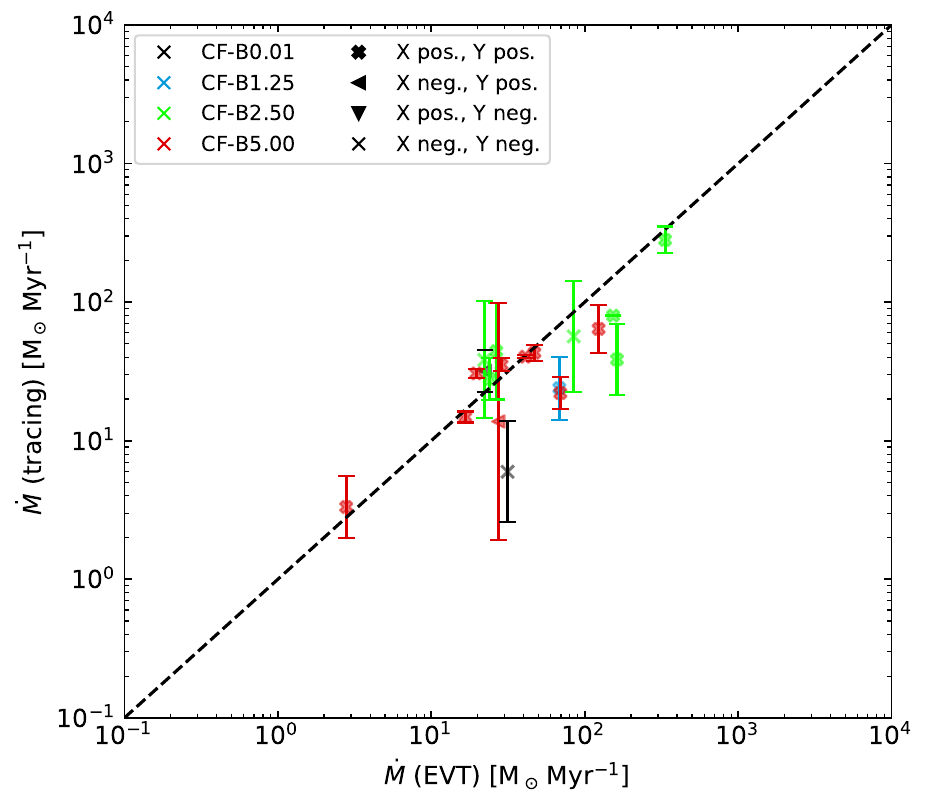}
  \caption{
  Comparison of the mass accretion rate, ${\dot{M}}$, of trackable \simcos{} at \T{20},
  as obtained from object tracing (see Fig.~\ref{fig:TimeDep/Traces}) vs. from the EVT derivation.
  We include a one-to-one line (dashed line) to guide the eye.
  }
  \label{fig:md-comparison}
\end{figure}

Fig. \ref{fig:md-comparison} shows a comparison of the mass accretion rates $\dot{M}$ of trackable \simcos{}, as obtained from the mass evolution, versus measured in the Eulerian frame for the same objects, at \T{20}.
For most of the trackable \simcos{}, both values are of similar size, with a tendency to overestimate $\dot{M}$ when using the value from the Eulerian frame.

\begin{figure}
  \centering
  \includegraphics[width=\columnwidth]{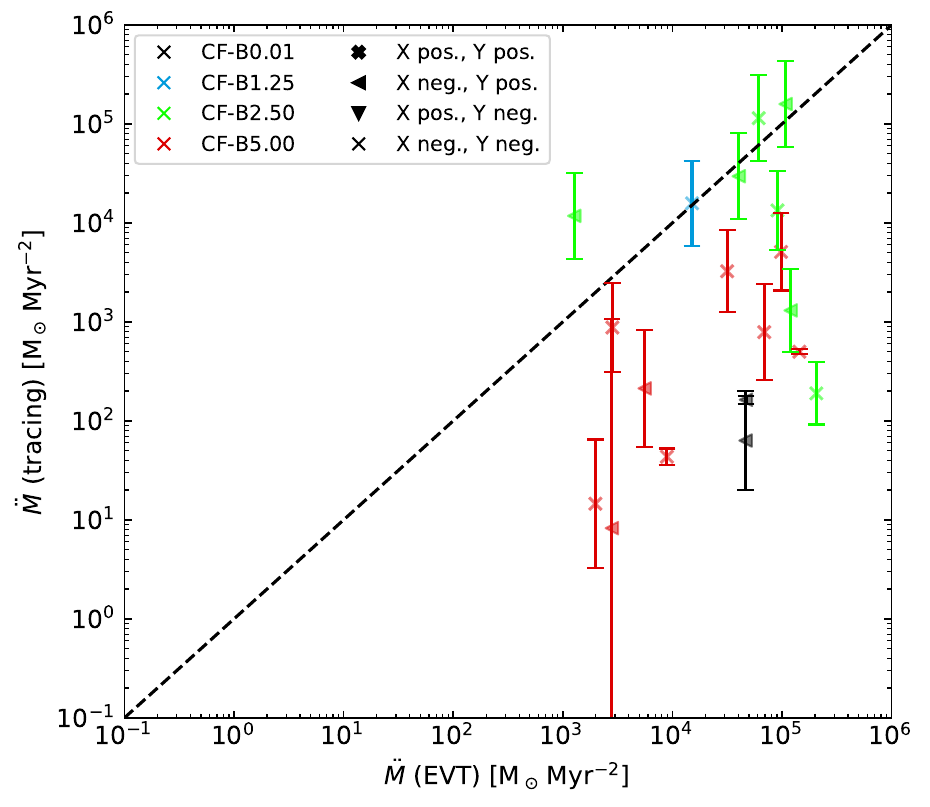}
  \caption{
  Comparison of rate of change of the mass accretion rate, ${\ddot{M}}$, of trackable \simcos{} at \T{20},
  as obtained from object tracing (see Fig.~\ref{fig:TimeDep/Traces}) vs. from the EVT derivation.
  }
  \label{fig:mdd-comparison}
\end{figure}

Fig. \ref{fig:mdd-comparison} shows a comparison of the rates of change of the mass accretion rate of trackable \simcos{}, as obtained from the mass evolution by object tracking, versus measured in the Eulerian frame, at \T{20}.
We note that while only the absolute values are displayed in the Figure, the shape of the symbol indicates the sign of both the Eulerian value (\dir{x}-axis) and the mass evolution value (\dir{x}-axis).
The values are largely uncorrelated, with even the sign disagreeing for some \simcos{}.

\section{Gallery of \simcos{}}
\begin{figure}
  \centering
  \includegraphics[width=\columnwidth]{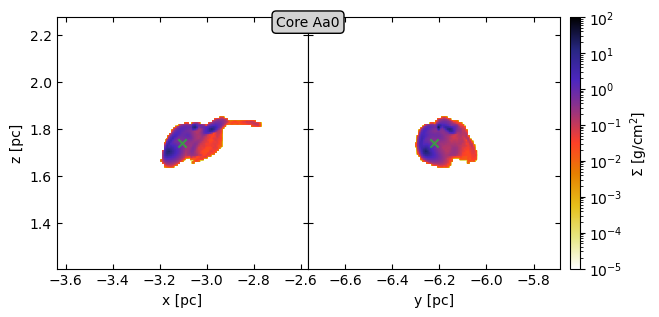}
  \includegraphics[width=\columnwidth]{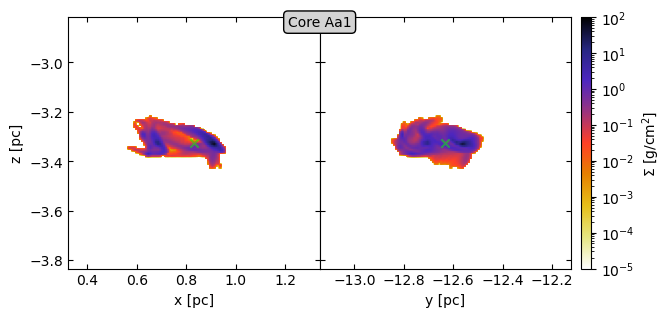}
  \includegraphics[width=\columnwidth]{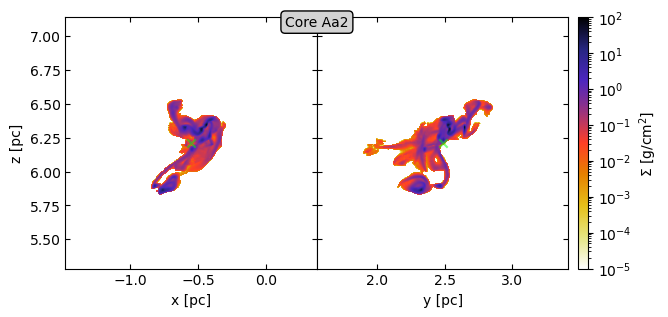}
  \includegraphics[width=\columnwidth]{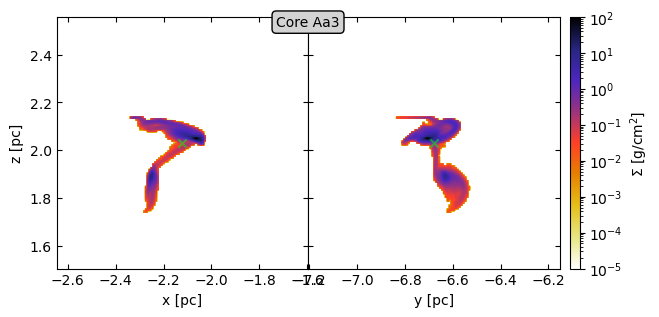}
  \includegraphics[width=\columnwidth]{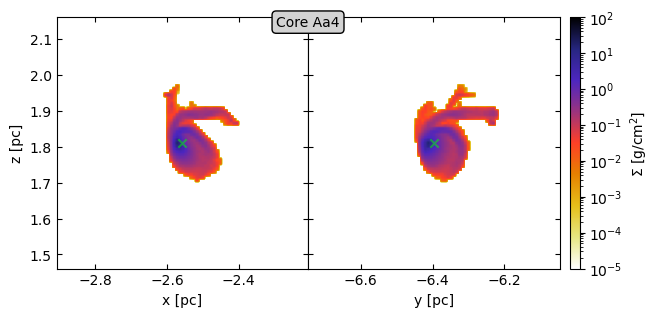}
  \caption{
    Projection of \simcos{} Aa0--Aa4, Simulation \CF{0.01}a. The green marker indicates the position of the core's centre of mass.
  }
  \label{fig:cores-A0}
\end{figure}

\begin{figure}
  \centering
  \includegraphics[width=\columnwidth]{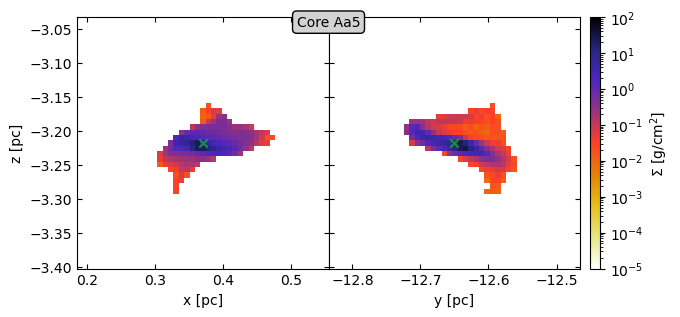}
  \includegraphics[width=\columnwidth]{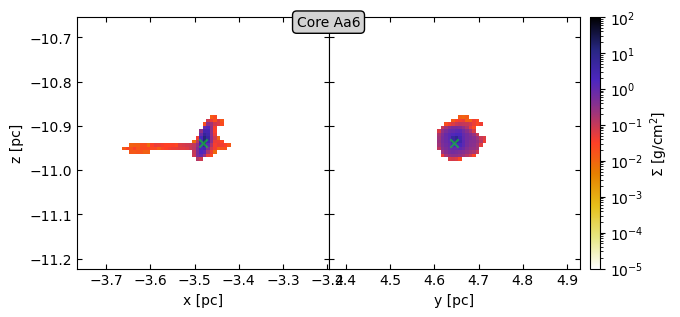}
  \includegraphics[width=\columnwidth]{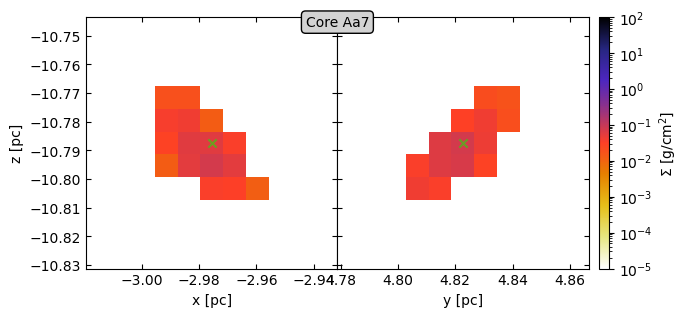}
  \includegraphics[width=\columnwidth]{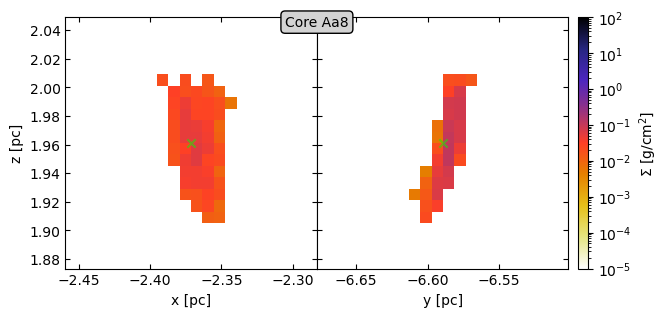}
  \includegraphics[width=\columnwidth]{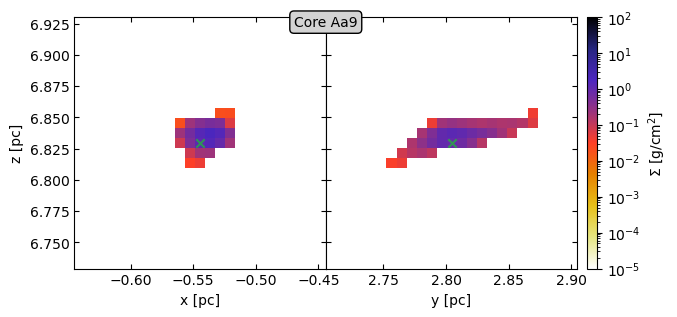}
  \caption{
    Projection of \simcos{} Aa5--Aa9, Simulation \CF{0.01}a. The green marker indicates the position of the core's centre of mass.
  }
  \label{fig:cores-A5}
\end{figure}

\begin{figure}
  \centering
  \includegraphics[width=\columnwidth]{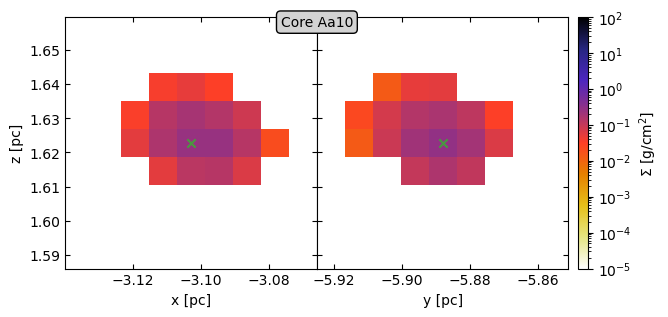}
  \includegraphics[width=\columnwidth]{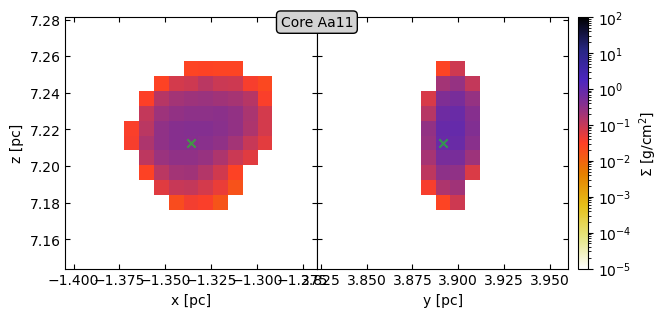}
  \includegraphics[width=\columnwidth]{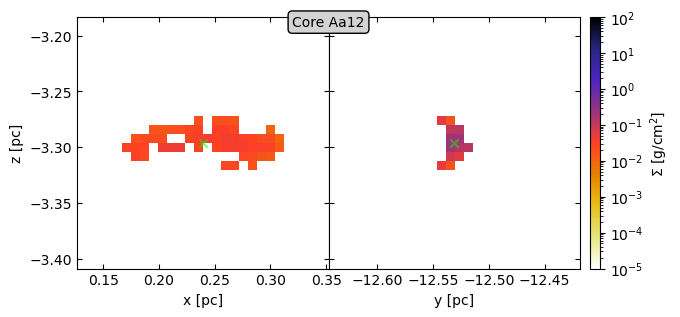}
  \caption{
    Projection of \simcos{} Aa10--Aa12, Simulation \CF{0.01}a. The green marker indicates the position of the core's centre of mass.
  }
  \label{fig:cores-A10}
\end{figure}

\begin{figure}
  \centering
  \includegraphics[width=\columnwidth]{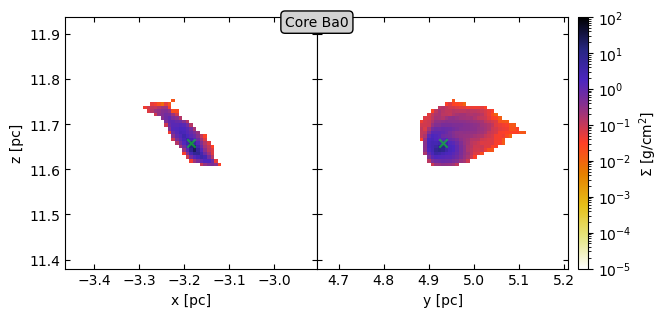}
  \includegraphics[width=\columnwidth]{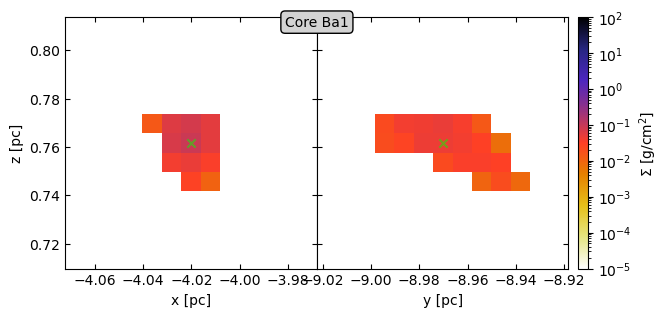}
  \caption{
    Projection of \simcos{} Ba0--Ba1, Simulation \CF{1.25}a. The green marker indicates the position of the core's centre of mass.
  }
  \label{fig:cores-B0}
\end{figure}

\begin{figure}
  \centering
  \includegraphics[width=\columnwidth]{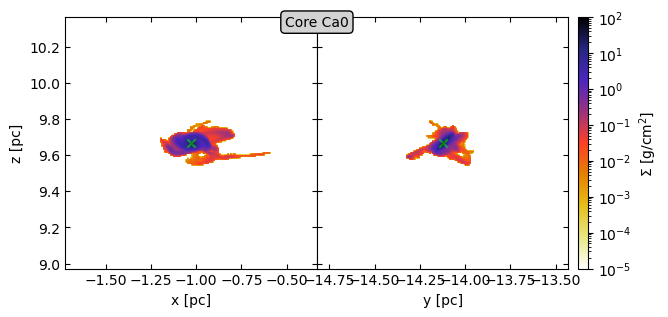}
  \includegraphics[width=\columnwidth]{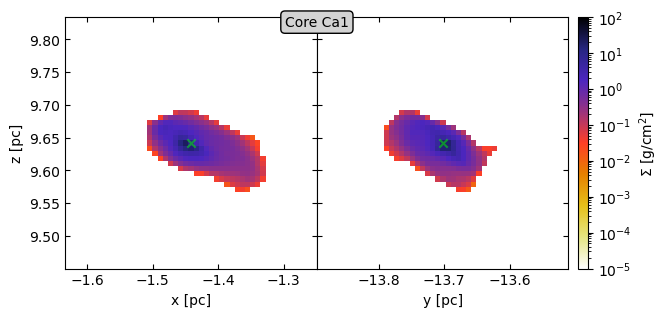}
  \includegraphics[width=\columnwidth]{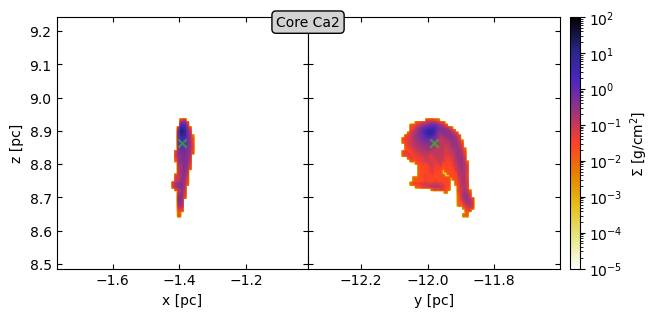}
  \includegraphics[width=\columnwidth]{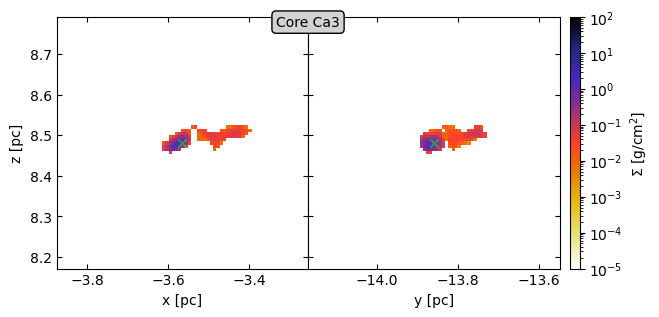}
  \includegraphics[width=\columnwidth]{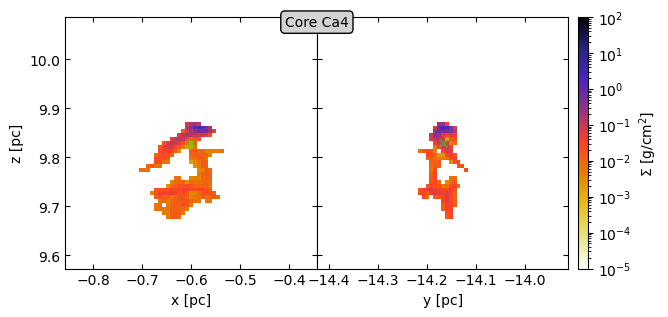}
  \caption{
    Projection of \simcos{} Ca0--Ca4, Simulation \CF{2.50}a. The green marker indicates the position of the core's centre of mass.
  }
  \label{fig:cores-C0}
\end{figure}

\begin{figure}
  \centering
  \includegraphics[width=\columnwidth]{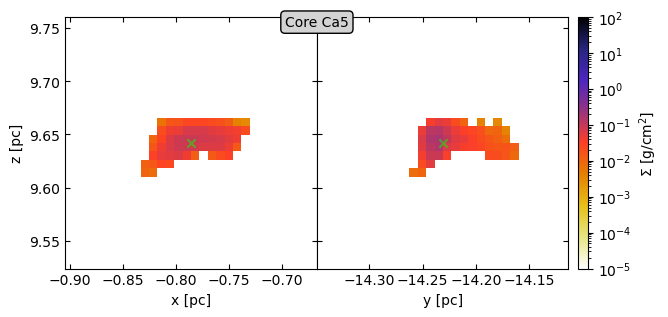}
  \includegraphics[width=\columnwidth]{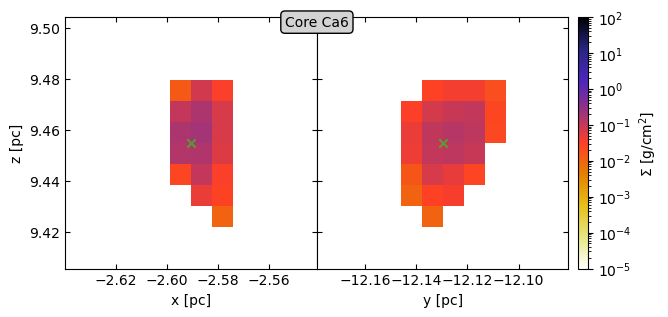}
  \includegraphics[width=\columnwidth]{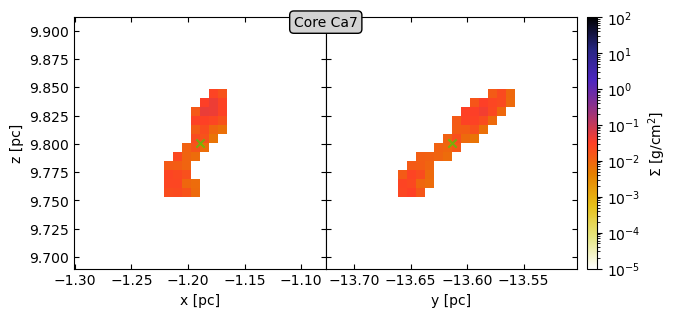}
  \includegraphics[width=\columnwidth]{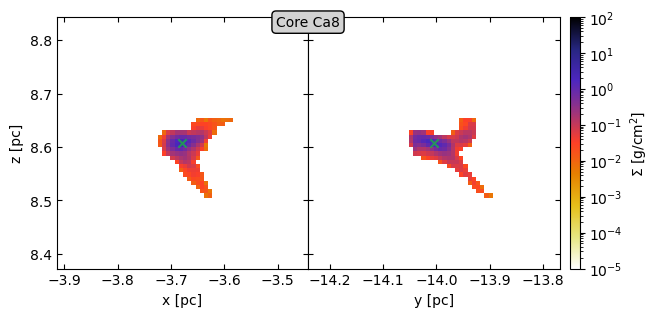}
  \includegraphics[width=\columnwidth]{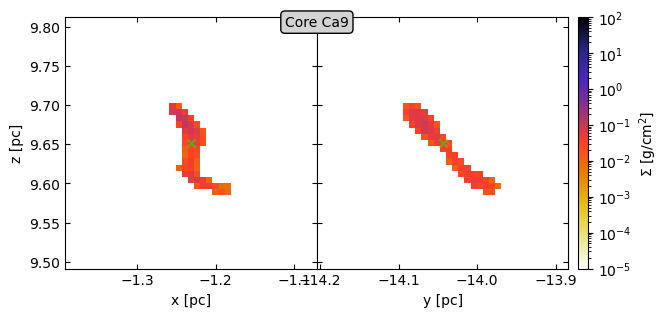}
  \caption{
    Projection of \simcos{} Ca5--Ca9, Simulation \CF{2.50}a. The green marker indicates the position of the core's centre of mass.
  }
  \label{fig:cores-C5}
\end{figure}

\begin{figure}
  \centering
  \includegraphics[width=\columnwidth]{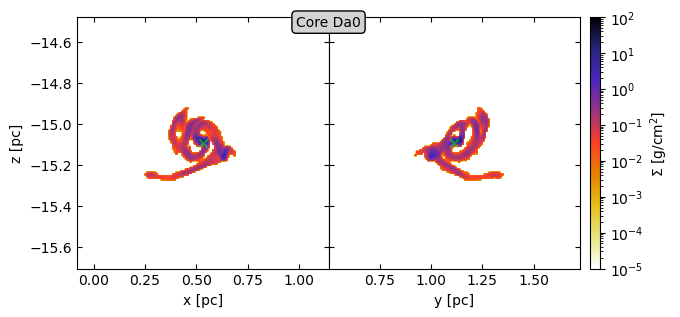}
  \includegraphics[width=\columnwidth]{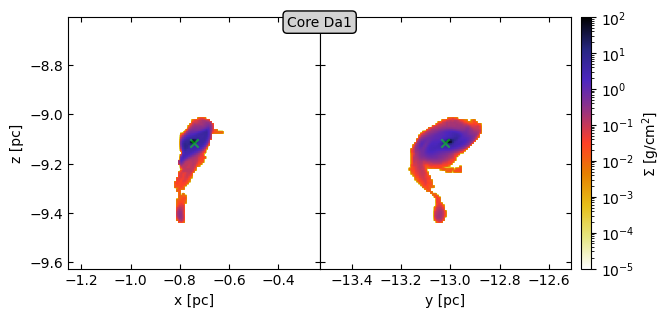}
  \includegraphics[width=\columnwidth]{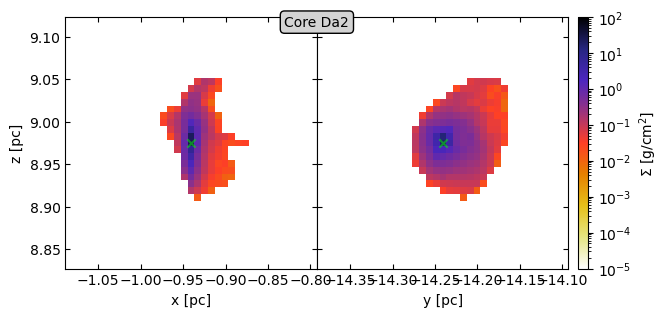}
  \includegraphics[width=\columnwidth]{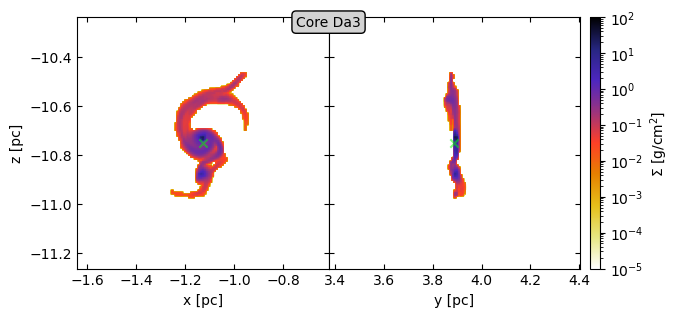}
  \includegraphics[width=\columnwidth]{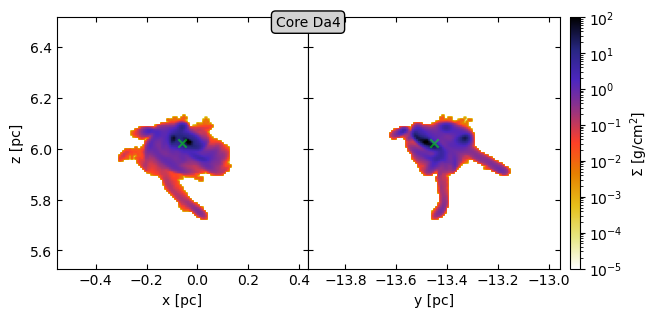}
  \caption{
    Projection of \simcos{} Da0--Da4, Simulation \CF{5.00}a. The green marker indicates the position of the core's centre of mass.
  }
  \label{fig:cores-D0}
\end{figure}

\begin{figure}
  \centering
  \includegraphics[width=\columnwidth]{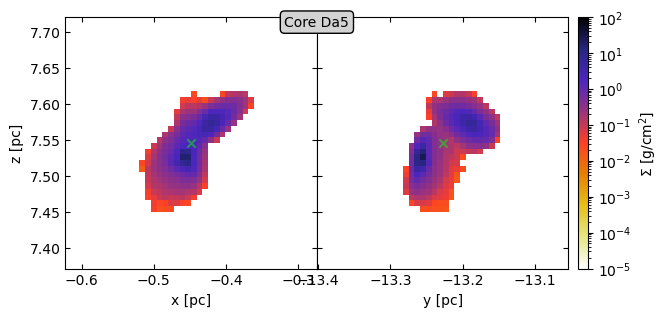}
  \includegraphics[width=\columnwidth]{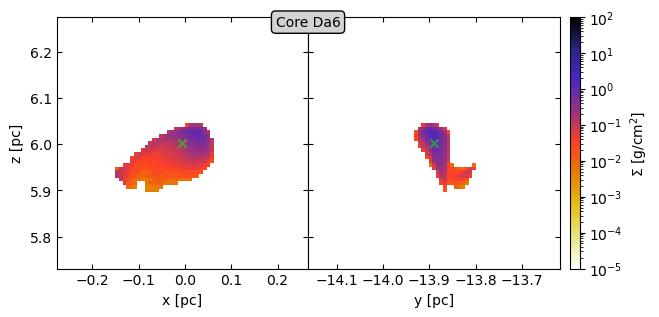}
  \includegraphics[width=\columnwidth]{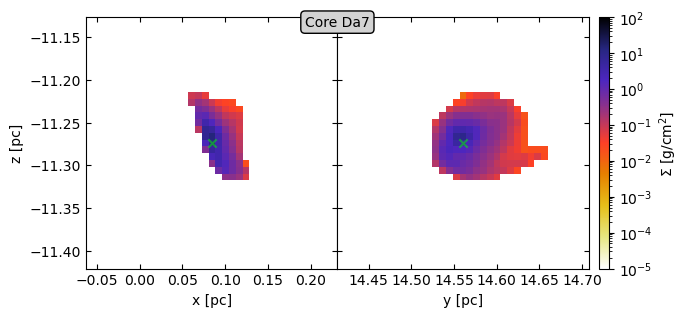}
  \includegraphics[width=\columnwidth]{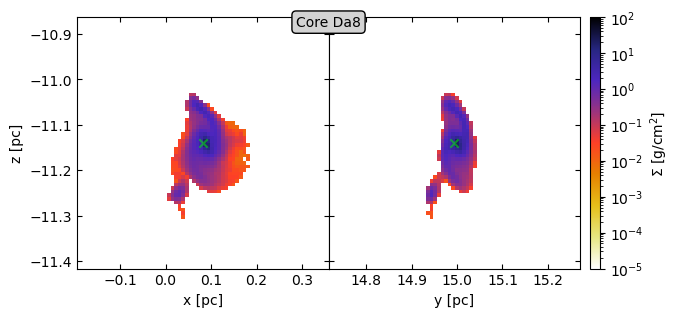}
  \includegraphics[width=\columnwidth]{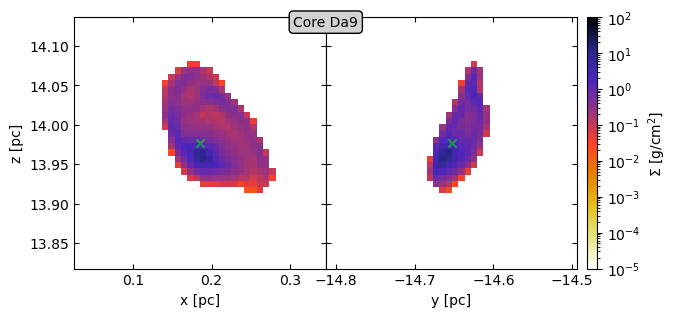}
  \caption{
    Projection of \simcos{} Da5--Da9, Simulation \CF{5.00}a. The green marker indicates the position of the core's centre of mass.
  }
  \label{fig:cores-D5}
\end{figure}

\begin{figure}
  \centering
  \includegraphics[width=\columnwidth]{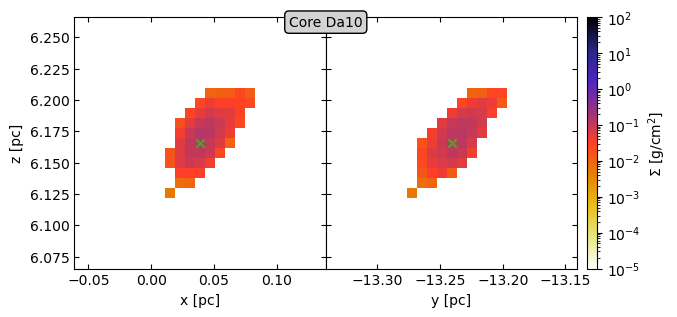}
  \includegraphics[width=\columnwidth]{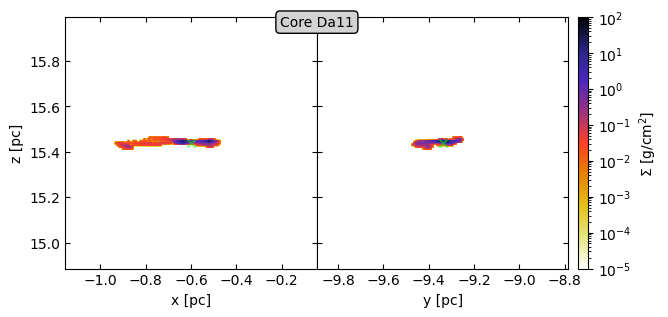}
  \includegraphics[width=\columnwidth]{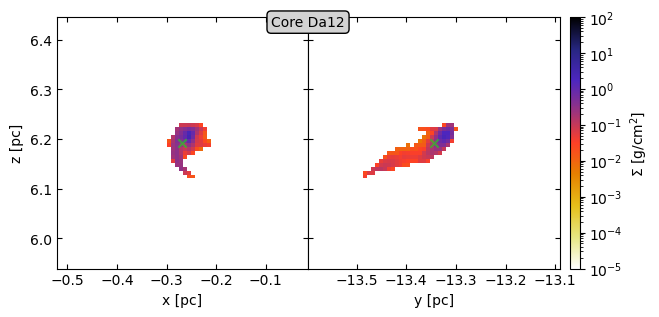}
  \includegraphics[width=\columnwidth]{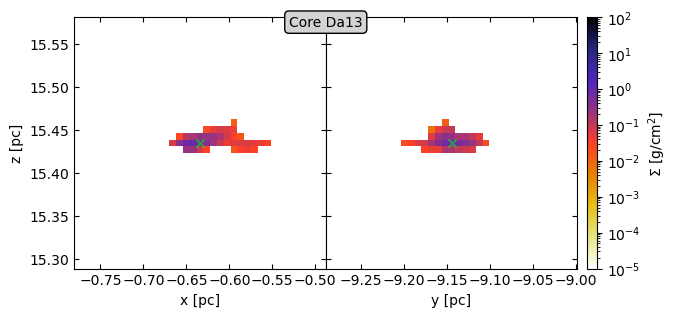}
  \includegraphics[width=\columnwidth]{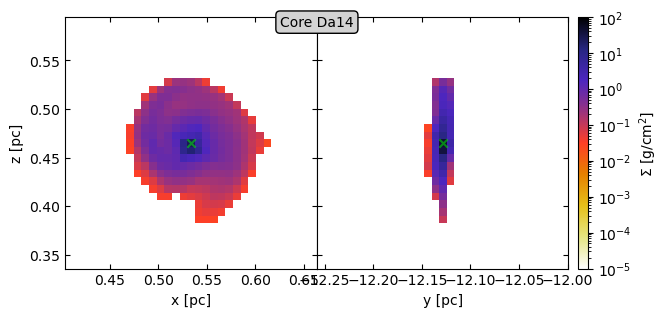}
  \caption{
    Projection of \simcos{} Da10--Da14, Simulation \CF{5.00}a. The green marker indicates the position of the core's centre of mass.
  }
  \label{fig:cores-D10}
\end{figure}

\begin{figure}
  \centering
  \includegraphics[width=\columnwidth]{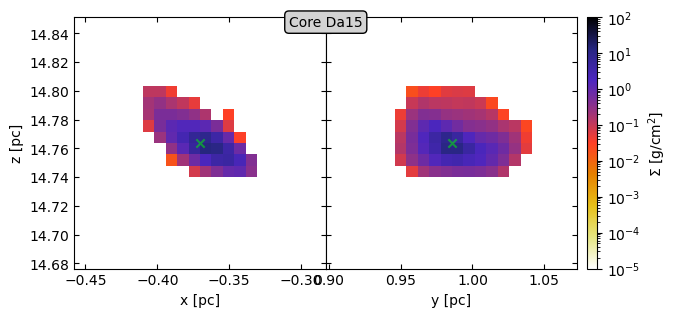}
  \includegraphics[width=\columnwidth]{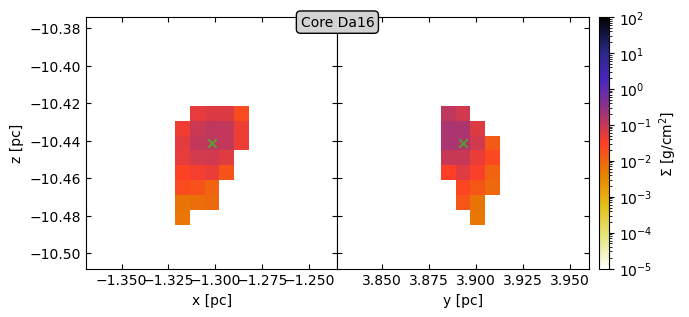}
  \includegraphics[width=\columnwidth]{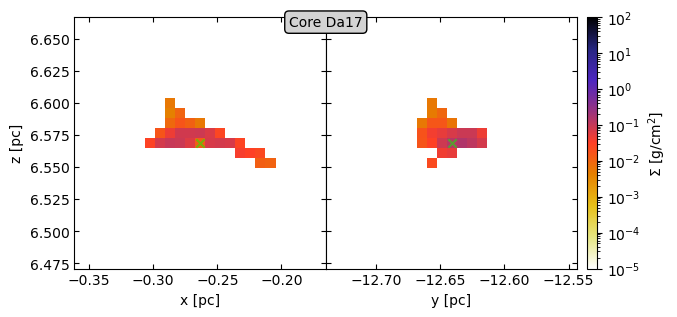}
  \includegraphics[width=\columnwidth]{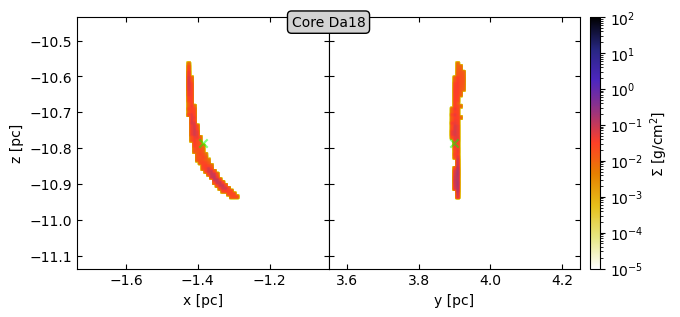}
  \includegraphics[width=\columnwidth]{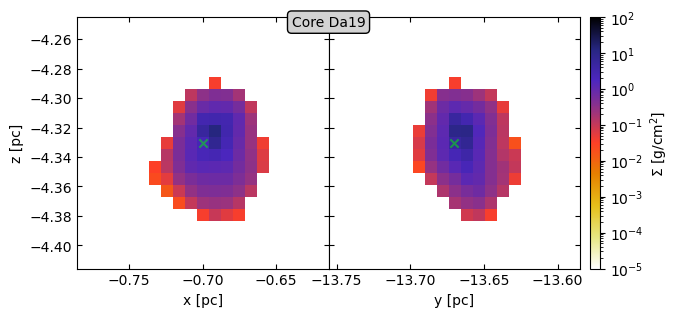}
  \caption{
    Projection of \simcos{} Da15--Da19, Simulation \CF{5.00}a. The green marker indicates the position of the core's centre of mass.
  }
  \label{fig:cores-D15}
\end{figure}

\begin{figure}
  \centering
  \includegraphics[width=\columnwidth]{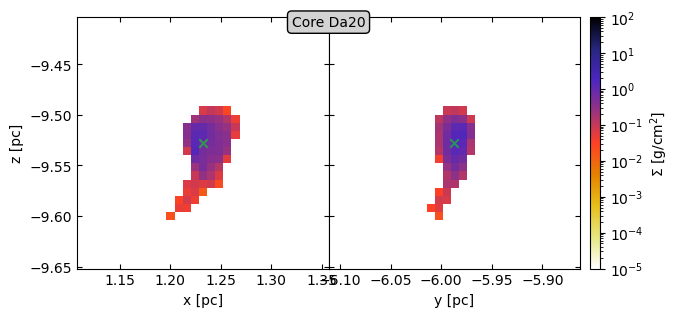}
  \includegraphics[width=\columnwidth]{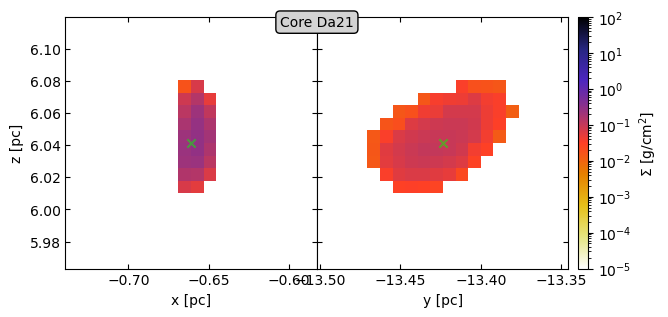}
  \includegraphics[width=\columnwidth]{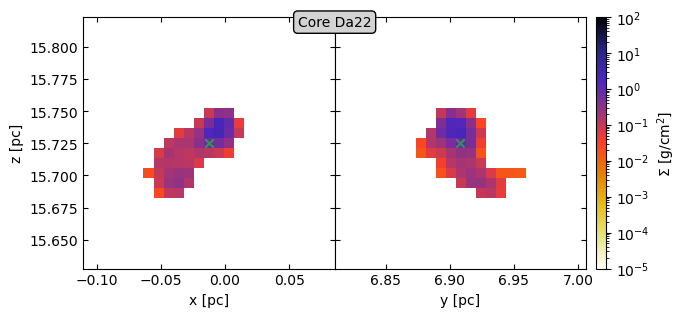}
  \includegraphics[width=\columnwidth]{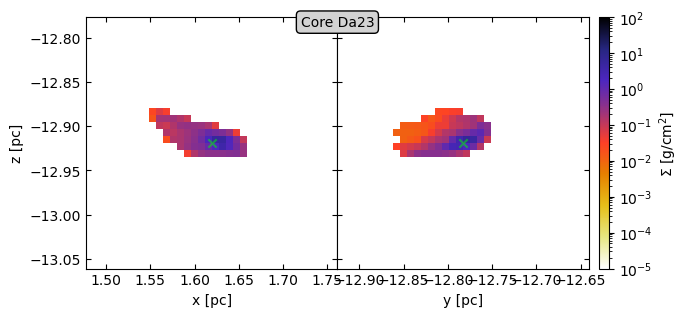}
  \includegraphics[width=\columnwidth]{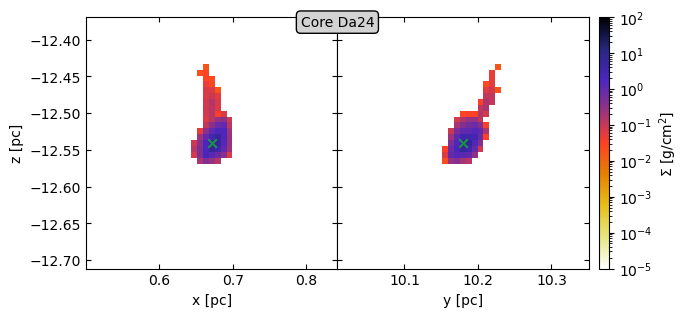}
  \caption{
    Projection of \simcos{} Da20--Da24, Simulation \CF{5.00}a. The green marker indicates the position of the core's centre of mass.
  }
  \label{fig:cores-D20}
\end{figure}

Fig.~\ref{fig:cores-A0}--\ref{fig:cores-D20} show \revc{surface density projections} of the \simcos{}.
The slice directions are perpendicular to the \dir{y} (left) and the \dir{z} directions.
The black outlines mark the in-slice cross sections of the \simcos{}.


\begin{figure}
  \centering
  \includegraphics[width=\columnwidth]{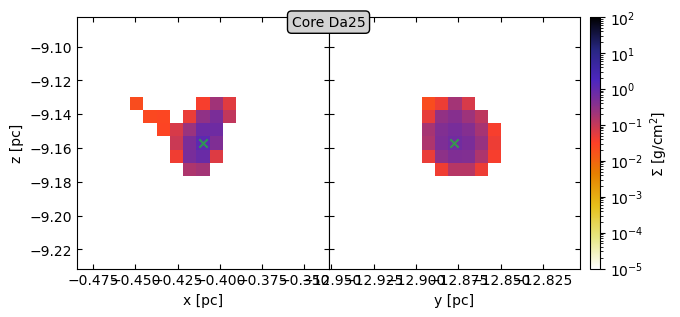}
  \caption{
    Projection of 3D core Da25, Simulation \CF{5.00}a. The green marker indicates the position of the core's centre of mass.
  }
  \label{fig:cores-D25}
\end{figure}


\bsp	
\label{lastpage}
\end{document}